\documentstyle[12pt,eqsection,indent,subeqnarray,epsf]{article}

\begin{document}

\font\sevenrm  = cmr7
\def\fancyplus{\hbox{+\kern-6.65pt\lower3.2pt\hbox{\sevenrm --}%
\kern-4pt\raise4.6pt\hbox{\sevenrm --}%
\kern-9pt\raise0.65pt\hbox{$\tiny\vdash$}%
\kern-2pt\raise0.65pt\hbox{$\tiny\dashv$}}}
\def\fancycross{\hbox{$\times$\kern-9.8pt\raise3.6pt\hbox{$\tiny \times$}%
\kern-1.2pt\raise3.6pt\hbox{$\tiny \times$}%
\kern-10.5pt\lower1.0pt\hbox{$\tiny \times$}%
\kern-1.2pt\lower1.0pt\hbox{$\tiny \times$}}}
\def\fancysquare{%
\hbox{$\small\Box$\kern-9.6pt\raise7.1pt\hbox{$\vpt\backslash$}%
\kern+3.9pt\raise7.1pt\hbox{$\vpt /$}%
\kern-11.5pt\lower3.0pt\hbox{$\vpt /$}%
\kern+3.9pt\lower3.0pt\hbox{$\vpt\backslash$}}}
\def\fancydiamond{\hbox{$\diamond$\kern-4.25pt\lower3.8pt\hbox{$\vpt\vert$}%
\kern-2.25pt\raise7pt\hbox{$\vpt\vert$}%
\kern-7.05pt\raise1.57pt\hbox{\vpt --}%
\kern+5.25pt\raise1.57pt\hbox{\vpt --}}}

\def\scrmtens{{\stackrel{\leftrightarrow}{\cal M}_T}}
\def\ttens{{\stackrel{\leftrightarrow}{T}}}
\def\scro{{\cal O}}
\def\smfrac#1#2{{\textstyle\frac{#1}{#2}}}
\def\smhalf{ {\smfrac{1}{2}} }

\def\btau{\mbox{\protect\boldmath $\tau$}}

\def\bn{{\bf n}}
\def\br{{\bf r}}
\def\bz{{\bf z}}
\def\bv{{\bf v}}
\def\bh{\mbox{\protect\boldmath $h$}}

\newcommand{\imag}{\mathop{\rm Im}\nolimits}
\newcommand{\real}{\mathop{\rm Re}\nolimits}

\def\vs{\vec{\mbox{$s$}}}
\def\vsigma{\vec{\mbox{$\sigma$}}}
\def\vtau{\vec{\mbox{$\tau$}}}
\def\va{\vec{\mbox{$a$}}}
\def\vb{\vec{\mbox{$b$}}}
\def\vJ{\vec{\mbox{$J$}}}

\newcommand{\taum}{\tau_{int,\,\vec{\cal M}}}
\newcommand{\taux}{\tau_{int,\,{\cal M}_V^2}}
\newcommand{\tauxT}{\tau_{int,\,{\cal M}_T^2}}
\newcommand{\tauA}{\tau_{int,\,A}}
\newcommand{\taue}{\tau_{int,\,{\cal E}}}
\newcommand{\taueV}{\tau_{int,\,{\cal E}_V}}
\newcommand{\taueT}{\tau_{int,\,{\cal E}_T}}
\newcommand{\taudele}{\tau_{int,\,({\cal E}-\overline{E})^2}}

\newcommand{\tauxexp}{\tau_{exp,{\cal M}_V^2}}

\newcommand{\plotdot}{\makebox(0,0){$\bullet$}}
\newcommand{\plotcross}{\makebox(0,0){{\Large $\times$}}}

\newcommand{\plota}{\makebox(0,0){$\circ$}}      
\newcommand{\plotb}{\makebox(0,0){$\star$}}      
\newcommand{\plotc}{\makebox(0,0){$\bullet$}}    
\newcommand{\plotd}{\makebox(0,0){{\scriptsize $+$}}}       
\newcommand{\plote}{\makebox(0,0){{\scriptsize $\times$}}}  
\newcommand{\plotf}{\makebox(0,0){$\ast$}}       

\newcommand{\plotA}{\makebox(0,0){$\triangleleft$}}   
\newcommand{\plotB}{\makebox(0,0){$\triangleright$}}  
\newcommand{\plotC}{\makebox(0,0){$\diamond$}}        
\newcommand{\plotD}{\makebox(0,0){{\scriptsize $\oplus$}}} 
\newcommand{\plotE}{\makebox(0,0){{\scriptsize $\otimes$}}}
\newcommand{\plotF}{\makebox(0,0){{\scriptsize $\ominus$}}}

\def\reff#1{(\ref{#1})}
\newcommand{\csd}{critical slowing-down}
\newcommand{\be}{\begin{equation}}
\newcommand{\ee}{\end{equation}}
\newcommand{\<}{\langle}
\renewcommand{\>}{\rangle}
\newcommand{\half}{ {{1 \over 2 }}}
\newcommand{\quarter}{ {{1 \over 4 }}}
\newcommand{\fourth}{\quarter}
\newcommand{\eighth}{ {{1 \over 8 }}}
\newcommand{\sixteenth}{ {{1 \over 16 }}}
\def\var{ \hbox{var} }
\newcommand{\HB}{ {\hbox{{\scriptsize\em HB}\/}} }
\newcommand{\MGMC}{ {\hbox{{\scriptsize\em MGMC}\/}} }
\newcommand{\gtilde}{ {\widetilde{G}} }
\newcommand{\longto}{\longrightarrow}

\def\spose#1{\hbox to 0pt{#1\hss}}
\def\ltapprox{\mathrel{\spose{\lower 3pt\hbox{$\mathchar"218$}}
 \raise 2.0pt\hbox{$\mathchar"13C$}}}
\def\gtapprox{\mathrel{\spose{\lower 3pt\hbox{$\mathchar"218$}}
 \raise 2.0pt\hbox{$\mathchar"13E$}}}

\newcommand{\scra}{{\cal A}}
\newcommand{\scrb}{{\cal B}}
\newcommand{\scrc}{{\cal C}}
\newcommand{\scrd}{{\cal D}}
\newcommand{\scre}{{\cal E}}
\newcommand{\scrf}{{\cal F}}
\newcommand{\scrg}{{\cal G}}
\newcommand{\scrh}{{\cal H}}
\newcommand{\scrk}{{\cal K}}
\newcommand{\scrl}{{\cal L}}
\newcommand{\scrm}{{\cal M}}
\newcommand{\scrmvec}{\vec{\cal M}}
\newcommand{\scrp}{{\cal P}}
\newcommand{\scrr}{{\cal R}}
\newcommand{\scrs}{{\cal S}}
\newcommand{\scru}{{\cal U}}

\def\bsigma{\mbox{\protect\boldmath $\sigma$}}
\def\bpi{\mbox{\protect\boldmath $\pi$}}
\def\bs{ {\bf s} }
\newcommand{\re}{\mathop{\rm Re}\nolimits}
\newcommand{\im}{\mathop{\rm Im}\nolimits}
\newcommand{\tr}{\mathop{\rm tr}\nolimits}
\newcommand{\CP}{ \hbox{\it CP\/} }
\def\T{{\rm T}}

\def\msbar{ {\overline{\hbox{\scriptsize MS}}} }
\def\normalmsbar{ {\overline{\hbox{\normalsize MS}}} }

\def\eff{ {\hbox{\scriptsize\em eff}} }
\def\HMN{ {\hbox{\scriptsize HMN}} }

\font\specialroman=msym10 scaled\magstep1  
\font\sevenspecialroman=msym7              
\def\zed{\hbox{\specialroman Z}}
\def\szed{\hbox{\sevenspecialroman Z}}
\def\R{\hbox{\specialroman R}}
\def\sR{\hbox{\sevenspecialroman R}}
\def\Z{\hbox{\specialroman Z}}
\def\N{\hbox{\specialroman N}}
\def\C{\hbox{\specialroman C}}
\renewcommand{\emptyset}{\hbox{\specialroman ?}}

\font\german=eufm10 scaled\magstep1     
\def\germang{\hbox{\german g}}
\def\germansu{\hbox{\german su}}
\def\germanso{\hbox{\german so}}

%
%
\newenvironment{sarray}{
          \textfont0=\scriptfont0
          \scriptfont0=\scriptscriptfont0
          \textfont1=\scriptfont1
          \scriptfont1=\scriptscriptfont1
          \textfont2=\scriptfont2
          \scriptfont2=\scriptscriptfont2
          \textfont3=\scriptfont3
          \scriptfont3=\scriptscriptfont3
        \renewcommand{\arraystretch}{0.7}
        \begin{array}{l}}{\end{array}}

\newenvironment{scarray}{
          \textfont0=\scriptfont0
          \scriptfont0=\scriptscriptfont0
          \textfont1=\scriptfont1
          \scriptfont1=\scriptscriptfont1
          \textfont2=\scriptfont2
          \scriptfont2=\scriptscriptfont2
          \textfont3=\scriptfont3
          \scriptfont3=\scriptscriptfont3
        \renewcommand{\arraystretch}{0.7}
        \begin{array}{c}}{\end{array}}

\title{\vspace*{-2cm} Multi-Grid Monte Carlo 
         via $XY$ Embedding  \\[4mm]
         I. General Theory and 
          Two-Dimensional $O(N)$-Symmetric Nonlinear $\sigma$-Models
      }
\author{
  \\[-0.5cm]
  \small  Tereza Mendes               \\[-0.2cm]
  \small\it Department of Physics     \\[-0.2cm]
  \small\it New York University       \\[-0.2cm]
  \small\it 4 Washington Place        \\[-0.2cm]
  \small\it New York, NY 10003 USA    \\[-0.2cm]
  \small {\tt MENDES@MAFALDA.PHYSICS.NYU.EDU} \\[-0.2cm]
  {\protect\makebox[5in]{\quad}}  
  \\[-0.35cm] \and
  {\small Andrea Pelissetto }        \\[-0.2cm]
  {\small\it Dipartimento di Fisica}        \\[-0.2cm]
  {\small\it Universit\`a degli Studi di Pisa}        \\[-0.2cm]
  {\small\it Pisa 56100, ITALIA}        \\[-0.2cm]
  {\small {\tt PELISSET@SUNTHPI1.DIFI.UNIPI.IT}}  \\[-0.2cm]
  {\protect\makebox[5in]{\quad}}  
   \\[-0.35cm] \and
  \small  Alan D. Sokal               \\[-0.2cm]
  \small\it Department of Physics     \\[-0.2cm]
  \small\it New York University       \\[-0.2cm]
  \small\it 4 Washington Place        \\[-0.2cm]
  \small\it New York, NY 10003 USA    \\[-0.2cm]
  \small {\tt SOKAL@NYU.EDU} \\[-0.2cm]
  {\protect\makebox[5in]{\quad}}  
   \\
}

\vspace{0.2cm}
\maketitle
\thispagestyle{empty}   

\vspace{-0.4cm}
\begin{abstract}
We introduce a variant of
the multi-grid Monte Carlo (MGMC) method,
based on the embedding of an $XY$ model into the target model,
and we study its mathematical properties for a variety of
nonlinear $\sigma$-models.
We then apply the method to the
two-dimensional $O(N)$-symmetric nonlinear $\sigma$-models
(also called $N$-vector models) with $N=3,4,8$
and study its dynamic critical behavior.
Using lattices up to $256 \times 256$,
we find dynamic critical exponents
$z_{int,{\cal M}^2} \approx 0.70 \pm 0.08$, $0.60 \pm 0.07$,
$0.52 \pm 0.10$ for $N=3,4,8$, respectively
(subjective 68\% confidence intervals).
Thus, for these asymptotically free models,
critical slowing-down is greatly reduced compared to local algorithms,
but not completely eliminated;
and the dynamic critical exponent does apparently vary with $N$.
We also analyze the static data for $N=8$ using a
finite-size-scaling extrapolation method.
The correlation length $\xi$ agrees with the four-loop
asymptotic-freedom prediction to within $\approx 1\%$
over the interval $12 \ltapprox \xi \ltapprox 650$.
\end{abstract}

\clearpage

%
%

\section{Introduction}

By now it is widely recognized
\cite{Sokal_Lausanne,Adler_LAT88,Wolff_LAT89,Sokal_LAT90}
that better simulation algorithms,
with strongly reduced critical slowing-down,
are needed for high-precision Monte Carlo studies of
statistical-mechanical systems near critical points
and of quantum field theories (such as QCD) near the continuum limit.
One promising class of such algorithms is
{\em multi-grid Monte Carlo}\/ (MGMC)
\cite{MGMC_0,MGMC_1,MGMC_2,MGMC_O4,mgmco4d1,%
Meyer-Ortmanns_85,Mack_Cargese,Mack-Meyer_90,%
Hasenbusch_LAT90,Hasenbusch-Meyer_PRL,Hasenbusch_CP3,%
Hulsebos_91,Laursen_91,Laursen_93,Grabenstein_92,Grabenstein_93a,%
Grabenstein_93b,Grabenstein_thesis,Grabenstein_94,Janke_multicanonical_MGMC}:
this is a collective-mode approach that introduces block updates
(of fixed shape but variable amplitude) on all length scales.
The basic ingredients of the method are\footnote{
   See \cite{MGMC_1} for details.
}:
 
1) {\em Interpolation operator:}\/
This is a rule specifying the shape of the block update.
The interpolations most commonly used are
{\em piecewise-constant}\/ (square-wave updates)
and {\em piecewise-linear}\/ (pyramidal-wave updates).
 
2) {\em Cycle control parameter $\gamma$:}\/
This is an integer number that determines the way in which the
different block sizes are visited.
In general, blocks of linear size $2^l$
are updated $\gamma^l$ times per iteration.
Thus, in the W-cycle ($\gamma=2$) more emphasis is placed
on large length scales than in the V-cycle ($\gamma=1$).

3) {\em Basic (smoothing) iteration:}\/
This is the local Monte Carlo update that is performed on each level.
Typically one chooses to use {\em heat-bath}\/ updating if the distribution
can be sampled in some simple way, and {\em Metropolis}\/ updating otherwise.
 
4) {\em Implementation:}\/
The computations can be implemented either in the
{\em recursive multi-grid}\/ style
using explicit coarse-grid fields
\cite{Hackbusch_85,Briggs_87,MGMC_0,MGMC_1,MGMC_2,MGMC_O4,mgmco4d1},
or in the {\em unigrid}\/ style using block updates acting directly
on the fine-grid fields
\cite{McCormick-Ruge,Mack_Cargese,Mack-Meyer_90,%
Hasenbusch_LAT90,Hasenbusch-Meyer_PRL,Hasenbusch_CP3}.
For a $d$-dimensional system of linear size $L$,
the computational labor per iteration is
\be
  \hbox{Work(MG)} \;\sim\;
  \cases{ L^d                & for $\gamma < 2^d$  \cr
          \noalign{\vskip 1mm}
          L^d \, \log L      & for $\gamma = 2^d$  \cr 
          \noalign{\vskip 1mm}
          L^{\log_2 \gamma}  & for $\gamma > 2^d$  \cr 
        }
\label{workMG_0}
\ee
for the recursive multi-grid approach, and
\be
  \hbox{Work(UG)} \;\sim\;
  \cases{ L^d \, \log L               & for $\gamma=1$ \cr
          \noalign{\vskip 1mm}
          L^{d \,+\, \log_2 \gamma}   & for $\gamma>1$ \cr
        }
\label{workUG}
\ee
for the unigrid approach.
Thus, the unigrid implementation is marginally more expensive
for a V-cycle, but prohibitively more expensive for a W-cycle.

The efficiency of the MGMC method can be analyzed rigorously in the case
of the Gaussian (free-field) model, for which it can be proven
\cite{MGMC_0,MGMC_1,Goodman-Sokal_unpub}
that critical slowing-down is completely eliminated.
That is, the {\em autocorrelation time}\/ $\tau$ is bounded as
the correlation length $\xi$ and the lattice size $L$ tend to infinity,
so that the {\em dynamic critical exponent}\/ $z$ is zero.\footnote{
   See \cite{Sokal_LAT90} for a pedagogical discussion of
   the various autocorrelation times and their associated
   dynamic critical exponents.
}
More precisely, the algorithm with piecewise-linear interpolation
exhibits $z = 0$ for both the V-cycle and the W-cycle, while the algorithm with
piecewise-constant interpolation has $z = 0$ only for the W-cycle
(the piecewise-constant V-cycle has $z = 1$).

One is therefore motivated to apply MGMC to ``nearly Gaussian''
systems; one might hope that critical slowing-down would likewise
be completely eliminated (possibly modulo a logarithm)
or at least greatly reduced compared to the
$z \approx 2$ of local algorithms.
In particular, we are interested in applying MGMC to
asymptotically free two-dimensional $\sigma$-models, such as
the $N$-vector models [also called $O(N)$-invariant $\sigma$-models] for $N>2$,
the $SU(N)$ and $SO(N)$
principal chiral models,
and the $RP^{N-1}$ and $\CP^{N-1}$ $\sigma$-models.\footnote{
   In \cite[Section IX.C]{MGMC_1} we conjectured that
   that for asymptotically free theories ---
   which are ``almost Gaussian'' except at very long distances ---
   the critical slowing-down should be completely eliminated
   except for a possible logarithm.
   However, when we made a numerical study of MGMC
   in the two-dimensional 4-vector model \cite{MGMC_O4},
   we found to our surprise that the dynamic critical exponent
   is {\em not}\/ zero:
   $z_{int,{\cal M}^2} = 0.60 \pm 0.07$ (subjective 68\% confidence
   interval).  We then produced \cite[Section 4.1]{MGMC_O4} a heuristic
   explanation of this fact, based on the logarithmically decaying
   deviations from Gaussianness in an asymptotically free theory.

   Shortly thereafter, Richard Brower (private communication)
   pointed out to us that if our explanation is correct, it follows
   that for a {\em one}\/-dimensional $\sigma$-model --- in which the
   deviations from Gaussianness decay much more rapidly, namely as a
   power law --- the critical slowing-down should be {\em completely}\/
   eliminated, i.e.\ we should have $z_{MGMC} = 0$ (with {\em no}\/
   logarithm).  We hardly doubted that this would be the case, but we
   decided to make an empirical test to settle the matter once and for all.
   We have recently \cite{mgmco4d1} reported numerical evidence suggesting
   that the critical slowing-down is {\em not}\/ in fact completely
   eliminated in the one-dimensional $O(4)$-symmetric $\sigma$-model,
   but is rather logarithmic:  $\tau_{int,{\cal M}^2} \sim \log\xi$.
   It goes without saying that we have no idea why this is the case.
}

In view of the rigorous results for the Gaussian case,
we want to investigate two questions:

1) Is $z=0$ for all asymptotically free two-dimensional $\sigma$-models?
If not, does $z$ vary from one asymptotically free model to another?

2) Is the algorithm with piecewise-constant interpolation and a W-cycle
as efficient (at least in order of magnitude)
as the one with piecewise-linear interpolation and a V-cycle,
i.e.\ is $z_{PC,W}$ equal to $z_{PL,V}$ for these models?


The key design choice in a MGMC algorithm is
that of the interpolation operator;
indeed, this choice determines most of the remaining ingredients.
If one chooses a ``smooth'' interpolation such as piecewise-linear,
then it is usually
impossible to implement true recursive MGMC,
as there is no simple form for the induced coarse-grid Hamiltonians.\footnote{
   In particular, this occurs for the nonlinear $\sigma$-models.
   The trouble is that concepts like ``an update of half the magnitude \ldots''
   have no natural meaning in a nonlinear manifold;
   and whatever precise definition one chooses, one is left with
   a horribly nonlinear expression for the coarse-grid Hamiltonian.
}
Therefore one is obliged to use the unigrid style, a V-cycle,
and Metropolis updating.
We call this the ``German'' approach
\cite{Meyer-Ortmanns_85,Mack_Cargese,Mack-Meyer_90,%
Hasenbusch_LAT90,Hasenbusch-Meyer_PRL,Hasenbusch_CP3}.
On the other hand, if one chooses a ``crude'' interpolation such as
piecewise-constant,
then one is obliged to use a W-cycle in order to have a chance at $z < 1$;
but piecewise-constant interpolation usually gives rise to a simple
coarse-grid Hamiltonian (typically a slight generalization of the
fine-grid Hamiltonian\footnote{
   For example, consider a model in which the fields $\{ g_x \}$
   are unitary matrices, and in which the block update is
   left-multiplication $g_x^{new} = h_y g_x^{old}$
   where $x$ is a fine-grid site, $y$ is a coarse-grid site,
   $h_y$ is the coarse-grid field,
   and $x \in B_y \equiv$ the fine-grid block corresponding to
   coarse-grid site $y$.
   Then, if the Hamiltonian on grid $l$ is
   $$
   H_l(g) \;=\; \sum_{\<xx'\>}  \real\tr (\alpha_{xx'} g_x^{\dagger} g_{x'})
   \;\mbox{,}
   $$
   the induced Hamiltonian on grid $l+1$ will be
   $$
   H_{l+1}(h) \;=\; \sum_{\<yy'\>}  \real\tr (\alpha'_{yy'} h_y^{\dagger}
   h_{y'})
   \;\mbox{,}
   $$
   where the coarse-grid couplings \{$\alpha'_{yy'}\}$
   are given in terms of fine-grid couplings $\{\alpha_{xx'}\}$ by
   $$
   \alpha'_{yy'} \;=\;
   \sum_{ \< xx' \> \colon\, x \in B_y,\,  x' \in B_{y'} }
     g_{x'} \alpha_{xx'} g_x^{\dagger}
   \;\mbox{.}
   $$
   (Remark:  In this paper grid 0 is the finest grid, and grid $l$
    maps to blocks of linear size $2^l$.  This notation is opposite to
    that used in \cite{MGMC_1,MGMC_2,MGMC_O4};
    it now seems to us to be more natural.)
}),
so that one can use the recursive multi-grid
style and, at least in principle, heat-bath updating.
This is the approach taken by our group
\cite{MGMC_0,MGMC_1,MGMC_2,MGMC_O4,mgmco4d1}.

The ``German'' version has the advantage of being easy to implement
for diverse models,
but its use of Metropolis updates introduces several free parameters
that have to be adjusted\footnote{
   For each block size $2^l$, one must choose the Metropolis hit size $t_l$
   and the number $N_l$ of Metropolis hits per site.
   As shown by Grabenstein and Pinn
   \cite{Grabenstein_92,Grabenstein_93a,Grabenstein_thesis},
   for smooth interpolations
   in Gaussian and ``near-Gaussian'' models (such as $\sigma$-models)
   it is natural to take $t_l = t_0 2^{-(d-2)l/2}$
   and $N_l = N_0$;
   this is expected to ensure an acceptance fraction roughly
   independent of $l$.
   So it suffices to choose $t_0$ and $N_0$.
},
making it more difficult to test systematically.
``Our'' version has no free parameters, but its implementation is
cumbersome and model-dependent, in the sense that the program
(and in particular the heat-bath subroutine)
has to be drastically rewritten for each distinct model.
For example, among the $N$-vector models, the only ones that can be
handled conveniently by our ``direct'' MGMC method are
$N=2$ \cite{MGMC_2} and $N=4$ \cite{MGMC_O4},
by exploiting the isomorphism with the
$U(1)$ and $SU(2)$ groups, respectively.\footnote{
   For general $N$-vector models, our direct MGMC method is still applicable,
   but one must first ``lift'' the theory from the manifold $S^{N-1}$
   to the group $SO(N)$, as described in \cite[Section IV]{MGMC_1} and
   \cite[Section 2]{MGMC_O4}.  One then has the problem of writing an
   efficient heat-bath subroutine for generating random variables
   $R \in SO(N)$ with probability density $\sim \exp\tr(AR)$,
   where $A$ is a general real $N \times N$ matrix.
   For $N \ge 3$ this seems to be quite messy.
}
 
With this problem in mind, we have developed a new implementation of
MGMC that combines some of the advantages of both methods;
in particular, it can be used conveniently for a large class of $\sigma$-models
with very little modification of the program.\footnote{
   We devised this approach after extensive discussions with
   Martin Hasenbusch and Steffen Meyer at the Lattice '92 conference
   in Amsterdam.
   In particular, the idea of $XY$ embedding is made explicit in their work:
   see equations (5)/(6) in \cite{Hasenbusch-Meyer_PRL}
   and equations (5)--(9) in \cite{Hasenbusch_CP3}.
}
The idea is to {\em embed}\/ angular variables $\{ \theta_x \}$
into the given $\sigma$-model, and then update the resulting
induced $XY$ model by our standard (piecewise-constant, W-cycle,
heat-bath, recursive) MGMC method.
We do not claim that this approach is superior in practice
to the ``German'' method --- that remains to be determined ---
but we do think that it is well suited for the systematic study of
the dynamic critical behavior of MGMC algorithms.

In the simplest cases, the induced $XY$ Hamiltonian takes the form
\be
{\cal H}_{embed} \;=\;  - \sum_{\<xx'\>}
\left[ \alpha_{xx'}\,\cos(\theta_x \,-\, \theta_{x'}) \,+\,
   \beta_{xx'}\,\sin(\theta_x \,-\, \theta_{x'}) \right]
   \;,
\label{ham_embed0}
\ee
where the induced couplings $\{\alpha_{xx'},\beta_{xx'}\}$
depend on the current configuration $\{ \varphi_x^{old} \}$
of the original model.
Such a ``generalized $XY$ Hamiltonian'' is easily simulated by MGMC;
indeed, the coarse-grid Hamiltonians in $XY$-model MGMC are inevitably
of the form \reff{ham_embed0},
even when the fine-grid Hamiltonian is the standard
$XY$ model $\alpha_{xx'} \equiv \alpha \ge 0$, $\beta_{xx'} \equiv 0$
\cite{MGMC_1,MGMC_2}.
So one may just as easily start from \reff{ham_embed0}
already on the finest grid.

For future reference, let us call the Hamiltonian \reff{ham_embed0}
{\em ferromagnetic}\/ if
$\alpha_{xx'} \ge 0$ and $\beta_{xx'} = 0$ for all bonds $\< xx' \>$;
let us call it {\em unfrustrated}\/ if there exists a configuration
$\{\theta_x\}$ that simultaneously minimizes the bond energy
$- [\alpha_{xx'} \cos(\theta_x - \theta_{x'}) +
    \beta_{xx'} \sin(\theta_x - \theta_{x'})]$
on all bonds $\< xx' \>$.

In models with isotensor or adjoint-representation couplings,
the induced $XY$ model will in some cases have also ``isospin-2'' terms:
\begin{eqnarray}
{\cal H}_{embed} &=&  - \sum_{\<xx'\>}
\left[ \alpha_{xx'}\,\cos(\theta_x \,-\, \theta_{x'}) \,+\,
   \beta_{xx'}\,\sin(\theta_x \,-\, \theta_{x'}) \right. \nonumber \\
& & \qquad\qquad +\,
    \left. \gamma_{xx'}\,\cos (2\theta_x \,-\, 2\theta_{x'}) \,+\,
\delta_{xx'}\,\sin (2\theta_x \,-\, 2\theta_{x'}) \right]
   \;.
\label{ham_embed2_0}
\end{eqnarray}
We shall call such a Hamiltonian ferromagnetic if
$\alpha_{xx'}, \gamma_{xx'} \ge 0$ and $\beta_{xx'} = \delta_{xx'} = 0$
for all bonds $\< xx' \>$;
we call it unfrustrated if there exists a configuration
$\{\theta_x\}$ that simultaneously minimizes the bond energy
$- [\alpha_{xx'} \cos(\theta_x - \theta_{x'}) +
    \beta_{xx'} \sin(\theta_x - \theta_{x'}) +
    \gamma_{xx'} \cos (2\theta_x - 2\theta_{x'}) +
    \delta_{xx'} \sin (2\theta_x - 2\theta_{x'}) ]$
on all bonds $\< xx' \>$.

Already in a previous paper, we noted in passing
\cite[pp.~516--517]{CEPS_swwo4c2}
the possibility of $XY$-embedding MGMC;
we argued, however, that it is {\em no better}\/ than direct MGMC,
because the updates in $XY$-embedding MGMC
are simply a subset of those available in direct MGMC.
This is true, but it is also true that
$XY$-embedding MGMC is likely to be {\em no worse}\/ than direct MGMC
(except for a constant factor),
at least for asymptotically free $\sigma$-models:
this is because, in such a model, the spin waves along different
axes in internal-spin space are weakly coupled.\footnote{
   It would be very interesting to study direct versus $XY$-embedding MGMC
   also in {\em non}\/-asymptotically-free $\sigma$-models,
   such as the {\em three}\/-dimensional $N$-vector model (e.g.\ with $N=4$).
   Of course, the dynamic critical behavior of MGMC in such a model
   is an open question even in the $XY$ case $N=2$.
}
And $XY$-embedding MGMC can be much more convenient to implement
than direct MGMC.

One might wonder why we have bothered studying MGMC algorithms for
two-dimensional $N$-vector models,
when Wolff's \cite{Wolff_89a} cluster algorithm
already succeeds in completely eliminating the critical slowing-down
for these models
\cite{Wolff_89a,Edwards_89,Wolff_90,CEPS_swwo4c2,CEPS_rpn_dynamic}.
The reason is that the success of generalized Wolff-type embedding algorithms
appears to be limited to $N$-vector models
(and possibly also $RP^{N-1}$ models) \cite{CEPS_swwo4c2},
while MGMC can be expected to behave qualitatively similarly
for all asymptotically free $\sigma$-models.\footnote{
   Additional difficulties might arise in models that have important
   discrete excitations (i.e.\ vortices) in addition to spin waves:
   for example, the $RP^{N-1}$ model, the $SO(N)$ chiral model,
   and the adjoint $SU(N)$ chiral model.
   These difficulties would be analogous to the mediocre performance of MGMC
   in the two-dimensional $XY$ model in the high-temperature (vortex) phase
   \cite{MGMC_2,Hulsebos_91}.
}
Therefore, we want to understand systematically the dynamic critical
behavior of MGMC in asymptotically free $\sigma$-models.
We have started with the $N$-vector models because of their simplicity,
but our eye is ultimately on more general $\sigma$-models,
such as those taking values in
$SU(N)$ \cite{Hasenbusch-Meyer_PRL,Mendes_LAT95,MGMC_SU3}
or $\CP^{N-1}$ \cite{Hasenbusch-Meyer_PRL,Hasenbusch_CP3}.

\bigskip

The plan of this paper is as follows:
In Section \ref{sec2} we introduce $XY$ embeddings
for a variety of nonlinear $\sigma$-models,
and we analyze their mathematical properties.
In Section \ref{sec3} we report our numerical data
on the $XY$-embedding MGMC algorithm for the two-dimensional $N$-vector
models with $N=3,4,8$.
In Section \ref{sec4} we analyze our static data
for the $N=8$ case, using a finite-size-scaling extrapolation method
\cite{Luscher_91,Kim_93,fss_greedy,o3_scaling_prl,fss_greedy_fullpaper}
and comparing the results to the four-loop asymptotic-freedom predictions
\cite{CP_4loops}.
In Section \ref{sec5} we analyze our dynamic data for $N=3,4,8$,
using traditional finite-size-scaling plots
to extract the dynamic critical exponents $z_{int, {\cal M}^2}$.
In Section \ref{sec6} we discuss our findings.

A brief preliminary version of this work appeared in \cite{Mendes_LAT95}.
In a subsequent paper \cite{MGMC_SU3}
we will report results for $XY$-embedding MGMC applied to the
two-dimensional $SU(3)$ principal chiral model.

\section{$XY$ Embedding}   \label{sec2}

In this section we introduce $XY$ embeddings for a variety of 
nonlinear $\sigma$-models, and we analyze their mathematical properties.
In particular, for each model and each choice of embedding, we want to 
answer the following two questions:
\begin{itemize}
\item[(a)] Is the induced $XY$ Hamiltonian of the {\em simple form}
\reff{ham_embed0} [as opposed to the ``mixed isospin-1/isospin-2''
form \reff{ham_embed2_0}]?
This question is relevant because it is much easier to write an MGMC
program (and in particular a heat-bath subroutine) for 
\reff{ham_embed0} than for \reff{ham_embed2_0}.
\item[(b)] Is the induced $XY$ Hamiltonian {\em unfrustrated}? This question 
{\em may} be relevant to the dynamic critical behavior of the resulting
algorithm (at any rate we want to {\em find out} whether it is relevant).
\end{itemize}

After making some general comments on embedding algorithms (Section
\ref{sec2.1}), we introduce the models and the embeddings and compute
the induced $XY$ Hamiltonians (Section 2.2). Next we ask whether or not
these induced $XY$ Hamiltonians are frustrated (Section 2.3). We also 
discuss some alternative embeddings (Section 2.4). Finally, we
summarize our results (Section 2.5).

\subsection{Generalities on Embedding Algorithms}   \label{sec2.1}

The general idea of embedding algorithms \cite{Sokal_LAT90,CEPS_swwo4c2}
is the following:
``Foliate'' the configuration space of the original model into ``leaves''
isomorphic to the configuration space of some ``embedded'' model
(which may be an Ising model, $XY$ model, etc.).
The Monte Carlo process then moves around the current leaf,
using any legitimate Monte Carlo algorithm for
simulating the conditional probability distribution of the original model
restricted to that leaf;
this conditional probability distribution
(which of course depends on the current configuration of the original model)
defines the induced Hamiltonian for the embedded model.\footnote{
   This same structure arises in the coarse-grid updates of MGMC algorithms;
   we have termed it ``partial resampling''
   \cite{Sokal_Lausanne,MGMC_0,MGMC_1}.
}
Of course, one must combine this procedure with other moves,
or with different foliations, in order to make the algorithm ergodic.

Clearly, the updates available in an embedding algorithm with a specified
foliation are a {\em subset}\/ of all the updates available in a general
Monte Carlo algorithm for the original model.  So embedding never
enlarges the space of possible algorithms;  if anything, it restricts it.
And the smaller the ``leaves'' of the foliation, the slower the algorithm
is likely to move through the configuration space.
The advantage of embedding is, rather, that the embedded model may be
much {\em simpler}\/ than the original model.
As a result, it may be easier to invent a good algorithm for updating
the embedded model than for updating the original model;
or even if good algorithms for both models can be invented,
the former algorithm may be easier to implement in practice.
Thus, embedding may even lead to a {\em loss}\/ in efficiency
compared to alternative algorithms;
but if this loss is not too great, it may be outweighed by
the {\em simplicity}\/ of programming and by the {\em flexibility}\/
of the method (which is important in case we want to apply the method
systematically to several similar models).

The dynamic critical behavior of an embedding algorithm
is determined by the combined effect of two {\em completely distinct}\/ issues:
\begin{itemize}
 \item[i)]  How well the chosen embedding
     succeeds in ``encoding'' the important large-scale collective modes
     of the original model into the embedded variables.
 \item[ii)]  How well the chosen algorithm for updating the embedded model
     succeeds in its task.
\end{itemize}
Thus, if the physically relevant large-scale collective motions of the original
model cannot be obtained by motions {\em within}\/ a leaf,
then the embedding algorithm will have severe critical slowing-down
{\em no matter what}\/ method is used to update the embedded variables.
On the other hand, if the embedding algorithm with a {\em particular}\/ choice
of updating method for the embedded model shows severe critical slowing-down,
this does {\em not}\/ necessarily mean that the embedding {\em per se}\/
works badly:  the poor performance might be due to slow decorrelation in the
inner-loop subroutine for updating the embedded model,
and could possibly be remedied by switching to a better inner-loop algorithm.

To study question (i) independently of question (ii),
it is conceptually useful to consider the
{\em idealized embedding algorithm\/} \cite{CEPS_swwo4c2},
in which a new configuration on the current leaf is chosen,
{\em independent of the old configuration\/},
with probabilities given by the conditional probability distribution
restricted to that leaf.
Of course, such an algorithm is not practical, but that is not its role.
Rather, it serves as a standard of comparison
(and presumed lower bound on the autocorrelation time)
for {\em all}\/ algorithms based on the given embedding.
To approximate in practice the idealized embedding algorithm,
we can update the embedded model
by $N_{hit}$ hits of some chosen algorithm,
and extrapolate to $N_{hit}=\infty$.\footnote{
   Preferably we would perform simulations for successively increasing
   values of $N_{hit}$ until the autocorrelation time is
   constant within error bars.  However, this may not always be feasible.
}
To be sure, this test procedure can be very time-consuming;
we have implemented it in our work on generalized Wolff-type embedding
algorithms \cite{CEPS_swwo4c2},
for which the embedded variables are Ising spins.

Here, by contrast, the embedded variables are $XY$ spins,
and we suspect \cite[p.~517]{CEPS_swwo4c2}
that the idealized embedding algorithm will have {\em no}\/
critical slowing-down (because of the weak coupling between
spin waves along different internal-spin axes in an
asymptotically free $\sigma$-model).
This prediction should, of course, be carefully tested;
but in this initial study we have for simplicity restricted attention
to the ``practical'' algorithm having $N_{hit}=1$.
We want to know whether the $XY$-embedding MGMC algorithm
lies in the same dynamic universality class as the direct MGMC algorithm
for the same model;  and we want to know whether (and how)
the dynamic critical exponents vary from one asymptotically free model
to another.

There is one situation in which one might worry that the
practical ($N_{hit}=1$) $XY$-embedding MGMC algorithm
could behave poorly even though the idealized ($N_{hit}=\infty$)
algorithm would behave well:
this is when the induced $XY$ Hamiltonian is strongly frustrated,
so that it has multiple ground states (or approximate ground states)
separated by large energy barriers.  In such a situation,
the MGMC moves, which are basically spin waves, would probably
be insufficient to surmount these energy barriers and to find the
distant ground states.\footnote{
   A similar problem occurs in MGMC for the ordinary two-dimensional
   $XY$ model just above the Kosterlitz-Thouless critical temperature:
   the MGMC moves are inefficient at creating and destroying vortices
   \cite{MGMC_2,Hulsebos_91}.
}
We have therefore taken care to investigate, for each of our
$XY$ embeddings, whether the induced $XY$ Hamiltonian is frustrated or not.
(This is a natural theoretical question in any case:
 one wants to know the properties of the induced $XY$ Hamiltonian.
 It is also relevant for {\em analytical}\/ applications of
 $XY$ embedding, such as the proof of correlation inequalities
 \cite{Mack_79,Frohlich_79,CEPS_RPN}.)
It turns out that in the case of $N$-vector models the
induced $XY$ Hamiltonian is unfrustrated,
while for $SU(N)$ chiral models it is generically frustrated.
Nevertheless, $XY$-embedding MGMC performs well
(dynamic critical exponents $z$ in the range 0.4--0.7)
in {\em both}\/ cases \cite{Mendes_LAT95,MGMC_SU3}.
We think this is because the induced $XY$ Hamiltonian in the $SU(N)$ case
is only {\em weakly}\/ frustrated as $\beta\to\infty$;
we intend to study this point in more detail in a subsequent paper
devoted specifically to $SU(N)$ chiral models \cite{MGMC_SU3}.

Let us now define our notation for nonlinear $\sigma$-models.
The original model will have field variables (``spins'') $\varphi_x$
living on lattice sites $x \in \Z^d$ and
taking values in a compact manifold (``target space'') $M$.\footnote{
   We shall use the notation $\varphi_x$ for generic discussions.
   When discussing specific models we shall use standard notations:
   $\bsigma_x$ for real unit vectors, $\bz_x$ for complex unit vectors,
   $U_x$ for $SU(N)$ matrices, etc.
}
The Hamiltonian $\scrh$ of this model will have a global symmetry group $G$,
i.e.\ $\scrh (\{\varphi_x\}) = \scrh (\{ g \varphi_x\})$
for all $g \in G$ and all field configurations $\{\varphi_x\}$.
We shall assume that $G$ is a compact Lie group with
Lie algebra $\germang$, and that $G$ acts transitively on $M$.

It is worth mentioning a useful heuristic principle for finding
candidates for ``good'' embedding algorithms \cite[p.~517]{CEPS_swwo4c2};
this principle includes as special cases
both the Wolff-type embedding of Ising spins
and the multi-grid-type embedding of spin waves.
The idea is to take an exact {\em global}\/ symmetry of the model
(global reflection in the case of Wolff,
 global rotation in the case of multi-grid)
and to apply it in an {\em inhomogeneous}\/ way
(constant on clusters in the case of Wolff,
 constant on blocks in the case of multi-grid).
In other words, one picks a subgroup $H \subset G$ (possibly all of $G$)
and then embeds a model taking values in $H$,
via the update $\varphi_x^{new} =  h_x \varphi_x^{old}$.
Because $H$ is an exact global symmetry of the original model,
the energy cost of such an update
is {\em zero}\/ whenever the field $\{h_x\}$ is constant (independent of $x$).
Therefore, if $\{h_x\}$ is ``almost constant''
(e.g.\ constant on a large block),
then the energy cost is only a ``surface term''.\footnote{
   The principle of looking for collective-mode updates
   whose energy cost is only a ``surface term''
   is also made explicit by Grabenstein and Pinn
   \cite{Grabenstein_thesis,Grabenstein_94}.
}

Let us mention, finally, that embedding algorithms can be written
in two alternate (but physically equivalent) parametrizations.
On the one hand, the embedded variables $\{h_x\}$ can be considered
as {\em updates}\/ to the current values $\{ \varphi_x^{old} \}$
of the original spins;  this is the interpretation taken in the
formula $\varphi_x^{new} =  h_x \varphi_x^{old}$.
On the other hand, in {\em some}\/ models the embedded variables
can be taken to be a {\em subset of the new values}\/ $\{ \varphi_x^{new} \}$
of the original spins:  for example, in the $N$-vector model
the embedded variable $\theta_x$ can be taken to be the angle of
certain components of $\varphi_x^{new}$ with respect to some fixed basis
[see \reff{decompose} ff.\ for details].
The first parametrization always exists, while the second one exists
only in some cases (e.g.\ in $N$-vector models but apparently
not in $SU(N)$ models).
The advantage of the second parametrization, when it does exist,
is that certain properties of the embedded Hamiltonian
(e.g.\ nonfrustration) are more manifest.
When both parametrizations exist,
they are related by a gauge transformation with gauge group $H$.

\subsection{$XY$ Embedding: Some Examples}   \label{sec2.2}

\subsubsection{Embedded Variables as Updates} \label{sec2.2.1}

The basic idea of our algorithm is to embed angular variables $\{ \theta_x \}$
into the given $\sigma$-model;  we then update the resulting
induced $XY$ model by our standard (piecewise-constant, W-cycle,
heat-bath, recursive) MGMC method.

Consider, therefore, a $\sigma$-model with target manifold $M$ and
global symmetry group $G$;
we assume that $G$ is a connected compact Lie group with
Lie algebra $\germang$, and that $G$ acts transitively on $M$.
The idea behind $XY$ embedding
is to choose randomly a $U(1)$ subgroup $H \subset G$,
and to apply a ``rotation'' $\theta_x$ in this subgroup
to the original spin variable $\varphi_x \in M$.
Thus, the angular variables $\theta_x$ are {\em updates}\/
to the original variables $\varphi_x$.
More precisely, we define the updated variable $\varphi_x^{new}$ by
\be
  \varphi_x^{new} \;=\; e^{i\theta_x RTR^{-1}} \varphi_x^{old}
                  \;=\; R e^{i\theta_x T} R^{-1} \varphi_x^{old}
  \;,
\label{update}
\ee
where $R$ is a random element of $G$,
and $T$ is a fixed nonzero element (to be specified later)
of the Lie algebra $\germang$.
The embedded $XY$ model consisting of the spins $\{ \theta_x \}$
is then simulated using the induced Hamiltonian
\be
   \scrh_{embed}( \{ \theta_x \} )   \;=\;
   \scrh( \{ \varphi_x^{new} \} )
   \;,
 \label{ham_induced}
\ee
with initial condition $\theta_x = 0$
(i.e.\ $\varphi_x^{new} = \varphi_x^{old}$) for all $x$.
At each iteration of the algorithm, a new random matrix $R$ is chosen.

In general the induced $XY$ Hamiltonian \reff{ham_induced}
can be extremely complicated (and thus impractical to simulate by
true recursive MGMC).  However, if the original Hamiltonian $\scrh$
is sufficiently ``nice'' {\em and}\/ one makes a clever choice of the
generator $T$, then in some cases the induced $XY$ Hamiltonian can be
reasonably simple.  Thus, we shall always choose the matrix $T$
to have all its eigenvalues in the set $\{-1,0,1\}$.
This implies that
\be
  e^{i\theta T} \;=\; T^2\, \cos \theta \,+\, iT\,\sin \theta
    \,+\, (I \,-\, T^2)
  \;,
 \label{T_eigenvalues}
\ee
where $I$ is the identity matrix.

Let us now illustrate the method in five models:
\begin{itemize}
   \item[1)]  The $N$-vector model:  $M=S^{N-1}$, $G=SO(N)$.
The original variables $\bsigma_x$ are unit vectors in $\R^N$,
and the original Hamiltonian is
\be
  {\cal H}_{N-vector} \;=\;
   -\beta \sum_{\<xx'\>} \bsigma_x \cdot \bsigma_{x'}
  \;.
\label{N-vector}
\label{ham_N-vector}
\ee

   \item[2)]  The $SU(N)$ principal chiral model:  $M=SU(N)$, 
       $G = SU(N) \times SU(N)$.
The original variables $U_x$ of this model are $SU(N)$ matrices,
and the original Hamiltonian is
\be
  {\cal H}_{SU(N)} \;=\;  -\beta  \sum_{\<xx'\>}
    \re \tr (U_x^{\dagger} \, U_{x'})
  \;.
\label{ham_SUN}
\ee
[Similar methods apply to other principal chiral models, e.g.\ $SO(N)$.]

   \item[3)]  The mixed isovector/isotensor model:   $M=S^{N-1}$, $G=SO(N)$.
This model generalizes the $N$-vector model:
the original variables $\bsigma_x$ are again unit vectors in $\R^N$,
but the Hamiltonian is now
\be
  \scrh_{mixed}(\{\bsigma\})   \;=\;
    -\, \beta_V \sum_{\< xx' \>} \bsigma_x \cdot \bsigma_{x'}
    \;
    -\, {\beta_T \over 2} \sum_{\< xx' \>} (\bsigma_x \cdot \bsigma_{x'})^2
  \;.
 \label{eq1.3}
\ee
[If $\beta_T = 0$, this is the $N$-vector model.
 If $\beta_V = 0$, this is the pure isotensor model,
 which has a local $Z_2$ gauge group;
 it can therefore be thought of as taking values in
 $M = RP^{N-1} \equiv S^{N-1}/Z_2$.\footnote{
      Note also the special case $N=4$ and $\beta_V = 0$:
      since $RP^3 \simeq SO(3)$ via the correspondence
      $$ (V_x)_{ij} \;=\;
         (2{\sigma_x^{(0)}}^2 - 1) \delta_{ij}  \,+\,
          2 \sigma_x^{(i)} \sigma_x^{(j)}  \,+\,
          2 \sigma_x^{(0)} \varepsilon_{ijk} \sigma_x^{(k)}  \;, $$
      where $\bsigma_x \in S^3$ and $\,V_x \in SO(3)$,
      we have
      $\tr(V_x^\T V_{x'})   =  4(\bsigma_x \cdot \bsigma_{x'})^2 - 1$,
      so that the $RP^3$ model (with {\em this}\/ choice of lattice action)
      is equivalent to the $SO(3)$ principal chiral model.
\label{RP3_equiv_SO3}
}]

   \item[4)]  The complex mixed isovector/isotensor model:
      $M=S^{2N-1}$, $G= U(N) \subset SO(2N)$.
The original variables $\bz_x$ are unit vectors in $\C^N$, and the
original Hamiltonian is
\be
  \scrh_{mixed} (\{\bz\})   \;=\;
    -\, \beta_V \sum_{\< xx' \>} \real (\bz_x^* \cdot \bz_{x'})
    \;
    -\, {\beta_T \over 2} \sum_{\< xx' \>} |\bz_x^* \cdot \bz_{x'}|^2
  \;.
\label{ham:CPN}
\ee
[If $\beta_T = 0$, this is $2N$-vector model written in complex notation.
 If $\beta_V = 0$, this is the pure complex isotensor model, which has a
 global $SU(N)$ symmetry group and a local $U(1)$ gauge group;
 it can therefore be thought of as taking values in
 $M = \CP^{N-1} \equiv S^{2N-1}/U(1)$.\footnote{
      Note also the special case $N=2$ and $\beta_V = 0$:
      since $\CP^1 \simeq S^2$ via the correspondence
      $\bsigma_x = \bz_x^* \btau \bz_x$ where $\btau$ are the Pauli matrices,
      we have
      $|\bz_x^* \cdot \bz_{x'}|^2   =
          (1 + \bsigma_x \cdot \bsigma_{x'}) / 2$,
      so that the $\CP^1$ model (with {\em this}\/ choice of lattice action)
      is equivalent to the standard 3-vector model.
\label{CP1_equiv_O3}
}]

   \item[5)]  The mixed fundamental/adjoint $SU(N)$ principal chiral model:
$M=SU(N)$, $G = SU(N) \times SU(N)$.
Again the original variables $U_x$ of this model are $SU(N)$ matrices,
and the Hamiltonian is
\be
  \scrh_{SU(N)} (\{U\})   \;=\;
    -\, \beta_F \sum_{\< xx' \>} \real\tr(U_x^{\dagger} U_{x'})
    \;
    -\, \beta_A \sum_{\< xx' \>} \tr_{A}(U_x^{\dagger} U_{x'})
    \;,
 \label{ham:SUN_mixed}
\ee
where of course
\be
   \tr_{A}(U)   \;=\;  |\tr(U)|^2  - 1   \;.
\ee
[If $\beta_A = 0$, this is the usual $SU(N)$ chiral model.
 If $\beta_F = 0$, this is the pure adjoint chiral model,
 which has a global $SU(N) \times SU(N)$ symmetry group and
 a local $Z_N$ gauge group;
 it can therefore be thought of as taking values in 
$M = SU(N)/Z_N$.\footnote{Note also the special case $N=2$: since 
$SU(2)\simeq S^3$ via the correspondence
$U_x=\sigma_x^{(0)} I + i\,\vsigma_x\cdot\vtau$ where 
$\bsigma_x\equiv (\sigma_x^{(0)},\vsigma_x) \in S^3$ and $\vtau$ are
the Pauli matrices, we have $\tr(U_x^{\dagger} U_{x'}) = 2 \,
\bsigma_x\cdot\bsigma_{x'}$, so that the $SU(2)$ mixed
fundamental/adjoint model is equivalent to the $N=4$ mixed
isovector/isotensor model.}]
\end{itemize}

\bigskip

For models 1 and 3, the symmetry group is $SO(N)$,
and the update is
\be
  \bsigma_x^{new} \;=\; e^{i\theta_x RTR^{-1}} \bsigma_x^{old}
                  \;=\; R e^{i\theta_x T} R^{-1} \bsigma_x^{old}
  \;,
\label{N-vector_update}
\ee
where $R$ is a random $SO(N)$ matrix
and $iT$ is a fixed antisymmetric $N \times N$ real matrix.
As mentioned above, we want $T$ to have eigenvalues in $\{\pm 1, 0\}$.
It follows that $T$ has $k$ eigenvalues $+1$,
$k$ eigenvalues $-1$, and $N-2k$ eigenvalues 0,
for some integer $k$ satisfying $1 \le k \le \lfloor N/2 \rfloor$.
Here we shall choose $k=1$;
the subgroup $H$ generated by $RTR^{-1}$
is then rotation in a single plane in $\R^N$.
(Other choices of $T$ will be discussed in Section~\ref{sec2.4.1} below.)
Without loss of generality (because of the random choice of $R$)
we can take
\be
T \;=\;  \left( \begin{array}{cccc}
           0 & i & 0 & \\
           -i & 0 & 0 & \cdots \\
           0 & 0 & 0 & \\
            & \vdots & & \ddots
    \end{array} \right)
   \;.
\label{TNvect}
\ee
We then have the following induced Hamiltonians:

\medskip

{\em Model 1 ($N$-vector model):}\/
The induced $XY$ Hamiltonian is
\be
{\cal H}_{embed} \;=\;  - \sum_{\<xx'\>}
\left[ \alpha_{xx'}\,\cos(\theta_x \,-\, \theta_{x'}) \,+\,
   \beta_{xx'}\,\sin(\theta_x \,-\, \theta_{x'}) \right]
   \,+\, {\rm const}
   \;,
\label{ham_embed}
\ee
where the induced couplings $\{\alpha_{xx'},\beta_{xx'}\}$
are given by
\begin{subeqnarray}
\alpha_{xx'} &=& \beta \, \bsigma_{x} \cdot R T^2 R^{-1} \bsigma_{x'}     \\
\beta_{xx'} &=& \beta \, \bsigma_{x} \cdot R (-iT) R^{-1} \bsigma_{x'}
\label{general}
\end{subeqnarray}
(and for simplicity we have dropped the superscript {\em old}\/
on the $\bsigma$'s).
For the $k=1$ update with the explicit choice \reff{TNvect} for $T$, we have
\begin{subeqnarray}
\alpha_{xx'} &=& \beta \;
            \left[ (R^{-1}\bsigma_x)_1\,(R^{-1}\bsigma_{x'})_1 \,+\,
                   (R^{-1}\bsigma_x)_2\,(R^{-1}\bsigma_{x'})_2 \right] \\
\beta_{xx'} &=& \beta \; 
            \left[ (R^{-1}\bsigma_x)_1\,(R^{-1}\bsigma_{x'})_2 \,-\,
                   (R^{-1}\bsigma_x)_2\,(R^{-1}\bsigma_{x'})_1 \right]
 \label{eq2.12}
\end{subeqnarray}
Note that these couplings are in general {\em disordered}\/
and {\em non-ferromagnetic}\/.  However, we shall show below that
they are {\em unfrustrated}\/:  that is, they are
$U(1)$-gauge-equivalent to a ferromagnetic set of couplings.


\medskip

{\em Model 3 (mixed isovector/isotensor model):}\/
The induced $XY$ Hamiltonian is
\begin{eqnarray}
{\cal H}_{embed} &=&  - \sum_{\<xx'\>}
\left[ \alpha_{xx'}\,\cos(\theta_x \,-\, \theta_{x'}) \,+\,
   \beta_{xx'}\,\sin(\theta_x \,-\, \theta_{x'}) \right. \nonumber \\
& & \qquad\qquad +\,
    \left. \gamma_{xx'}\,\cos (2\theta_x \,-\, 2\theta_{x'}) \,+\,
\delta_{xx'}\,\sin (2\theta_x \,-\, 2\theta_{x'}) \right]
   \,+\, {\rm const}
   \;, \qquad
\label{ham_embed2}
\end{eqnarray}
with the induced couplings $\{\alpha_{xx'},\beta_{xx'},\gamma_{xx'},
\delta_{xx'}\}$
given by
\begin{subeqnarray}
\alpha_{xx'} &=& (\beta_V \,+\, \beta_T\,c_{xx'}) a_{xx'}     \\
\beta_{xx'}  &=& (\beta_V \,+\, \beta_T\,c_{xx'}) b_{xx'}    \\
\gamma_{xx'} &=& \frac{\beta_T}{4} \, (a_{xx'}^2 \,-\, b_{xx'}^2)      \\
\delta_{xx'} &=& \frac{\beta_T}{2} \, a_{xx'}\,b_{xx'}
\label{mixed_couplings}
\end{subeqnarray}
where we have defined
\begin{subeqnarray}
a_{xx'} &=&  \bsigma_x \cdot R T^2 R^{-1} \bsigma_{x'}     \\
b_{xx'} &=&  \bsigma_x \cdot R (-iT) R^{-1} \bsigma_{x'}   \\
c_{xx'} &=&  \bsigma_x \cdot R (I - T^2) R^{-1} \bsigma_{x'}
\end{subeqnarray}
For the $k=1$ update with the explicit choice \reff{TNvect} for $T$, we get
\begin{subeqnarray}
a_{xx'} &=& \left[ (R^{-1}\bsigma_x)_1\,(R^{-1}\bsigma_{x'})_1 \,+\,
                   (R^{-1}\bsigma_x)_2\,(R^{-1}\bsigma_{x'})_2 \right] \\
b_{xx'} &=& \left[ (R^{-1}\bsigma_x)_1\,(R^{-1}\bsigma_{x'})_2 \,-\,
                   (R^{-1}\bsigma_x)_2\,(R^{-1}\bsigma_{x'})_1 \right] \\
c_{xx'}  &=& \sum_{i=3}^N \, (R^{-1}\bsigma_x)_i\,(R^{-1}\bsigma_{x'})_i
\slabel{cxx'}
\label{eq2.17}
\end{subeqnarray}
Note that for $N \ge 3$ we have in general $c_{xx'} \neq 0$,
so that $\alpha_{xx'},\beta_{xx'} \neq 0$
even in the pure isotensor model.\footnote{
\label{T2=1_mixed}
   However, if $N$ is even we may choose the {\em alternate}\/ embedding
   with $k=N/2$:  this yields $T^2 = I$ and hence $c_{xx'} = 0$.
   With {\em this}\/ embedding, the pure isotensor model ($\beta_V=0$)
   has $\alpha_{xx'} = \beta_{xx'} = 0$.
   For further discussion, see Section \ref{sec2.4.1}.}

\bigskip

For models 2 and 5,
the symmetry group is $SU(N)_{left} \times SU(N)_{right}$,
but we shall exploit here only the left-multiplication subgroup.
Thus, we shall take as our update
\be
  U_x^{new} \;=\; e^{i\theta_x RTR^{-1}} U_x^{old}
                  \;=\; R e^{i\theta_x T} R^{-1} U_x^{old}
  \;,
\label{SUN_update}
\ee
where $R$ is a random $SU(N)$ matrix and
$T$ is a fixed traceless Hermitian matrix
with eigenvalues in $\{ \pm 1, 0 \}$.
It follows that $T$ has $k$ eigenvalues $+1$,
$k$ eigenvalues $-1$, and $N-2k$ eigenvalues 0,
where $1 \le k \le \lfloor N/2 \rfloor$.
As before, we shall choose $k=1$
(see Section \ref{sec2.4.1} for other choices).
Without loss of generality we can take
\be
T  \;=\;    \left( \begin{array}{cccc}
           1 & 0 & 0 & \\
           0 & -1 & 0 & \cdots \\
           0 & 0 & 0 & \\
            & \vdots & & \ddots
    \end{array} \right)
\;.
\label{TSUN}
\ee
We then have the following induced Hamiltonians:

\medskip

{\em Model 2 ($SU(N)$ principal chiral model):}\/
The induced $XY$ Hamiltonian is again of the form \reff{ham_embed},
with couplings
\begin{subeqnarray}
\alpha_{xx'} &=& \beta\, \re\tr ( U_x^{\dagger} R T^2 R^{-1} U_{x'})   \\
\beta_{xx'} &=&  \beta\, \re\tr ( U_x^{\dagger} R (-iT) R^{-1} U_{x'})
           \;=\; \beta\, \im\tr ( U_x^{\dagger} R T R^{-1} U_{x'})
\label{suncouplings}
\end{subeqnarray}
For the $k=1$ update with the explicit choice \reff{TSUN} for $T$, we have
\begin{subeqnarray}
\alpha_{xx'} &=& \beta\, 
    \left[ \re (R^{-1}\,U_{x'}\,U_x^{\dagger}\,R)_{11} \,+\,
           \re (R^{-1}\,U_{x'}\,U_x^{\dagger}\,R)_{22} \right]   \\
\beta_{xx'} &=& \beta\, 
    \left[ \im (R^{-1}\,U_{x'}\,U_x^{\dagger}\,R)_{11} \,-\,
           \im (R^{-1}\,U_{x'}\,U_x^{\dagger}\,R)_{22} \right]   
  \label{SUN_couplings_explicit}
\end{subeqnarray}
Note that these couplings are in general {\em disordered}\/
and {\em non-ferromagnetic}\/;
moreover, as will be seen in the next subsection,
they are also generically {\em frustrated}\/,
in contrast to what happens for the $N$-vector models.

\medskip

{\em Model 5 (mixed fundamental/adjoint $SU(N)$ chiral model):}\/
The induced $XY$ Hamiltonian is again of the form \reff{ham_embed2},
with couplings
\begin{subeqnarray}
\alpha_{xx'} &=& \re \,[(\beta_F \,+\,2\,\beta_A \,c_{xx'}^*)\,a_{xx'}]  \\
\beta_{xx'} &=& \re \,[(\beta_F  \,+\,2\,\beta_A \,c_{xx'}^*)\,b_{xx'}]  \\
\gamma_{xx'} &=& \frac{\beta_A}{2} \, ( |a_{xx'}|^2 \,-\, |b_{xx'}|^2 )  \\
\delta_{xx'} &=& \beta_A \, \re (a_{xx'}\,b_{xx'}^*)
\label{mixed_adjoint_couplings}
\end{subeqnarray}
where we have defined
\begin{subeqnarray}
a_{xx'} &=&  \tr [U_x^{\dagger}  R T^2 R^{-1} U_{x'} ]    \\
b_{xx'} &=& \tr [ U_x^{\dagger}  R (-iT) R^{-1} U_{x'}]   \\
c_{xx'} &=& \tr [ U_x^{\dagger}  R (I - T^2) R^{-1} U_{x'}]
\end{subeqnarray}
Note that for $N=2$ we have $T^2 = I$ and hence $c_{xx'} = 0$.
On the other hand, for $N \ge 3$ we have in general $c_{xx'} \neq 0$,
so that $\alpha_{xx'},\beta_{xx'} \neq 0$ even in the pure adjoint 
model.\footnote{ However, if $N$ is even we may choose the {\em alternate}
embedding with $k=N/2$: this yields $T^2=I$ and hence $c_{xx'}=0$.
With {\em this} embedding, the pure adjoint model ($\beta_F=0$) has 
$\alpha_{xx'}\,=\,\beta_{xx'}\,=\,0$.}
\medskip

{\em Model 4 (complex mixed isovector/isotensor model):}\/
The symmetry group is $U(N)$, and we take the update to be
\be
  \bz_x^{new} \;=\; e^{i\theta_x RTR^{-1}} \bz_x^{old}
                  \;=\; R e^{i\theta_x T} R^{-1} \bz_x^{old}
  \;,
\ee
where $R$ is a random $SU(N)$ matrix and
$T$ is a fixed Hermitian matrix
with eigenvalues in $\{ \pm 1, 0 \}$.\footnote{
   Note that $T$ is {\em not}\/ required to be traceless,
   even in the pure $\CP^{N-1}$ model ($\beta_V  = 0$).
   Of course, in this latter model the $U(1)$ subgroup of $U(N)$
   is a local gauge group, so only the traceless part of $T$
   [i.e.\ $T - (N^{-1} \tr T) I$] plays an interesting dynamical role.
   In particular, two update matrices $T$ and $T'$ that differ by a 
   multiple of the identity give rise to physically equivalent 
   embeddings (differing only by a gauge transformation) and
   identical induced $XY$ Hamiltonians.
\label{footnote_traceless_CPN}
}
It follows that $T$ has $k$ eigenvalues $+1$,
$l$ eigenvalues $-1$, and $N-k-l$ eigenvalues 0, for suitable integers
$k$ and $l$.
Here we shall choose $k=1$, $l=0$ (see also \cite{Hasenbusch_CP3});
without loss of generality we can take
\be
T  \;=\;    \left( \begin{array}{cccc}
           1 & 0 & 0 & \\
           0 & 0 & 0 & \cdots \\
           0 & 0 & 0 & \\
            & \vdots & & \ddots
    \end{array} \right)
\;.
\label{TUN}
\ee
The evaluation of the embedded Hamiltonian is analogous to model 3;
it is of the form \reff{ham_embed2}, with couplings
\begin{subeqnarray}
\alpha_{xx'} &=& \re [ (\beta_V \,+\, \beta_T \,c_{xx'}^*) \, a_{xx'}]   \\
\beta_{xx'}  &=& \re [ (\beta_V \,+\, \beta_T \,c_{xx'}^*) \, b_{xx'}]   \\
\gamma_{xx'} &=& \frac{\beta_T}{4} \, ( |a_{xx'}|^2 \,-\, |b_{xx'}|^2 )  \\
\delta_{xx'} &=& \frac{\beta_T}{2} \, \re (a_{xx'}\,b_{xx'}^*)
\label{zcouplings}
\end{subeqnarray}
where we have defined
\begin{subeqnarray}
a_{xx'} &=&  \bz_x^* \cdot R T^2 R^{-1} \bz_{x'}     \\
b_{xx'} &=&  \bz_x^* \cdot R (-iT) R^{-1} \bz_{x'}   \\
c_{xx'} &=&  \bz_x^* \cdot R (I - T^2) R^{-1} \bz_{x'}
\slabel{complex_cxx'}
\label{z_coeffs}
\end{subeqnarray}
Note that because of the particular choice \reff{TUN} of $T$,
we have $T^2 = T$ and hence $b_{xx'} = -i a_{xx'}$,
which implies that $\gamma_{xx'} = \delta_{xx'} = 0$.
So, perhaps surprisingly,
the embedded Hamiltonian is of the simple form \reff{ham_embed}
in spite of the isotensor couplings in \reff{ham:CPN}.
This happens whenever all the complex rotations are in the same direction,
i.e.\  either $l=0$ (hence $T^2 = T$) or $k=0$ (hence $T^2 = -T$).
In these cases we have 
\be
\bz_x^* \cdot \bz_{x'} \,=\, A_{xx'} e^{\pm i(\theta_x - \theta_{x'})}
     \,+\, B_{xx'}
\; \mbox{,}
\ee
without double-angle terms.\footnote{There is another interesting
possibility: If we take $k+l=N$, we have $T^2 = I$ and hence
$c_{xx'}=0$. With this embedding, the pure isotensor model ($\beta_V=0$)
has $\alpha_{xx'}=\beta_{xx'}=0$. Of course, this embedding is 
gauge-equivalent to one with $k'=k$, $l'=0$ and $\theta_x$ replaced 
by $2\theta_x$ (see footnote \ref{footnote_traceless_CPN}). 
For further discussion, see Section 2.4.1.
\label{gauge-equivalent_CPN}
}

\subsubsection{Alternate Interpretation} \label{sec2.2.2}

For the two models taking values in $S^{N-1}$ --- viz.\ the $N$-vector
and mixed isovector/isotensor models --- there is a simpler and more
natural way of specifying the embedding of angular variables.
The embedding is clearly given by choosing randomly a two-dimensional
subspace $P \subset \R^N$
and then rotating only the component of $\bsigma_x$ in $P$.
[From \reff{N-vector_update}/\reff{TNvect} we see that $P$ is
 the image of the 1--2 plane under the rotation $R$.]
To be precise, let us decompose each vector $\bsigma_x$ into its parts
parallel and perpendicular to the plane $P$:
\be
   \bsigma_x \;=\; \bsigma_x^{\parallel}  \,+\,  \bsigma_x^{\perp}
   \;.
\label{decompose}
\ee
We then fix $\{ \bsigma_x^{\perp} \}$ at their current values,
and consider only $\{ \bsigma_x^{\parallel} \}$
as dynamical variables.
Substituting \reff{decompose} into the original $N$-vector Hamiltonian
\reff{N-vector}, we get
\be
{\cal H}_{N-vector} \;=\; 
 -\beta \, \sum_{\<xx'\>} \bsigma_x
\cdot \bsigma_{x'} \;=\; -\beta \,  \sum_{\<xx'\>}
     \bsigma_x^{\parallel} \cdot \bsigma_{x'}^{\parallel} \,+\,
    {\rm const}
\;\mbox{.}
\ee
The angular variables $\{ \theta_x \}$
can then be defined in either of two convenient ways:
\begin{itemize}
\item[1)] Let $\{\theta_x\}$ be the angles of $\bsigma_x^{\parallel}$ with
respect to the some fixed basis vectors in the plane $P$.  In this case
$\bsigma_x^{\parallel} \cdot \bsigma_{x'}^{\parallel}
 = |\bsigma_x^{\parallel}|\,|\bsigma_{x'}^{\parallel}|
   \,\cos(\theta_x - \theta_{x'})$,
so that the embedding has the form \reff{ham_embed}, with couplings
\begin{subeqnarray}
\alpha_{xx'} &=& \beta\, |\bsigma_x^{\parallel}|\,|\bsigma_{x'}^{\parallel}| \\
\beta_{xx'} &=& 0
\label{ferromag}
\end{subeqnarray}
and the embedded model is simulated starting from the present configuration
for the angles (call it $\{\overline{\theta}_x\}$).
These couplings are clearly ferromagnetic (provided that $\beta \ge 0$).
This is the embedding implementation that is used in our MGMC program.

\item[2)]
Alternatively, we can take $\{\theta_x\}$ to be the angles of
$\bsigma_x^{\parallel}$ with respect to its present orientation
(given by $\{\overline{\theta}_x\}$ in the previous coordinates).
In this case the couplings are given by
\begin{subeqnarray}
\alpha_{xx'} &=& \;\;\beta\,
            |\bsigma_x^{\parallel}|\,|\bsigma_{x'}^{\parallel}| \;
                \cos(\overline{\theta}_x \,-\, \overline{\theta}_{x'})   \\
\beta_{xx'} &=& - \beta\,
             |\bsigma_x^{\parallel}|\,|\bsigma_{x'}^{\parallel}| \;
                \sin(\overline{\theta}_x \,-\, \overline{\theta}_{x'}) 
\label{disordered}
\end{subeqnarray}
and the embedded model is simulated starting from $\theta_x \equiv 0$.
This is the approach taken in \reff{N-vector_update}--\reff{eq2.12} above;
and of course \reff{disordered} is identical to \reff{eq2.12}.
\end{itemize}

The sets of couplings \reff{ferromag} and \reff{disordered} are 
related by a $U(1)$ gauge transformation:
namely, the angle $\,\theta_x^{(1)}\,$ in the
``first'' variables is equal to $\,\theta_x^{(2)} + 
\overline{\theta}_x\,$ in the ``latter'' variables, for all $x$. 
Therefore, the couplings \reff{eq2.12}/\reff{disordered},
though in general non-ferromagnetic, are nevertheless {\em unfrustrated}\/,
since they are $U(1)$-gauge-equivalent to the
ferromagnetic couplings \reff{ferromag}.

For the mixed isovector/isotensor model we can use the same
decomposition \reff{decompose}, but the embedded Hamiltonian
becomes somewhat more complicated.
In interpretation \#2, the induced $XY$ Hamiltonian has already been given
in \reff{ham_embed2}--\reff{eq2.17}.
In interpretation \#1, we get
\begin{eqnarray}
{\cal H}_{embed} &\equiv&  - \sum_{\<xx'\>}
\biggl[ (\beta_V \,+\,\beta_T \, \bsigma_x^{\perp} \cdot \bsigma_{x'}^{\perp})
 \, |\bsigma_x^{\parallel}|\,|\bsigma_{x'}^{\parallel}| \,
 \cos(\theta_x \,-\, \theta_{x'})                            \nonumber \\
& & \qquad\qquad \,+\, \frac{\beta_T}{4} \,
 |\bsigma_x^{\parallel}|^2\,|\bsigma_{x'}^{\parallel}|^2
    \,\cos (2\theta_x \,-\, 2\theta_{x'}) \biggr]  \,+\, {\rm const}
 \;.
\label{mixed_ham_embed}
\end{eqnarray}
The couplings are clearly ferromagnetic if $\beta_V \ge \beta_T \ge 0$.


\medskip

{\bf Remark.}  A similar representation can be employed for the
{\em complex}\/ mixed isovector/isotensor model with the choice
\reff{TUN} of $T$.
The embedding is specified by choosing a random complex direction
$\bv \in \C^N$ (in fact $\bv = R {\bf e}_1$) and then rotating
only the component of $\bz_x$ along $\bv$.
That is, we decompose $\bz_x = \bz_x^\parallel + \bz_x^\perp$ with
\be
   \bz_x^\parallel   \;=\;   (\bz_x \cdot \bv^{*}) \, \bv  \;.
\ee
The analogue of interpretation \#1 is obtained by defining
\be
   \theta_x   \;=\;   \arg(\bz_x \cdot \bv^{*})   \;.
\ee
The induced $XY$ Hamiltonian is of the form \reff{ham_embed} (``single-angle''
couplings only), with
\begin{subeqnarray}
\alpha_{xx'} &=&
    [\beta_V  \,+\,   \beta_T \real (\bz_x^{\perp *} \cdot \bz_{x'}^\perp)]
    \, |\bz_x^\parallel| \, |\bz_{x'}^\parallel|  \\
\beta_{xx'} &=&  - \beta_T \imag (\bz_x^{\perp *} \cdot \bz_{x'}^\perp)
    \, |\bz_x^\parallel| \, |\bz_{x'}^\parallel|
\label{zinterp_1}
\end{subeqnarray}
However, these couplings are not in general ferromagnetic,
even if $\beta_V \ge \beta_T \ge 0$,
since $\bz_x^{\perp *} \cdot \bz_{x'}^\perp$ need not be real.
In fact, we shall show in the next section that they are in 
general frustrated, except in the pure $\CP^1$ model ($N=2$, $\beta_V=0$).
In interpretation  \#2 we have the expression \reff{zcouplings}, with
\begin{subeqnarray}
a_{xx'} &=& i\, b_{xx'} \;=\; 
 \bz_x^{\parallel *} \cdot \bz_{x'}^\parallel
\;=\; |\bz_x^\parallel| \, |\bz_{x'}^\parallel| 
    \,e^{-i(\overline{\theta}_x \,-\, \overline{\theta}_{x'})}  \\
c_{xx'} &=& \bz_x^{\perp *} \cdot \bz_{x'}^\perp
\end{subeqnarray}
and hence
\begin{subeqnarray}
\alpha_{xx'} &=& \re [ (\beta_V \,+\, 
\beta_T \,\bz_x^{\perp} \cdot \bz_{x'}^{\perp *}) \, 
\bz_x^{\parallel *} \cdot \bz_{x'}^\parallel]   \\
\beta_{xx'}  &=& \im [ (\beta_V \,+\, 
\beta_T\,\bz_x^{\perp} \cdot \bz_{x'}^{\perp *}) \, 
\bz_x^{\parallel *} \cdot \bz_{x'}^\parallel]   \\
\gamma_{xx'} &=& \delta_{xx'} \;=\;0
\label{zinterp_2}
\end{subeqnarray}


\subsection{Frustration}   \label{sec2.3}

Although our original models \reff{ham_N-vector}--\reff{ham:SUN_mixed}
are ferromagnetic and translation-invariant,
we have seen that the corresponding induced $XY$ Hamiltonians
are generically non-ferro\-mag\-net\-ic and non-translation-invariant:
they contain both
$\cos(\theta_x - \theta_{x'})$ and $\sin(\theta_x - \theta_{x'})$ terms,
with coefficients of arbitrary sign.
This raises the question of whether the induced $XY$ Hamiltonians
may be frustrated.  Recall that a Hamiltonian\footnote{
   More precisely, frustration is not a property of a Hamiltonian
   {\em per se}\/, but of a specific {\em decomposition}\/ of that
   Hamiltonian into ``elementary bonds''.
   Here we shall always consider the elementary terms in the
   Hamiltonian to be the {\em complete}\/ energy associated with
   a {\em single}\/ bond $\< xx' \>$.
}
is said to be {\em unfrustrated}\/ if there exists a configuration
of the field variables that simultaneously minimizes the energy
on each bond.

One case is easy:  any {\em ferromagnetic}\/ $XY$ Hamiltonian
is obviously unfrustrated, since any configuration
$\theta_x \equiv {\rm constant}$ (independent of $x$)
simultaneously minimizes the energy on each bond.
Furthermore, any $XY$ Hamiltonian that is $U(1)$-gauge-equivalent
to a ferromagnetic Hamiltonian is also unfrustrated.
Thus, the induced $XY$ Hamiltonian for the $N$-vector model is
unfrustrated:  it is either ferromagnetic (interpretation \#1)
or gauge-equivalent to a ferromagnetic one (interpretation \#2).
This is also true for the mixed isovector/isotensor model with
$\beta_V \ge \beta_T \ge 0$ because of \reff{mixed_ham_embed}.\footnote{For
a more complete discussion of the mixed case, see Section 2.4.1.
}

For the $SU(N)$ principal chiral models, however, the situation is
less clear, as no manifestly ferromagnetic representation appears to be
available.  Here we shall show that no such representation can exist:
the induced $XY$ Hamiltonians are in general {\em frustrated}\/
for all $N \ge 2$.  We shall also locate the underlying cause of
this frustration.

{\bf Remark.} When we say that the induced $XY$ Hamiltonians are 
``in general frustrated'', what we mean is that there {\em exist}
configurations $\{\varphi_x^{old}\}$ of the original model
(which we shall present explicitly) for which the induced $XY$
Hamiltonians are frustrated. We conjecture that in fact a much
stronger statement is true, namely that the induced $XY$ Hamiltonians are
{\em generically} frustrated, in the sense that frustration
occurs for {\em all} configurations $\{\varphi_x^{old}\}$ except those lying 
on a lower-dimensional submanifold of the configuration space.
We expect that this can be proven by suitable modifications of our
examples, but we have not attempted to do so.

To test whether the generalized $XY$ model \reff{ham_embed}
is frustrated, we consider the
product around a plaquette of the quantity $\alpha_{xx'} - i \beta_{xx'}$.
If this product is real and positive for {\em all}\/ plaquettes,
then the model is unfrustrated.\footnote{
   In periodic boundary conditions this statement is not quite correct:
   one should also insist that this product be real and positive for
   ``Polyakov loops'' that wind around the lattice by periodicity.
}
Conversely, if for some plaquette the product is {\em not}\/ real and
positive, then the model is frustrated.
This is because we can rewrite \reff{ham_embed} as  
\be
{\cal H}_{embed} \;=\;  - \sum_{\<xx'\>}
\re \left[ (\alpha_{xx'}\,-\,i\,\beta_{xx'}) \,
         e^{i\,(\theta_x \,-\, \theta_{x'})} \right]
    \;=\; - \sum_{\<xx'\>}
\re \left[ \rho_{xx'}\, e^{i\,(\phi_{xx'}\,+\,
    \theta_x \,-\, \theta_{x'})} \right]
\;\mbox{,}
\ee
where $\rho_{xx'} \equiv (\alpha_{xx'}^2 + \beta_{xx'}^2)^{1/2} \ge 0$
and $\phi_{xx'} = \arg(\alpha_{xx'} - i \beta_{xx'})$;
to minimize the energy on bond $\<xx'\>$ we need
\be
\theta_{x'} \;=\; \theta_x \,+\, \phi_{xx'}
\ee
(provided that $\rho_{xx'} \neq 0$).
This can be done simultaneously for all bonds $\<xx'\>$
if and only if for each plaquette
we have
\be
\sum_{\Box} \, \phi_{xx'} \;=\; 0\,(\mbox{mod}\, 2\pi)
\;\mbox{,}
\ee
or, equivalently,
\be
\prod_{\Box} \, (\alpha_{xx'}\,-\,i\,\beta_{xx'}) \; 
\mbox{ real and positive. }
\label{prod_plaq}
\ee

Let us now use the criterion \reff{prod_plaq} to test for frustration
in our various models:
\bigskip

{\em Model 2 (SU(N) principal chiral model):}
We shall prove frustration for all $N\geq 2$, by showing
that \reff{prod_plaq} does {\em not}\/ in general hold
for the embedding \reff{SUN_update}--\reff{SUN_couplings_explicit}.
Without loss of generality we can take $R=I$.
We begin with the simplest case $N=2$.


In the case of $SU(2)$ matrices, it is easy to show that
for our choice of $T$ we get from \reff{suncouplings} the simple
form
\be
\alpha_{xx'}\,-\,i\,\beta_{xx'} \;=\; 2\,\beta
\, {(U_x\,U_{x'}^{\dagger})}_{11}
\;\mbox{.}
\ee
Then we want to show that around a plaquette with sites $A,B,C,D$
the product 
\be
{(U_A\,U_B^{\dagger})}_{11} \, {(U_B\,U_C^{\dagger})}_{11} \, 
{(U_C\,U_D^{\dagger})}_{11} \, {(U_D\,U_A^{\dagger})}_{11} 
\label{su2_plaq}
\ee
is not necessarily real and positive. It suffices to take
$U_A\,=\,U_D\,=\,I$ and general $U_B$ and $U_C$:
\begin{subeqnarray}
U_B &=& \left( \begin{array}{cc}
             \alpha & \beta \\
             -\beta^{*} & \alpha^{*}
          \end{array} \right)
\\[2mm]
U_C &=& \left( \begin{array}{cc}
             \gamma & \delta \\
             -\delta^{*} & \gamma^{*}
          \end{array} \right) 
\end{subeqnarray}
where $|\alpha|^2 + |\beta|^2 = |\gamma|^2 + |\delta|^2  = 1$.
We get for the product in \reff{su2_plaq}
\be
|\alpha|^2 \, |\gamma|^2 \,+\, \alpha^{*} \beta \gamma \delta^{*}
\;.
\ee
Although the first term is real and positive, the second is free to have an
imaginary part (for example, take $\alpha=\cos\theta$, 
$\beta=\sin\theta$, $\gamma=\cos\theta'$, $\delta=i\sin\theta'$),
and therefore the couplings are in general frustrated.

This example can be immediately extended to $SU(N)$ for $N \ge 3$:
it suffices to take matrices $U$ which are block-diagonal,
in which the uppermost $2\times 2$ block is as above,
and the lowermost $(N-2) \times (N-2)$ block is an arbitrary $SU(N-2)$ matrix.
Thus, for all $N \ge 2$ our $SU(N)$ algorithm leads in general
to a {\em frustrated}\/ induced $XY$ model.
 
The presence of frustration even in the $SU(2)$ case may seem surprising,
as there is an isomorphism relating the
$SU(2)$ chiral model to the 4-vector model,
and we know that the embedding \reff{N-vector_update}/\reff{TNvect}
for $N$-vector models is unfrustrated.  The explanation is that the
$SU(2)$ embedding \reff{SUN_update} does {\em not}\/ map into the
4-vector embedding \reff{N-vector_update}/\reff{TNvect},
but rather into a {\em different}\/ embedding!
Let us work this out in detail, using the isomorphism
\be
   U  \;=\; s_0 \,I \,+\, i \vs \cdot \vsigma
\label{U4vec}
\ee
between $U \in SU(2)$ and $\bs=(s_0,s_1,s_2,s_3) \in S^3 \subset \R^4$;
here $I$ is the $2\times2$ identity matrix,
$\vsigma$ represents the Pauli matrices, and
we use the symbol $\vs$ to denote the $3$-vector $(s_1,s_2,s_3)$.
The Hamiltonian \reff{ham_SUN} for the $SU(2)$ chiral model
corresponds to the Hamiltonian \reff{ham_N-vector} for the 4-vector model,
with the identification $\beta_{4-vector} = 2\beta_{SU(2)}$.

Let us now work out the update at a single site $x$
(for simplicity we suppress the subscript $x$ from the notation).
We have
\be
   RTR^{-1} \;=\; \va \cdot \vsigma
 \label{RTR}
\ee
for some unit vector $\va \in \R^3$.
Substituting \reff{U4vec} and \reff{RTR} in \reff{SUN_update}, we get
\be
 s_0^{new} I \,+\, i\,\vs^{new}\cdot \vsigma  \;\,=\;\,
(\cos\theta\,+\, i\,\va \cdot \vsigma \, \sin\theta) \,
 (s_0^{old} I \,+\, i\,\vs^{old}\cdot \vsigma)
\ee
and hence
\begin{subeqnarray}
s_0^{new} &=& (\cos\theta)\,s_0^{old} \,-\, (\sin\theta)\,\va \cdot \vs^{old} \\
\vs^{new} &=& (\cos\theta)\,\vs^{old}\,+\, (\sin\theta)
         \,[s_0^{old}\,\va \,-\, (\va \times \vs^{old})]
\label{SU2_S3_update}
\end{subeqnarray}
where we have used
\be
(\va \cdot \vsigma)(\vs\cdot \vsigma)\;=\; \va \cdot \vs \,+\,
i\,(\va \times \vs)\cdot \vsigma
\;.
\ee
For example, if we choose $\va=(1,0,0)$, we have
\begin{subeqnarray}
s_0^{new} &=& (\cos\theta)\,s_0^{old} \,-\,(\sin\theta) \, s_1^{old} \\
s_1^{new} &=& (\sin\theta)\,s_0^{old} \,+\,(\cos\theta) \, s_1^{old} \\
s_2^{new} &=& (\cos\theta)\,s_2^{old} \,+\,(\sin\theta) \, s_3^{old} \\
s_3^{new} &=& -(\sin\theta)\,s_2^{old} \,+\,(\cos\theta) \, s_3^{old}
 \label{SU2_update}
\end{subeqnarray}
and we can see that this corresponds to {\em two}\/ coupled rotations
for the $4$-vector $\bs$: a counterclockwise rotation on
the plane defined by the 0-direction and $\va$, and a clockwise rotation on the
plane orthogonal to this one.
By contrast, our standard 4-vector update \reff{N-vector_update},
with $T$ given by the $k=1$ choice \reff{TNvect},
corresponds to a rotation in one plane only.
Clearly, the $SU(2)$ update \reff{SU2_update}
corresponds to the 4-vector update \reff{N-vector_update}
with $T$ given instead by the $k=2$ choice
\be
T \;=\;    \left( \begin{array}{cccc}
           0 & i & 0 & 0 \\
           -i & 0 & 0 & 0 \\
           0 & 0 & 0 & -i \\
           0 & 0 & i & 0
    \end{array} \right)
    \;\,.
  \label{T_k=2}
\ee

The induced $XY$ Hamiltonian is then \reff{ham_embed} with the couplings
\begin{subeqnarray}
\alpha_{xx'} &=& \;\;2\,\beta\,\left[
            |\bsigma_x^{\parallel}|\,|\bsigma_{x'}^{\parallel}| \;
\cos(\overline{\theta}_x^{\parallel} \,-\, \overline{\theta}_{x'}^{\parallel})
\,+\; |\bsigma_x^{\perp}|\,|\bsigma_{x'}^{\perp}| \;
  \cos(\overline{\theta}_x^{\perp} \,-\, \overline{\theta}_{x'}^{\perp})\right]
\\
\beta_{xx'} &=& - 2\,\beta\,\left[
             |\bsigma_x^{\parallel}|\,|\bsigma_{x'}^{\parallel}| \;
\sin(\overline{\theta}_x^{\parallel} \,-\, \overline{\theta}_{x'}^{\parallel})
\,-\; |\bsigma_x^{\perp}|\,|\bsigma_{x'}^{\perp}| \;
\sin(\overline{\theta}_x^{\perp} \,-\, \overline{\theta}_{x'}^{\perp})\right]
\label{frust_couplings}
\end{subeqnarray}
where $\overline{\theta}_x^{\parallel}$ and
$\overline{\theta}_x^{\perp}$ are the initial angles
of $\bsigma_x^{\parallel} \equiv (s_0,s_1)$
and $\bsigma_x^{\perp} \equiv (s_2,s_3)$, respectively.
These couplings are clearly frustrated, since although there exist gauge
transformations that will make the ``parallel'' or the ``perpendicular''
part of the couplings become ferromagnetic separately, we cannot in general
make them {\em simultaneously}\/ ferromagnetic (unless by chance we had
$\overline{\theta}_x^{\parallel} = \overline{\theta}_x^{\perp}$
for all $x$, or some similar degenerate situation).
For example, consider a plaquette $ABCD$ with 
angles ($\overline{\theta}_x^{\parallel},\overline{\theta}_x^{\perp}$)
given by $(\theta,0),(0,0),(0,0),(0,0)$, respectively.
Then the product of
\be
\alpha_{xx'}\,-\,i\beta_{xx'} \;=\;
2\beta\,\left[|\bsigma_x^{\parallel}|\,|\bsigma_{x'}^{\parallel}|
\,e^{i(\overline{\theta}_x^{\parallel}\,-\,\overline{\theta}_{x'}^{\parallel})}
\,+\, |\bsigma_x^{\perp}|\,|\bsigma_{x'}^{\perp}| \,
e^{-i(\overline{\theta}_x^{\perp} \,-\, \overline{\theta}_{x'}^{\perp})}\right]
\ee
around the plaquette will have a complex part proportional to
\be
i \sin\theta \,|\bsigma_A^{\parallel}|\,|\bsigma_A^{\perp}|
\,(|\bsigma_B^{\parallel}|\,|\bsigma_D^{\perp}| \,-\,
|\bsigma_D^{\parallel}|\,|\bsigma_B^{\perp}|)
\;,
\ee
which can clearly be nonzero.

{\bf Remark.}
The $k=2$ 4-vector update with both rotations clockwise, namely
\be
T \;=\;    \left( \begin{array}{cccc}
           0 & i & 0 & 0 \\
           -i & 0 & 0 & 0 \\
           0 & 0 & 0 & i \\
           0 & 0 & -i & 0
    \end{array} \right)
    \;\,,
\label{T_right_update}
\ee
corresponds to the {\em right} $SU(2)$ update
\be
U^{new}\;=\;U^{old}\,e^{i\theta \,
 \vec{\scriptsize\it a}\cdot \vec{\scriptsize \sigma}}
\;.
\ee
\bigskip

{\em Model 5 (mixed fundamental/adjoint SU(N) chiral model)}:
Let us now show, using \reff{mixed_adjoint_couplings}, that
frustration occurs also for the mixed fundamental/adjoint
$SU(2)$ chiral model whenever $\beta_A\geq0$ (and $\beta_F,\beta_A$ are
not both zero). The point is that for $SU(2)$ matrices we have 
$c_{xx'}=0$ and $a_{xx'},b_{xx'}$ both real; it follows that
\begin{subeqnarray}
\alpha_{xx'}\,-\,i\beta_{xx'} &=& \beta_F\,(a_{xx'}\,-\,ib_{xx'})
\slabel{SU2_ab} \\
\gamma_{xx'}\,-\,i\delta_{xx'}&=& \frac{\beta_A}{2}
       \,(a_{xx'}\,-\,ib_{xx'})^2
\end{subeqnarray}
Hence, if we write the induced $XY$ Hamiltonian in terms of 
``frustration angles'',
\begin{subeqnarray}
{\cal H}_{embed} &=&  - \sum_{\<xx'\>}
\re \left[ (\alpha_{xx'}\,-\,i\,\beta_{xx'}) \,
         e^{i\,(\theta_x \,-\, \theta_{x'})}
\,+\, (\gamma_{xx'}\,-\,i\,\delta_{xx'}) \,
        e^{2i\,(\theta_x \,-\, \theta_{x'})} \right] \\
    &=& - \sum_{\<xx'\>}
\re \left[ \rho_{xx'}\, e^{i\,(\phi_{xx'}\,+\,
    \theta_x \,-\, \theta_{x'})}
\;+\;  \rho'_{xx'}\, e^{i\,(\phi'_{xx'}\,+\,
2\,\theta_x \,-\, 2\,\theta_{x'})} \right]
\;,
\label{frust_angles}
\end{subeqnarray}
we have $\phi'_{xx'}=2\phi_{xx'}$ (mod $2\pi$) provided that $\beta_A\geq0$.
It follows that the presence or absence of frustration can be determined
solely by looking at $\prod\protect_{\Box} (\alpha_{xx'}\,-\,i\beta_{xx'})$.
And as we have already seen, for certain configurations this product has 
a nonzero imaginary part. 

To extend this proof to $SU(N)$ with $N\geq2$,
we take (when $R=I$) the matrices $U$ to be block-diagonal, in which the
uppermost $2\times2$ block is as above, and the lowermost $(N-2)\times(N-2)$
block is the {\em identity matrix}. We then have each $c_{xx'}=N-2$, so 
that the only change in the foregoing argument is that, in \reff{SU2_ab},
the prefactor $\beta_F$ is replaced by $\beta_F+2\beta_A(N-2)$.

\bigskip 

{\em Model 4 (complex mixed isovector/isotensor model):}
Likewise, we can show that the
complex mixed isovector/isotensor model with the choice
\reff{TUN} of $T$ leads to a frustrated induced $XY$ model
in the following two cases:
\begin{itemize}
\item[(i)] For all $N \ge 2$,
provided that $\beta_V, \beta_T \neq 0$.
Without loss of generality we can take $R=I$, hence
\be
   \bv \;=\; (1,0,0, \ldots)   \;.
\ee
Consider the product of $\alpha_{xx'} \,-\, i\beta_{xx'}$
around a plaquette. We have, from \reff{zinterp_2},
\be
\alpha_{xx'} \,-\, i\beta_{xx'} \;=\; 
   [\beta_V  \,+\,   \beta_T (\bz_x^{\perp *} \cdot \bz_{x'}^\perp)]
   \, \bz_x^{\parallel}\cdot \bz_{x'}^{\parallel *}
   \;.
\label{alfa-ibeta_complex_mixed}
\ee
Now consider, on the plaquette {\em ABCD}\/, the configuration
\begin{subeqnarray}
\bz_A &=& \bz_D \;=\;  (1,0,0, \ldots)    \\
\bz_B &=& (\cos\lambda, \sin\lambda, 0 , \ldots) \\
\bz_C&=& (\cos\tau, i\sin\tau, 0 , \ldots)
\label{z_config}
\end{subeqnarray}
We get
\be
   \prod_{\Box} (\alpha_{xx'} \,-\, i\beta_{xx'}) \;=\;
\cos^2\lambda\,\cos^2\tau\,\beta_V^3\,
   (\beta_V \,+\, i\beta_T \,  \sin\lambda \, \sin\tau)
\;,
\ee
which shows that the embedding is {\em frustrated}\/
if we take $\sin\lambda,\sin\tau, \cos\lambda, \cos\tau \neq 0$.

%

\item[(ii)] For all $N\geq 3$, whenever $\beta_T\neq0$.
Consider, on the plaquette
$ABCD$ the configuration
\begin{subeqnarray}
\bz_A &=& \bz_D \;=\;  (\cos \lambda, \sin\lambda,0,0, \ldots)    \\
\bz_B &=& \frac{1}{\sqrt3} \, (1, 1, 1, 0 , \ldots) \\[1mm]
\bz_C&=&  \frac{1}{\sqrt3}\, (1, 1,i, 0 , \ldots)
\label{z_config2}
\end{subeqnarray}
Then, using \reff{alfa-ibeta_complex_mixed} we get
\begin{eqnarray}
\prod_{\Box} (\alpha_{xx'}\,-\,i\beta_{xx'}) &=&
\frac{\cos^4 \lambda}{9}\,(\beta_V\,+\,\beta_T\,\sin^2\lambda)\,
\left(\beta_V\,+\,\beta_T\,\frac{\sin \lambda}{\sqrt 3}\right)^2\,\times
\nonumber \\[2mm]
& & \hskip 3cm \left[\beta_V\,+\,\frac{\beta_T}{3}\,(1\,+\,i)\right]
\;.
\end{eqnarray}
This has a nonzero imaginary part whenever $\beta_T\neq0$,
provided we choose $\lambda$ so that the first three factors are
nonvanishing; therefore the embedding is frustrated.
\end{itemize}
\par\noindent
The remaining cases are unfrustrated:
\begin{itemize}
\item[(iii)] If $\beta_T=0$, this is equivalent to the $2N$-vector model with 
our standard $k=1$ update.
\item[(iv)] If $N=2$ and $\beta_V=0$, this is the $\CP^1$ model, which is
equivalent to the 3-vector model (see footnote \ref{CP1_equiv_O3});
and it is easily seen that the update corresponds to our standard $k=1$
update for this latter model. The nonfrustration can also be seen directly
from \reff{alfa-ibeta_complex_mixed}, once one notes that 
$\bz_x^{\parallel}$ and $\bz_x^{\perp}$ are both one-dimensional.
\end{itemize}

\subsection{Alternative $XY$ Embeddings}   \label{sec2.4}

\subsubsection{Embeddings with $k \neq 1$}  \label{sec2.4.1}


Thus far we have been concentrating on the embeddings in which the
matrix $T$ has only one eigenvalue equal to $+1$, i.e.\  the cases
$k=1$. Here we shall examine the cases with $k>1$; in particular we
shall investigate the occurrence of frustration.

Note first that we must always have $N\geq 2k$ (or $N\geq k+l$
in the case of the complex mixed isovector/isotensor model).
For each model, therefore, we shall begin by investigating the minimal
cases $N=2k$ (or $N=k+l$). Once frustration has been proven for such
a case, the proof can be immediately extended to all higher values of
$N$ for the same fixed $k$ (and $l$): For the vector models, it suffices
to extend the chosen configuration with zeros in the last $N-2k$
(or $N-k-l$) components; then $a_{xx'}$ and $b_{xx'}$ are
unchanged, while $c_{xx'}\equiv0$, so that the couplings are
unchanged (even if $\beta_T\neq0$). For the $SU(N)$ matrix models,
we can take (when $R=I$) the matrices $U$ to be block-diagonal, with
the uppermost $2k\times2k$ block being as previously constructed,
and the lowermost $(N-2k)\times(N-2k)$ block being the {\em identity matrix};
then $a_{xx'}$ and $b_{xx'}$ are
unchanged, while $c_{xx'}\equiv N-2$, so that the only change
in the reasoning is to replace $\beta_F$ by $\beta_F+2\beta_A (N-2)$
in the formulae for $\alpha_{xx'}$ and $\beta_{xx'}$.

Let us now consider separately the various models:

\medskip
{\em Model 1 ($N$-vector model):}\/
For $k=1$ we have shown that the embedding is unfrustrated. For $k=2$
the minimal case is $N=4$: we have already discussed this case in the
preceding subsection, and have shown that the induced $XY$ Hamiltonian 
\reff{frust_couplings} is in general {\em frustrated} because the 
couplings are a sum of contributions written in terms of unrelated
angles. The same argument applies for general $k\geq2$ and $N=2k$:
the update consists of $k$ plane rotations and leads to couplings that are
a sum of $k$ contributions written in terms of unrelated angles.
Finally, as explained above, the proof of frustration immediately
extends to $N>2k$. Hence {\em all} cases with $k\geq2$ lead to frustration.

\medskip
{\em Model 2 ($SU(N)$ principal chiral model):}\/
For $k=1$ we have shown frustration for all $N\geq2$. Next consider 
$k\geq2$ and $N=2k$. If we look at configurations that are block-diagonal
(in the same basis in which $T$ is block-diagonal) in which each of the
$k$ blocks is an $SU(2)$ matrix, we get couplings that are the sum of $k$ 
terms each of which is as in \reff{frust_couplings}. In other
words, the couplings are a sum of $2k$ contributions written in terms
of unrelated angles, so they are in general frustrated. Finally,
as explained above, the proof immediately extends to $N>2k$.
Hence {\em all} cases lead to frustration.

\medskip
{\em Model 3 (mixed isovector/isotensor model):}\/
For $k=1$ we have shown that the couplings are unfrustrated
if  $\beta_V \geq \beta_T \geq0$.
For general $k$, the couplings \reff{mixed_couplings} lead to
\begin{subeqnarray}
\alpha_{xx'}\,-\,i\beta_{xx'} &=& (\beta_V\,+\,\beta_T\,c_{xx'})
\,(a_{xx'}\,-\,ib_{xx'}) \\
\gamma_{xx'}\,-\,i\delta_{xx'}&=& \frac{\beta_T}{4}
       \,(a_{xx'}\,-\,ib_{xx'})^2
\end{subeqnarray}
Hence, if we write the induced $XY$ Hamiltonian in terms of ``frustration
angles'' as in \reff{frust_angles}, we have 
$\phi'_{xx'}=2\phi_{xx'}$ (mod $2\pi$) provided that $\beta_T\geq0$.
Hence the presence or absence of frustration can be determined solely
by looking at  $\prod\protect_{\Box} (\alpha_{xx'}\,-\,i\beta_{xx'})$.
This product consists of two factors: the product of
$a_{xx'}\,-\,i\,b_{xx'}$ around the plaquette,
which we have already studied in connection with the pure $N$-vector
model; and the product of $\beta_V\,+\,\beta_T\,c_{xx'}$, which is 
always real but may in some cases be negative.
The cases with $N=2k$ are simple, since we have $T^2=I$ and hence 
$c_{xx'}=0$, so that the problem is 
reduced to that of frustration for the pure $N$-vector model:
frustration occurs whenever $k\geq2$. The proof  extends immediately 
to $N>2k$, as explained before.

The only cases left to investigate are the mixed models with
$\beta_T>\beta_V\geq0$ for $k=1$ and $N\geq3$. Now, for
$k=1$ we have already seen that $\prod (a_{xx'}\,-\,i\,b_{xx'})$ is
real and positive, so that the model will be unfrustrated if and only if
$\prod (\beta_V\,+\,\beta_T\,c_{xx'})$ is positive. Let us treat
separately the cases $\beta_V=0$ and $\beta_V>0$:

{\em Case (a): $\beta_V=0$ (pure $RP^{N-1}$ model).} The 
criterion for nonfrustration is
\be
\prod_{\Box} c_{xx'} \;=\; \prod_{\Box}
\bsigma_x^{\perp}\cdot\bsigma_{x'}^{\perp}
\;\geq\;0 \;.
\label{frust_criterion}
\ee
Two distinct situations then arise:
\begin{itemize}
\item[(i)] $N=3$: Here the vectors $\bsigma_x^{\perp}$ are one-dimensional, 
so that the product \reff{frust_criterion} is manifestly nonnegative
(each $\bsigma_x^{\perp}$ occurs exactly twice in the product). Therefore,
the model is unfrustrated.

\item[(ii)] $N\geq4$: For these cases one can easily find configurations with
$\prod\bsigma_x^{\perp}\cdot\bsigma_{x'}^{\perp}<0$: take, for example,
on a plaquette $ABCD$ the configuration
\begin{subeqnarray}
\bsigma_A^{\perp}&=& r_A\,(1,0,0,\ldots) \\
\bsigma_B^{\perp}&=& r_B\,(\cos\theta_{B},\sin\theta_{B},0,\ldots) \\
\bsigma_C^{\perp}&=& r_C\,(\cos\theta_{C},\sin\theta_{C},0,\ldots) \\
\bsigma_D^{\perp}&=& r_D\,(\cos\theta_{D},\sin\theta_{D},0,\ldots) 
\end{subeqnarray}
with all $r_i>0$. Then, we have frustration whenever the product
\be
\cos\theta_{B}\, \cos(\theta_{C}-\theta_{B}) \, \cos(\theta_{D}-\theta_{C})
\, \cos\theta_{D} 
\ee
is negative, e.g.\  for $\theta_{B}=\pi/3$,
$\theta_{C}=2\pi/3$ and $\theta_{D}=\pi$. We thus prove that all
$RP^{N-1}$ models with $N\geq4$ are frustrated when $k=1$.
\end{itemize}

{\em Case (b): $\beta_V>0$.} For any $N\geq3$ and any values
$\beta_T>\beta_V>0$,
we can find configurations such that $\prod(\beta_V\,+\,
\beta_T c_{xx'})$ is negative. For
example, take a configuration such that 
$\sigma_A^{\perp}$,$\sigma_B^{\perp}$,$\sigma_C^{\perp}$,$\sigma_D^{\perp}$
are all colinear, with $\sigma_A^{\perp}$,$\sigma_B^{\perp}$,
$\sigma_C^{\perp}\,>0$ and $\sigma_D^{\perp}<0$ relative to some fixed
orientation of this line. Then, whenever
$\sigma_A^{\perp}>\sigma_C^{\perp}$ and
\be
\frac{\beta_V}{\beta_T\,\sigma_A^{\perp}} \;<\;
|\sigma_D^{\perp}| \;<\; \frac{\beta_V}{\beta_T\,\sigma_C^{\perp}}
\;,
\ee
we have $\prod(\beta_V\,+\,\beta_T c_{xx'})<0$.

In summary, the mixed isovector/isotensor models are frustrated except
for two cases with $k=1$: (a) $\beta_V\geq\beta_T\geq0$ for any $N$, and
(b) the pure $RP^2$ model.


{\bf Remark.}
As mentioned in footnote \protect\ref{RP3_equiv_SO3}, the
case $N=4$ and $\beta_V=0$ (namely the $RP^3$ model) is equivalent to the
$SO(3)$ principal chiral model. It is interesting to 
write the $k=2$ $RP^3$ update in $SO(3)$ language.
Recall that a 4-vector $s=(s_0,\vs)\in S^3$ maps into a matrix
$V\in SO(3)$ via
\be
       (V_x)_{ij} \;=\;
         (2{s_0}^2 - 1) \delta_{ij}  \,+\,
          2 s_i s_j  \,+\,
          2 s_0 \varepsilon_{ijk} s_k
\;.
\label{map_S3_SO3}
\ee
Now, one case of the $k=2$ update for the $RP^3$ model is given by
\reff{SU2_S3_update}. Using \reff{map_S3_SO3}, we see that this translates 
into
\be
V^{new}\;=\;e^{2\,i\,\theta\,\vec{\scriptsize\it a}\cdot\vec{\scriptsize\it J}}
\,V^{old}
\;,
\ee
where $\vJ$ are the generators of the $\germanso(3)$ algebra,
i.e.\  $(J_i)_{jk}\,=\,-i\,\varepsilon_{ijk}$. This is the standard
left-update for the $SO(3)$ chiral model [analogous to \reff{SUN_update} 
for the $SU(N)$ chiral model] except that the angle is doubled.
This explains why the induced $XY$ Hamiltonian has only 
``isospin-2'' terms (see footnote \ref{T2=1_mixed}). 

\medskip
{\em Model 4 (complex mixed isovector/isotensor model):}\/
Let us start by considering the embeddings with $l=0$, so that
$T^2= T$. (The cases with $k=0$, $T^2=-T$ are trivially equivalent.) 
All such embeddings have $\gamma_{xx'}=\delta_{xx'}=0$ and
\be
\alpha_{xx'} \,-\,i\beta_{xx'} \;=\; 
[\beta_V \,+\, \beta_T \, (\bz_x^{\perp *}\cdot \bz_{x'}^{\perp})]
\; \bz_x^{\parallel}\cdot \bz_{x'}^{\parallel *}
\; ,
\ee
where $\bz_x^{\parallel}$ (resp. $\bz_x^{\perp}$) denotes the first
$k$ (resp. last $N-k$) components of $\bz_x$. We can prove
frustration in three cases:
\begin{itemize}
\item[(i)] If $N\geq k\geq2$ and $\beta_V\neq0$. It suffices to choose
configurations with $\bz_x^{\perp}\equiv 0$. Then the couplings are
those of a $k\geq2$ update for the $2k$-vector model, hence frustrated.
\item[(ii)] If $N\geq k+2$ and $\beta_T\neq0$. The proof is identical 
to that of case (ii) in Section 2.3, if in \reff{z_config2}
we insert $k-1$ zeros between the first and second coordinates.
\item[(iii)] If $N=k+1\geq3$, $\beta_V=0$ and $\beta_T\neq 0$.
This is a pure $\CP^{N-1}$ model, and adding a multiple of the identity to 
$T$ corresponds to a $U(1)$ gauge transformation. So the case $k=N-1$,
$l=0$ is gauge-equivalent to $k=0$, $l=1$, which was proven to be frustrated
for $N\geq3$ in case (ii) of Section 2.3.
\end{itemize}
The only remaining case with $k\geq2$ is $N=k$, $\beta_V=0$:
this is a mere gauge transformation of the $\CP^{N-1}$ model, and the
induced $XY$ Hamiltonian is zero.

%

For  cases with $k,l\neq0$ we no longer have $\gamma_{xx'},\delta_{xx'}=0$,
and a more careful analysis is needed. We have been able to obtain a
result only for the pure $\CP^{N-1}$ model ($\beta_V=0$). Consider first
the minimal case $N=k+l$: here we have $T^2=I$ and hence $c_{xx'}=0$,
so that $\alpha_{xx'}=\beta_{xx'}=0$. The criterion for frustration is
then given in terms of
\be
\gamma_{xx'}-i\delta_{xx'}\;=\;\beta_T\,
(\bz_x^+\cdot \bz_{x'}^{+ *})(\bz_x^{- *}\cdot \bz_{x'}^-) \;,
\ee
where $\bz_x^+$ (resp. $\bz_x^-$) denotes the first $k$ (resp. the last $l$)
components of $\bz_x$. For $k=l=1$ this is manifestly unfrustrated,
as $\prod (\gamma_{xx'}-i\delta_{xx'})\,=\,
\prod |\bz_x^+|^2\,|\bz_x^-|^2\geq 0$; but for all cases with $k\geq2$
and/or $l\geq2$ it is easy to devise frustrated configurations, by the 
same method as was used for the $RP^{N-1}$ models with 
$N\geq4$.\footnote{ {\em Alternate proof}: The $\CP^{N-1}$ model with
$N=k+l$ is gauge-equivalent to one with $k'=k$, $l'=0$ and $\theta_x$
replaced by $2\theta_x$ (see footnotes \ref{footnote_traceless_CPN}
and \ref{gauge-equivalent_CPN}). Hence this is
unfrustrated for $k=1$ and frustrated for $k\geq2$, as proven in 
Section 2.3.}
Finally, the proof of frustration extends to $N>k+l$, as explained
earlier.

The remaining cases of the $\CP^{N-1}$ models are $k=1$, $l=1$ with 
$N\geq 3$. Let $z_x^+$ (resp. $z_x^-$) denote the first (resp. second)
component of $\bz_x$, and let $\bz_x^{\perp}$ denote the last $N-2$
components.
\begin{itemize}
\item[(a)]
Suppose we choose a configuration with $z_x^-\equiv 0$. From \reff{z_coeffs}
we then have $b_{xx'} = -i a_{xx'}$, so that $\gamma_{xx'}=\delta_{xx'}=0$
and 
\be
\alpha_{xx'} \,-\,i \beta_{xx'} \;=\; \beta_T \,
(\bz_x^{\perp *}\cdot\bz_{x'}^{\perp})\,(z_x^+ \, z_{x'}^{+ *}) \; .
\ee
For $N\geq 4$ it is easy to devise frustrated configurations, by the
same method as was used for the $RP^{N-1}$ models with $N\geq 4$.
\item[(b)]
For $N=3$ we can construct a frustrated configuration as follows:
consider on a plaquette $ABCD$ the configuration
\begin{subeqnarray}
\bz_A&=& \frac{1}{\sqrt{2}}\,(1,0,e^{i\mu}) \\[2mm]
\bz_B&=& \frac{1}{\sqrt{3}}\,(1,1,e^{i\nu}) \\[2mm]
\bz_C&=& \frac{1}{\sqrt{3}}\,(1,1,1) \\[2mm]
\bz_D&=& \frac{1}{\sqrt{3}}\,(1,1,e^{i\sigma})
\end{subeqnarray}
with $\mu,\nu,\sigma$ real and $-\pi/2<\nu,\sigma<\pi/2$.
Then for the links $BC$ and $CD$ we have from \reff{zcouplings}/\reff{z_coeffs}
(noting that $T$ is the diagonal matrix $\{1,-1,0\}$)
that $\beta_{xx'}=\delta_{xx'}=0$ and $\alpha_{xx'},\gamma_{xx'}>0$,
so that the unique minimum of the Hamiltonian for these links occurs at
$\theta_B=\theta_C=\theta_D$. For the links $DA$ and $AB$ we have that
$\gamma_{xx'}=\delta_{xx'}=0$ and
\be
(\alpha_{DA}\,-\,i\beta_{DA})\,(\alpha_{AB}\,-\,i\beta_{AB}) \;=\; 
\frac{\beta_T^2}{36}\,e^{i(\nu-\sigma)} \; .
\ee
Therefore, this configuration is frustrated unless $\nu=\sigma$.
\end{itemize}
So the $\CP^{N-1}$ models with $k=l=1$ are frustrated for all $N\geq3$.


\medskip
{\em Model 5 (mixed fundamental/adjoint $SU(N)$ chiral model):}\/
For $k=1$ all mixed models are frustrated. Therefore, for $k\geq2$
all mixed models will be likewise frustrated, since for $N=2k$ we can take 
configurations that are block-diagonal formed by $SU(2)$ matrices;
and we can extend the proof to $N>2k$ as before.

\subsubsection{Two-Sided Update for $SU(N)$}   \label{sec2.4.2}

In view of the results in Sections \ref{sec2.3} and 2.4.1
concerning the frustration in the $SU(N)$ case,
one might ask:  Can we find a {\em different}\/ update for
the $SU(N)$ chiral model that would yield unfrustrated couplings
for the induced $XY$ model?
One possibility is to make a simultaneous rotation on both the
left and the right:
\be
   U_x^{new}  \;=\;  R e^{i\theta_x T} R^{-1} U_x^{old}  S e^{i\theta_x T}
S^{-1}
  \label{symSUNupdate}
\ee
where $R,S$ are random $SU(N)$ matrices and
$T$ is a fixed traceless Hermitian matrix
with eigenvalues in $\{ \pm 1, 0 \}$.
In general $T$ has $k$ eigenvalues $+1$,
$k$ eigenvalues $-1$, and $N-2k$ eigenvalues 0,
for some integer $k$ satisfying $1 \le k \le \lfloor N/2 \rfloor$.

Inserting \reff{symSUNupdate} into the chiral-model Hamiltonian \reff{ham_SUN}
and using \reff{T_eigenvalues}, we find that the induced $XY$ Hamiltonian
is of the ``isovector/isotensor'' form \reff{ham_embed2}, with couplings
\begin{subeqnarray}
\alpha_{xx'} &=& \beta\,
   \{ \re\tr [ V_x^{\dagger}  T^2  V_{x'} (I-T^2) ] \,+\,
      \re\tr [ V_x^{\dagger}  (I-T^2)  V_{x'} T^2 ] \}  \\[2mm]
\beta_{xx'} &=& \beta\, 
   \{ \im\tr [ V_x^{\dagger}  T  V_{x'}(I-T^2) ] \,+\,
      \im\tr [ V_x^{\dagger}  (I-T^2)  V_{x'} T ] \}  \\[2mm]
\gamma_{xx'} &=& \frac{\beta}{2} \, 
     [\re\tr ( V_x^{\dagger}  T^2  V_{x'} T^2 ) \,+\,
      \re\tr ( V_x^{\dagger}  T  V_{x'} T ) ]  \\[2mm]
\delta_{xx'} &=& \frac{\beta}{2} \, 
     [\im\tr ( V_x^{\dagger}  T  V_{x'} T^2 ) \,+\,
      \im\tr ( V_x^{\dagger}  T^2  V_{x'} T ) ]
\end{subeqnarray}
where
\be
   V_x  \;\equiv\;  R^{-1} U_x^{old} S
   \;.
\ee

One interesting case is that in which $N$ is even and $k=N/2$:
then $T$ has only eigenvalues $\pm 1$ (and not also 0),
so that
\be
T^2 \;=\; I
\;.
 \label{Tsquared}
\ee
It follows that $\alpha_{xx'} = \beta_{xx'} = 0$,
so that the induced $XY$ Hamiltonian becomes again,
after a relabeling $\theta_x \to \theta_x/2$,
a ``pure isovector'' $XY$ Hamiltonian \reff{ham_embed}.

For $N=2$ (and hence $k=1$)
we can rewrite the update \reff{symSUNupdate} in 4-vector notation.
With $T$ given by \reff{TSUN}, let $RTR^{-1} = \va \cdot \vsigma$
and $STS^{-1} = \vb \cdot \vsigma$.
We have
\begin{subeqnarray}
s_0^{new} &=& s_0^{old} \,(\cos^2\theta \,-\, \va\cdot\vb \, \sin^2\theta)
            \,-\, \vs^{old} \cdot [(\va\,+\,\vb) \, \sin\theta\,\cos\theta
            \,+\, \va\times\vb \sin^2\theta]   \qquad \\[2mm]
\vs^{new} &=& s_0^{old}[(\va\,+\,\vb) \, \sin\theta\,\cos\theta \,-\,
           \va\times\vb \sin^2\theta] \,+\, \vs^{old}\,
            (\cos^2\theta \,+\, \va\cdot\vb \, \sin^2\theta)
\nonumber \\[2mm]
        & & \hskip 1cm 
      \,-\, [(\vs^{old} \cdot \va )\,\vb \,+\, (\vs^{old} \cdot \vb)\,\va ]
      \,\sin^2\theta \,+\, \vs^{old}\times(\va\,-\,\vb)
 \, \sin\theta\,\cos\theta   \qquad
\end{subeqnarray}
With the special choice $\va=\vb$
(which can be obtained from $R=S$ among other choices), we get
\begin{subeqnarray}
s_0^{new} &=& s_0^{old} \,\cos 2\theta \,-\, 
                  \vs^{old} \cdot\va \, \sin 2\theta  \\[2mm]
\vs^{new} &=& s_0^{old}\,\va \,(\sin 2\theta)   \,+\,
           (\vs^{old} \cdot \va )\,\va \,\cos 2\theta
           \,+\,  [\vs^{old} \,-\, (\vs^{old} \cdot \va )\,\va]  \quad
\end{subeqnarray}
which corresponds to a single rotation of angle $2\theta$ in the
plane defined by the 0-direction and $\va$.
Likewise, the choice $\va = -\vb$ yields
\begin{subeqnarray}
s_0^{new} &=& s_0^{old}   \\[2mm]
\vs^{new} &=& [\vs^{old} \,-\, (\vs^{old} \cdot \va )\,\va]\,
   \cos 2\theta \,+\, (\vs^{old} \times \va )\, \sin 2\theta \,+\,
   (\vs^{old} \cdot \va )\,\va 
\end{subeqnarray}
which corresponds to a single rotation of angle $2\theta$ in the
plane perpendicular to the 0-direction and to $\va$.
With more general choices of $R$ and $S$,
we obtain rotations of angle $2\theta$ in an arbitrary
two-dimensional subspace $P \subset \R^4$.
This (after a relabeling $\theta_x \to \theta_x/2$)
is our standard 4-vector update, and is therefore {\em unfrustrated}\/.

In retrospect we can see what is going on here.
We've already seen that the {\em left}\/ update \reff{SUN_update}
for $SU(2)$ corresponds in the 4-vector language to
an update with $T$ given by \reff{T_k=2}.
Likewise, a {\em right}\/ update for $SU(2)$ corresponds
in the 4-vector language to an update with $T$ given by 
\reff{T_right_update}.
Now, a {\em simultaneous}\/ left and right update for $SU(2)$ corresponds
to the {\em sum}\/ of these two $T$ matrices:
we thus get {\em twice}\/ the usual entry in the upper block,
and zero in the lower block.
This explains why we get a single rotation with a doubled angle.

Unfortunately, the update \reff{symSUNupdate}/\reff{Tsquared}
for $N=4,6,8,\ldots$ is generically frustrated.
Let us choose the configuration $\{U_x^{old}\}$ so that the
matrices $V_x$ are block-diagonal with $SU(2)$ matrices in each
block.
Then, the expressions for the couplings will be just
sums of $SU(2)$-type couplings,
which can each be written in terms of $4$-vectors as
in \reff{disordered} [since the update in each $SU(2)$ subspace
corresponds to a single plane rotation in 4-vector language].
We have therefore $k=N/2$ independent angles, so that the couplings are
frustrated whenever $k \ge 2$ and $N=2k$.

This example can be immediately extended to the cases $k \ge 2$ with $N > 2k$:
it suffices to choose the matrices $V_x$ to be block-diagonal,
with the uppermost $2k \times 2k$ block as above,
and the lowermost $(N-2k) \times (N-2k)$ block being an arbitrary
$SU(N-2k)$ matrix. (For such a {\em special} configuration one has 
$\alpha_{xx'}=\beta_{xx'}=0$, even though this does not
{\em in general} happen for $N>2k$.)

The only case left to investigate is $k=1$, $N =3$.
In this case frustration can be proven by making use of configurations 
taking values in the group's center:
\be
U_x \;=\; e^{(2\pi i/3) k_x} I
\label{center2}
\ee
where $k_x \in \{0,1,2\}$ and $I$ is the identity matrix.
Then the couplings \reff{suncouplings} become
\begin{eqnarray}
\alpha_{xx'} &=& \beta_{xx'} \;=\; \delta_{xx'} \;=\; 0 \nonumber \\
\gamma_{xx'} &=& 2\beta\,\cos\left[
          \frac{2\pi}{3}\,(k_x\,-\,k_{x'})\right] 
          \;=\; \beta\,(3\delta_{k_x k_{x'}} \,-\, 1)
\end{eqnarray}
The $\gamma_{xx'}$ are thus {\em real}\/ numbers,
and we will have frustration if their product around a plaquette is negative.
But this occurs whenever three of
the four sites have different values of $k_x$, for example $\{0,0,1,2\}$.

In conclusion, the two-sided $SU(N)$ embedding leads to a frustrated
$XY$ Hamiltonian in all cases {\em except}\/ the exceptional case $N=2$, $k=1$.

\subsection{Summary}

In conclusion, we have a {\em simple} induced Hamiltonian in the following
cases:
\begin{itemize}
\item[(i)] $N$-vector model, any $k$
\item[(ii)] $SU(N)$ principal chiral model, any $k$
\item[(iii)] complex mixed isovector/isotensor model, $k=0$ or $l=0$
\item[(iv)] $RP^{N-1}$ model, $N$ even, $k=N/2$
\item[(v)] pure adjoint $SU(N)$ principal chiral model, $N$ even, $k=N/2$
\item[(vi)] $\CP^{N-1}$ model, $k+l=N$
\item[(vii)] $SU(N)$ principal chiral model, $N$ even, $k=N/2$ two-sided
update
\end{itemize}
In the first three cases the induced $XY$ Hamiltonian has only
isospin-1 terms, while in the latter four cases it has only
isospin-2 terms.

Furthermore, we have an {\em unfrustrated} induced $XY$ Hamiltonian 
for the mixed isovector/isotensor models with $\beta_V\geq\beta_T\geq0$ 
and $k=1$, as well as for models equivalent to one of these. We also have an
unfrustrated Hamiltonian for the $RP^2$ model with $k=1$.
All other cases appear to lead in general to {\em frustrated} Hamiltonians.

\section{Numerical Results} \label{sec:results}  \label{sec3}

In this section we present our numerical results for the two-dimensional
$N$-vector models \reff{N-vector} with $N=3,4,8$.

\subsection{Observables to be Measured}   \label{sec3.1}

We wish to study various correlation functions of
the isovector field $\bsigma_x$ and the symmetric traceless isotensor
field
\be
   T_x^{(\alpha\beta)}   \;\equiv\;   \sigma_x^{(\alpha)} \sigma_x^{(\beta)}
                             \,-\,  {1 \over N} \delta^{\alpha\beta}   \;.
\ee
Our interest in the isotensor sector arose initially from our work on
mixed isovector/isotensor models \cite{CEPS_PRL,CEPS_LAT93,CEPS_RPN};
but even in the pure $N$-vector model
it is of some interest to measure isotensor observables.
We thus define the isovector and isotensor 2-point correlation functions
\begin{subeqnarray}
   G_V(x-y)   & = &   \< \bsigma_x \cdot \bsigma_y \>             \\[2mm]
   G_T(x-y)   & = &   \< \ttens_x \cdot \ttens_y \>
      \;\equiv\;   \sum\limits_{\alpha,\beta = 1}^N
     \< T_x^{(\alpha\beta)} T_y^{(\alpha\beta)} \>             \nonumber \\
      & = &   \< (\bsigma_x \cdot \bsigma_y)^2 \> - {1 \over N}
\end{subeqnarray}
Note that $SO(N)$ invariance
(which cannot be spontaneously broken in dimension $d \le 2$)
determines the 1-point and 2-point correlation
functions of $\bsigma$ and $\ttens$ in terms of $G_V$ and $G_T$:
\begin{subeqnarray}
   \< \sigma_x^{(\alpha)} \>    & = &    0                        \\[2mm]
   \< \sigma_x^{(\alpha)} \sigma_y^{(\beta)} \>    & = &
         {1 \over N} \, \delta^{\alpha\beta} \, G_V(x-y)          \\[2mm]
 \slabel{eq6.3c}
   \< T_x^{(\alpha\beta)} \>    & = &    0                        \\[2mm]
 \slabel{eq6.3d}
   \< T_x^{(\alpha\beta)} T_y^{(\kappa\lambda)} \>    & = &
         {1 \over N^2 + N - 2}  \,
         (\delta^{\alpha\kappa} \delta^{\beta\lambda} +
          \delta^{\alpha\lambda} \delta^{\beta\kappa} -
            {2 \over N}
          \delta^{\alpha\beta} \delta^{\kappa\lambda} )   \,
         G_T(x-y) \quad
\end{subeqnarray}

All our numerical work will be done on an $L \times L$ lattice
with periodic boundary conditions.
We are interested in the following quantities:
\begin{itemize}
\item  The isovector and isotensor energies\footnote{
   Actually $E_V$ (resp.\ $E_T$) is $-1$ (resp.\ $-2$) times
   the mean energy per {\em link}\/;
   we have chosen this normalization in order to have $0 \le E_{V,T} \le 1$,
   where $E_{V,T} = 1$ for a totally ordered state.
   Several other normalizations are in use in the literature.
}
\begin{subeqnarray}
   E_V   & = &
     \< \bsigma_0 \cdot \bsigma_{\bf e} \>
     \;=\;    G_V({\bf e})                                           \\[2mm]
   E_T   & = &
     \< (\bsigma_0 \cdot \bsigma_{\bf e})^2 \>
     \;=\;   G_T({\bf e}) + {1 \over N}
\end{subeqnarray}
where ${\bf e}$ stands for any nearest neighbor of the origin.
%
%
\item  The isovector and isotensor magnetic susceptibilities
\begin{eqnarray}
   \chi_\#   & = &   \sum\limits_x   G_\#(x)   \;,
\end{eqnarray}
where $\# = V \hbox{ or } T$.
\item  The isovector and isotensor correlation functions at the
smallest nonzero momentum:
\begin{eqnarray}
   F_\#   & = &   \sum\limits_x e^{ip_0 \cdot x} \,   G_\#(x)   \;,
\end{eqnarray}
where $p_0 = (\pm 2\pi/L, 0) \hbox{ or } (0,\pm 2\pi/L)$.
\item  The isovector and isotensor second-moment correlation lengths
\begin{eqnarray}
  \xi^{(2nd)}_\#  & = &  {(\chi_\#/F_\# \,-\, 1)^{1/2} \over 2\sin(\pi/L)}
  \;.
 \label{corr_len_2mom}
\end{eqnarray}
\item  The isovector and isotensor exponential correlation lengths
\begin{eqnarray}
  \xi^{(exp)}_\#  & = &   \lim\limits_{|x| \to\infty}
                           {-|x|  \over  \log G_\#(x)}
 \label{corr_len_exp}
\end{eqnarray}
and the corresponding mass gaps $m_{\#} = 1/ {\xi^{(exp)}_{\#}}$.
[These quantities make sense only if the lattice is essentially infinite
 (i.e.\ $L \gg \xi^{(exp)}_\#$) in at least one direction.
 We will not {\em measure}\/ any exponential correlation lengths in this work;
 but we will use $\xi^{(exp)}_\#$ as a theoretical standard of comparison.]
\end{itemize}

All these quantities except $\xi^{(exp)}_\#$
can be expressed as expectations involving the following observables:
\begin{subeqnarray}
   \scrm_V^2  & = &  \left( \sum_x \bsigma_x \right)^2       \\[2mm]
   \scrm_T^2  & = &  \left( \sum_x  \ttens_x \right)^2       \\[2mm]
   \scrf_V    &=&  \half \left[
                      \left| \sum_x e^{2\pi i x_1/L} \bsigma_x \right| ^2
                        \,+\,
                      \left| \sum_x e^{2\pi i x_2/L} \bsigma_x \right| ^2
                   \right]                           \\[2mm]
   \scrf_T    &=&  \half \left[
                      \left| \sum_x e^{2\pi i x_1/L} \ttens_x \right| ^2
                        \,+\,
                      \left| \sum_x e^{2\pi i x_2/L} \ttens_x \right| ^2
                   \right]                           \\[2mm]
   \scre_V    &=&  \sum_{\< xy \>}  \bsigma_x \cdot \bsigma_y    \\[2mm]
   \scre_T    &=&  \sum_{\< xy \>}  (\bsigma_x \cdot \bsigma_y)^2
\end{subeqnarray}
Thus,
\begin{subeqnarray}
   E_\#     & = &   \smhalf V^{-1}  \< \scre_\# \>              \\[2mm]
   \chi_\#  & = &   V^{-1}  \< \scrm_\#^2 \>            \\[2mm]
   F_\#     & = &   V^{-1}  \< \scrf_\# \>         
\end{subeqnarray}
where $\# = V \hbox{ or } T$;
here $V=L^2$ is the number of sites in the lattice.

\subsection{Autocorrelation Functions and Autocorrelation Times} \label{sec3.2}

Let us now define the quantities ---
autocorrelation functions and autocorrelation times ---
that characterize the Monte Carlo dynamics.
Let $A$ be an observable
(i.e.\ a function of the field configuration $\{\varphi_x\}$).
We are interested in the evolution of $A$ in Monte Carlo time,
and more particularly in the rate at which the system ``loses memory''
of the past.
We define, therefore, the {\em unnormalized autocorrelation function}\footnote{
   In the mathematics and statistics literature, this is called the
   {\em autocovariance function}\/.
}
\be
  C_{A}(t)  \;=\;   \< A_s A_{s+t} \>   -  \< A \> ^2  \,,
\ee
where expectations are taken {\em in equilibrium\/}.
The corresponding {\em normalized autocorrelation function\/} is
\be
  \rho_{A}(t)  \;=\;  C_{A}(t) / C_{A}(0) \,.
\ee
We then define the {\em integrated autocorrelation time}
\begin{eqnarray}
\tau_{int,A}  & =&
 \half \sum_{{t} \,=\, - \infty}^{\infty} \rho_{A} (t)\nonumber\\
 &=&  \half \ +\  \sum_{{t} \,=\, 1}^{\infty} \  \rho_{A} (t)
\end{eqnarray}
[The factor of $\half$ is purely a matter of convention;  it is
inserted so that $\tau_{int,A} \approx \tau$ if
$\rho_{A}(t) \approx e^{-|t|/ \tau}$ with $\tau \gg 1$.] 
Finally,
the {\em exponential autocorrelation time}\/ for the observable $A$ is
defined as
\be
\tau_{{\rm exp},A}  \;=\;
   \limsup_{{t}  \to \infty}  {|t| \over -  \log  | \rho_{A}(t)|}
   \;,
 \label{def_tau_exp}
\ee
and the exponential autocorrelation time (``slowest mode'')
for the system as a whole is defined as
\be
\tau_{{\rm exp}} \;=\; \sup_A \,  \tau_{{\rm exp},A}  \;.
\ee
Note that $\tau_{{\rm exp}} = \tau_{{\rm exp},A}$
whenever the observable $A$ is not orthogonal to the
slowest mode of the system.

The integrated autocorrelation time controls the statistical error
in Monte Carlo measurements of $\< A \>$.  More precisely,
the sample mean
\begin{equation}
\bar A \ \ \equiv\ \ {1 \over n }\  \sum_{t=1}^n \ A_t
\end{equation}
has variance
\begin{subeqnarray}
\var( \bar A )  &= &
  {1 \over n^2} \ \sum_{r,s=1}^n \ C_{A} (r-s) \\
 &=& {1 \over n }\ \sum_{{t} \,=\, -(n-1)}^{n-1}
  (1 -  {{|t| \over n }} ) C_{A} (t) \label{var_observa}  \\
 &\approx&  {1 \over n }\ (2 \tau_{int,A} ) \ C_{A} (0)
   \qquad {\rm for}\ n\gg \tau \label{var_observb} 
\end{subeqnarray}
Thus, the variance of $\bar{A}$ is a factor $2 \tau_{int,A}$
larger than it would be if the $\{ A_t \}$ were
statistically independent.
Stated differently, the number of ``effectively independent samples''
in a run of length $n$ is roughly $n/2 \tau_{int,A}$.
The autocorrelation time $\tau_{int,A}$ (for interesting observables $A$)
is therefore a ``figure of (de)merit'' of a Monte Carlo algorithm.

The integrated autocorrelation time $\tau_{int,A}$ can be estimated
by standard procedures of statistical time-series analysis
\cite{Priestley_81,Anderson_71}.
These procedures also give statistically valid {\em error bars}\/
on $\< A \>$ and $\tau_{int,A}$.
For more details, see \cite[Appendix C]{Madras_88}.
In this paper we have used a self-consistent truncation window of width
$c \tau_{int,A}$, where $c$ is a constant.
We have here used $c=6$;  this choice is reasonable whenever the
autocorrelation function $\rho_{A}(t)$ decays roughly exponentially.

\subsection{Summary of our Runs} \label{section:mgon_results}

We performed runs on the $N$-vector model with $N=3,4,8$,
using the $k=1$ $XY$-embedding MGMC algorithm described in Section~\ref{sec2.2},
on lattices of size $L = 32, 64, 128, 256$.
We updated the induced $XY$ model \reff{ham_embed}
using our standard $XY$-model MGMC program \cite{MGMC_2}
with $\gamma = 2$ (W-cycle)
and $m_1 = m_2 = 1$ (one heat-bath pre-sweep and one heat-bath post-sweep).
In all cases the coarsest grid is taken to be $2 \times 2$.
All runs used an ordered initial configuration (``cold start'').
The results of these computations are shown in
Tables~\ref{o3_staticdata}--\ref{o8_staticdata} (static data)
and \ref{o3_dyndata}--\ref{o8_dyndata} (dynamic data).

For the $N=4$ case we made only a few runs,
in order to compare $XY$-embedding MGMC
to our previous detailed study of the direct MGMC algorithm for this model
\cite{MGMC_O4}.
For $N=3$ and $N=8$, by contrast, we made a reasonably extensive
set of runs, enough to permit an {\em ab initio}\/ dynamic finite-size-scaling
analysis. Run lengths were in all cases between $\approx 5000\tau$
and $\approx 50000\tau$.

Our static data for $N=3, 4, 8$ are in good agreement with data
from previous simulations of these models
\cite{o3_scaling_prl,CEPS_RPN,o3_scaling_fullpaper,MGMC_O4,Wolff_O4_O8}.
To check this quantitatively, we made comparisons at each pair $(\beta,L)$
for which both ``old'' and ``new'' data are available,
and summed the resulting $\chi^2$ values.
The results are:
\begin{itemize}
\item $N=3$, comparison with
             \cite{o3_scaling_prl,CEPS_RPN,o3_scaling_fullpaper}: \\
degrees of freedom: 30 \\
$\chi^2$ for isovector susceptibility = 38.73, confidence level = 13\% \\
$\chi^2$ for isovector correlation length = 34.29, confidence level = 27\% \\
$\chi^2$ for isovector energy = 36.18, confidence level = 20\% \\
$\chi^2$ for isotensor susceptibility = 40.22, confidence level = 10\% \\
$\chi^2$ for isotensor correlation length = 38.23, confidence level = 14\% \\
$\chi^2$ for isotensor energy = 37.02, confidence level = 18\%

\item $N=4$, comparison with \cite{MGMC_O4} (isovector sector only)\footnote{
  We note that the energy defined in \cite{MGMC_O4} is a factor of $2$
  larger than the one employed here.
}: \\
degrees of freedom: 14 \\
$\chi^2$ for isovector susceptibility = 18.75, confidence level = 17\% \\
$\chi^2$ for isovector correlation length = 21.74, confidence level = 8\% \\
$\chi^2$ for isovector energy = 13.70, confidence level = 47\%

\item $N=8$, comparison with \cite{Wolff_O4_O8} (isovector sector only): \\
degrees of freedom: 4 \\
$\chi^2$ for isovector susceptibility = 1.57, confidence level = 81\% \\
$\chi^2$ for isovector correlation length = 2.58, confidence level = 63\% \\
$\chi^2$ for isovector energy = 6.53, confidence level = 16\%
\end{itemize}
These confidence levels are all individually acceptable,
but taken together they are slightly lower than would be expected
{\em a priori}\/.  This may be due to a very slight
(e.g.\ $\sim\! 10\%$) underestimation of the error bars in either
the ``old'' or the ``new'' data (or both).

\subsection{Computational Work}   \label{sec3.4}

In Table~\ref{cpu_timings} we show the CPU time per iteration
for our $XY$-embedding MGMC program
with $\gamma=2$ (W-cycle) and $m_1 = m_2 = 1$,
running on one processor of a Cray C-90,
for $L=32,64,128,256$ and $N=3,4,8$.
We also show, for comparison, the corresponding CPU time for
the $N=4$ {\em direct}\/ MGMC algorithm, using our old program \cite{MGMC_O4}
recompiled for the Cray C-90.
The first timing is for Monte Carlo iterations alone;
the second timing includes measurement of observables
(isovector only in the old program,
 isovector and isotensor in the new program).
We see that:

(a) For each model these timing data grow {\em sublinearly}\/ in the volume,
in contrast to the theoretical prediction \reff{workMG_0},
because the vectorization is more effective on the larger lattices.\footnote{
   Note that the heat-bath subroutine uses von Neumann rejection to
   generate the desired random variables
   \protect\cite[Appendix A]{MGMC_2}.
   The algorithm is vectorized by gathering all the sites of one sublattice
   (red or black) into a single Cray vector, making one trial of the
   rejection algorithm, scattering the ``successful'' outputs,
   gathering and recompressing the ``failures'', and repeating until
   all sites are successful.
   Therefore, although the original vector length in this subroutine is
   $L^2/2$,
   the vector lengths after several rejection steps are much smaller.
   It is thus advantageous to make the original vector length
   as large as possible.

   Note also that our old $N=4$ direct MGMC program \cite{MGMC_O4}
   used vectors of length $L^2/4$ instead of $L^2/2$.
   As a result, the ratio time(old program)/time(new program for $N=4$)
   is slightly higher on the small lattices than on the large lattices.
}
However, the ratio $\hbox{time}(2L)/\hbox{time}(L)$
is increasing with $L$, and does appear to be approaching
the theoretical value of 4 as $L \to\infty$.

(b)  At each $L$, the CPU time for Monte Carlo iterations grows roughly as
$a + bN$ with $b/a \approx 0.03$.  This is because the bulk of the CPU time
is spent in updating the induced $XY$ model, which is independent of $N$;
while a small fraction of the time is spent computing the induced $XY$
Hamiltonian and updating the $N$-vector spins, which are operations of
order $N$.

(c)  At each $L$, the CPU time for measurement of observables
(i.e.\ the difference between the two timings in Table~\ref{cpu_timings})
grows roughly as $aN + bN^2$ with $b/a$ of order 1.
This is because the measurement of isovector (resp.\ isotensor)
observables takes a CPU time of order $N$ (resp.\ $N^2$).
Our old program \cite{MGMC_O4}, by contrast, measured only
isovector observables;  the time spent in measurement is a factor
of $\approx\! 4$ smaller.

\medskip

The running speed on the Cray C-90
for our $XY$-embedding MGMC program at $L=256$ was about 329 MFlops
for the $N=3$ case, 322 MFlops for $N=4$, and 317 MFlops for $N=8$.
Our old $N=4$ direct MGMC program runs on the C-90 at 270 MFlops.

The total CPU time for the runs reported here was about 2000 Cray C-90 hours,
of which about 2/3 were devoted to $N=8$ and about 1/3 to $N=3$.
The CPU time for $N=4$ was comparatively small.

The statistical efficiency of a Monte Carlo algorithm
is inversely proportional
to the integrated autocorrelation time for the observable(s) of interest,
{\em measured in CPU units}\/.
We can compare our largest values of $\taux$ for the cases $N=3,4,8$
at lattice size $L=128$:
we get $\taux = 2.8,\, 3.0,\, 6.0$ Cray C-90 seconds,
respectively, for $N=3,4,8$ (times for Monte Carlo iterations only).
Thus, the CPU time to generate one ``effectively independent''
configuration grows (not surprisingly) with $N$,
perhaps slightly less than linearly.
Note, finally, that for $N=4$ the direct MGMC method is about
three times as efficient as the $XY$-embedding method:
while the CPU time per iteration is roughly the same for the two programs
(see Table~\ref{cpu_timings}),
the autocorrelation time $\taux$ is smaller by a factor of $\approx\! 3$
for the direct algorithm (see Section \ref{sec5}).

\section{Finite-Size-Scaling Analysis: Static Quantities}
\label{sec:fss_static}  \label{sec4}

\subsection{Preliminaries}

We shall analyze our static data using a
finite-size-scaling extrapolation method due originally to
L\"uscher, Weisz and Wolff \cite{Luscher_91}
and elaborated recently by our group
\cite{fss_greedy,o3_scaling_prl,fss_greedy_fullpaper}
(see also \cite{Kim_93}).
An exhaustive study of the $N=3$ case has been made elsewhere,
using much more precise and extensive data
\cite{o3_scaling_prl,CEPS_RPN,o3_scaling_fullpaper};
and our data for $N=4$ are too sparse to support this type of analysis.
Therefore, our analysis here of the static quantities will be restricted
to the case $N=8$.\footnote{
   A preliminary version of this analysis was reported in
   \cite{Caracciolo_LAT95}.
}

We wish to test the asymptotic-freedom predictions
\begin{eqnarray}
\xi_\#(\beta)    & = &   \widetilde{C}_{\xi_\#} \, \Lambda^{-1}
         \left[ 1 + {a_1 \over \beta} + {a_2 \over \beta^2} + \cdots \right]
                                           \label{xi_predicted2}  \\[3mm]
\chi_V(\beta)   & = &   \widetilde{C}_{\chi_V} \, \Lambda^{-2} \,
       \left( {2\pi\beta \over N-2} \right)^{\! -(N-1)/(N-2)}  \,
         \left[ 1 + {b_1 \over \beta} + {b_2 \over \beta^2} + \cdots \right]
                                           \label{chiV_predicted2} \\[3mm]
\chi_T(\beta)   & = &   \widetilde{C}_{\chi_T} \, \Lambda^{-2} \,
          \left( {2\pi\beta \over N-2} \right)^{\! -2N/(N-2)}  \,
         \left[ 1 + {d_1 \over \beta} + {d_2 \over \beta^2} + \cdots \right]
                                           \label{chiT_predicted2}
\end{eqnarray}
as $\beta \to \infty$, where
\be
   \Lambda  \;\equiv\;
      e^{-2\pi\beta/(N-2)} \,
      \left( {2\pi\beta \over N-2} \right)^{\! 1/(N-2)}
      2^{5/2} \exp\!\left[ {\pi \over 2(N-2)}  \right]
  \label{lambda-parameter}
\ee
is the fundamental mass scale\footnote{
   In \reff{lambda-parameter}, the exponential and power of $\beta$
   are universal;  the remaining factor is special to the
   standard nearest-neighbor action \reff{ham_N-vector},
   and comes from a one-loop lattice calculation
   \cite{Parisi_80,Shigemitsu_81}.
};
here
$\widetilde{C}_{\xi_\#}$, $\widetilde{C}_{\chi_V}$ and $\widetilde{C}_{\chi_T}$
are {\em universal}\/ (albeit non-perturbative) quantities characteristic
of the continuum theory (and thus depending only on $N$),
while the $a_k$, $b_k$ and $d_k$ are nonuniversal constants
that can be computed in weak-coupling perturbation theory on the lattice
at $k+2$ loops;
and $\xi_\#$ denotes any one of
$\xi_V^{(2nd)}, \xi_T^{(2nd)}, \xi_V^{(exp)}, \xi_T^{(exp)}$.
It is worth emphasizing that the {\em same}\/ coefficients $a_k$
occur in all four correlation lengths:
this is because the ratios of these correlation lengths
take their continuum-limit values plus corrections that are powers
of the mass $m = 1/\xi^{(exp)}$, hence exponentially small in $\beta$.

When analyzing the susceptibilities,
it is convenient to study instead the ratios
\begin{eqnarray}
{ \chi_V(\beta)   \over   \xi_\#(\beta)^2 }
   & = &
   { \widetilde{C}_{\chi_V}   \over   \widetilde{C}_{\xi_\#}^2 }
   \,
   \left( {2\pi\beta \over N-2} \right)^{\! -(N-1)/(N-2)}  \,
   \left[ 1 + {c_1 \over \beta} + {c_2 \over \beta^2} +
                                  {c_3 \over \beta^3} + \cdots \right]
  \label{chiVoverxiVsquared_predicted}   \\[3mm]
  { \chi_T(\beta)   \over   \xi_\#(\beta)^2 }
   & = &
   { \widetilde{C}_{\chi_T}   \over   \widetilde{C}_{\xi_\#}^2 }
   \,
   \left( {2\pi\beta \over N-2} \right)^{\! -2N/(N-2)}  \,
   \left[ 1 + {e_1 \over \beta} + {e_2 \over \beta^2} + \cdots \right]
  \label{chiToverxiTsquared_predicted}
\end{eqnarray}
The advantage of this formulation in the case of $\chi_V$
is that one additional term of perturbation theory is available
(i.e.\ $c_3$ but not $a_3$ or $b_3$).

For the standard nearest-neighbor action \reff{ham_N-vector},
the perturbative coefficients
$a_1$, $a_2$, $c_1$, $c_2$, $c_3$, $e_1$ and $e_2$
have been computed \cite{Falcioni_86,Luscher-Weisz_unpub,CP_3loops,CP_4loops}:
\begin{eqnarray}
   a_1  & = &  {1 \over 4} \,+\, {\pi \over 16} \,-\,2\pi G_1  \,+\,
         {  {1 \over 4} \,-\, {5\pi \over 48}   \over  N-2 }
\label{a1}
   \\[2mm]
   a_2  & = & \frac{0.0688 \,-\, 0.0028 \, N \,+\, 0.0107 \,N^2
      \,-\, 0.0129 \,N^3}{(N-2)^2}
\label{a2}
   \\[2mm]
   c_1  & = &  {\pi - 2  \over 4\pi} \, {N-1 \over N-2}
   \\[2mm]
   c_2  & = &  \frac{N-1}{(N-2)^2} \,\left[
     -\frac{3 N^2 \,-\, 23 N \,+\, 31}{96} \,+\,
           \frac{N-1}{8 \pi^2} 
 \,-\, \frac{2N-3}{8 \pi} \,+\, G_1\,  (N-2)^2 \right]  
   \\[2mm]
   c_3  & = &
 \frac{N-1}{(N-2)^3} \; (0.0198 + 0.0005 N - 0.0011 N^2
                     - 0.0102 N^3 + 0.0041 N^4)
   \\[2mm]
   e_1  & = &  {\pi - 2  \over 2\pi} \, {N \over N-2}
   \\[2mm]
   e_2  & = & \frac{(\pi-2)^2}{8\pi^2}\,
  \frac{N^2}{(N-2)^2}  \,+\, \frac{2 N}{N-2}
    \,\Bigl[  G_1\, (N-2) \,-\, \frac{1}{8 \pi}
   \,-\, \frac{3 N-14}{96} \Bigr] 
\end{eqnarray}
where
\be
G_1 \;\approx\; 0.0461636
\; .
\ee
From these expressions it is straightforward to obtain
\begin{subeqnarray}
   b_1  & = &  2a_1 + c_1   \\
   b_2  & = &  2a_2 + a_1^2 + 2a_1 c_1 + c_2 \\
   d_1  & = &  2a_1 + e_1   \\
   d_2  & = &  2a_2 + a_1^2 + 2a_1 e_1 + e_2
\end{subeqnarray}

Perturbation theory predicts trivially --- or rather, {\em assumes}\/ ---
that the lowest mass in the isotensor channel is the scattering state of
two isovector particles, i.e.\ there are no isotensor bound states:
\be
   \widetilde{C}_{\xi^{(exp)}_T}/\widetilde{C}_{\xi^{(exp)}_V}
   \;=\;
   \half \;.
 \label{mT=2mV}
\ee
The non-perturbative universal quantity
$\widetilde{C}_{\xi_V^{(exp)}} \equiv \Lambda_\msbar/m_V$
for the standard continuum $S^{N-1}$ $\sigma$-model
has been computed exactly by Hasenfratz, Maggiore and
Niedermayer (HMN) \cite{Hasenfratz-Niedermayer_1,Hasenfratz-Niedermayer_2,%
Hasenfratz-Niedermayer_3}
using the thermodynamic Bethe Ansatz:  it is
\be
   \widetilde{C}_{\xi_V^{(exp)}}
   \;=\;
   \widetilde{C}_{\xi_V^{(exp)}}^{\hbox{\scriptsize (HMN)}}
   \;\equiv\;
   \left( {e \over 8} \right) ^{1/(N-2)} \,
   \Gamma\biggl(1 + {1 \over N-2} \biggr)
   \;\,.
 \label{exact_Cxi}
\ee
For the rest, we know only their values at large $N$
\cite{Flyvbjerg_91c,Biscari_90,CP_4loops,CP_1overN}
\begin{eqnarray}
   \widetilde{C}_{\xi^{(2nd)}_V}/\widetilde{C}_{\xi^{(exp)}_V}   & = &
      1 \,-\,  {0.0032 \over N}  \,+\,  O(1/N^2)
 \label{xiV_1overN}           \\[2mm]
   \widetilde{C}_{\xi^{(2nd)}_T}/\widetilde{C}_{\xi^{(exp)}_T}   & = &
      \sqrt{2/3} \left[ 1 \,-\, {1.2031 \over N} \,+\, O(1/N^2) \right]
 \label{xiT_1overN}           \\[2mm]
   \widetilde{C}_{\chi_V}   & = &    2 \pi
      \left[ 1 \,+\,  {4 + 3\gamma_C - 3\gamma_E - 7 \log 2 \over N}
               \,+\,  O(1/N^2) \right]
 \label{chiV_1overN}       \\[2mm]
   \widetilde{C}_{\chi_T}   & = &
      \pi \left[ 1 \,+\,
                {-2 + 6 \log{\pi \over 4} - {36 \over \pi^2} \zeta'(2)
                 \over N}
             \,+\, O(1/N^2) \right]
 \label{chiT_1overN}
\end{eqnarray}
where $\gamma_E \approx 0.5772157$ is Euler's constant
and
\be
   \gamma_C   \;=\;
   \log\!\left(   {\Gamma(1/3) \Gamma(7/6) \over \Gamma(2/3) \Gamma(5/6)}
         \right)
   \;\approx\;   0.4861007  \;,
\ee
so that $4 + 3\gamma_C - 3\gamma_E - 7 \log 2 \approx -1.1254$ and
$-2 + 6 \log{\pi \over 4} - {36 \over \pi^2} \zeta'(2) \approx -0.0296$.
{}From \reff{xiV_1overN}/\reff{xiT_1overN} combined with \reff{mT=2mV},
we get
\be
 \widetilde{C}_{\xi^{(2nd)}_V}/\widetilde{C}_{\xi^{(2nd)}_T} \;=\;
\sqrt{6}  \left[ 1 \,+\, {1.1999 \over N} \,+\, O(1/N^2) \right]
\label{ratio_largeN}
\;\mbox{.}
\ee

We can also write asymptotic scaling in terms of an
``improved expansion parameter''
\cite{Martinelli_81,Samuel_85,Wolff_O4_O8,Lepage_93,CP_3loops,CEPS_RPN}
based on the energy $E_V = \< \bsigma_0 \cdot \bsigma_1 \>$.
First we invert the perturbative expansion \cite{Luscher_unpub,CP_3loops}
\be
   E_V(\beta) \;=\;  1 \,-\, {N-1 \over 4\beta} \,-\, {N-1 \over 32\beta^2}
                   \,-\, {0.005993 (N-1)^2 + 0.007270 (N-1)   \over   \beta^3}
                   \,+\, O(1/\beta^4)
\label{3-loop_energy}
\ee
and substitute into \reff{xi_predicted2};
this gives a prediction for $\xi$ as a function of $1-E_V$:
\be
\xi_\#(\beta)    \; = \;   \widetilde{C}'_{\xi_\#} \,
\exp\!\left[ {\pi (N-1) \over 2(N-2)}\,(1-E_V)^{-1}  \right]
      (1-E_V)^{\! 1/(N-2)}
         \left[ 1 + a_1' \,(1-E_V) + \cdots \right]
\label{xi_improved}
\ee
where
\begin{eqnarray}
\widetilde{C}'_{\xi_\#} & = & \widetilde{C}_{\xi_\#}
\,2^{-5/2}\,\exp\!\left[ -{\pi \over 4(N-2)} \right] \,
      \left[ {2(N-2) \over \pi(N-1)} \right]^{\! 1/(N-2)} \\[2mm]
a_1' & = & \frac{4}{N-1}\,a_1\,-\,\frac{1}{(N-1)(N-2)}\,
\left[ \half + \frac{\pi}{8} - 0.73086 - 0.602482 (N-1)  \right]
\end{eqnarray}
In a similar way we can obtain ``energy-improved'' expansions for
$\chi_V$ and $\chi_T$ and the ratios $\chi_V/\xi_\#^2$ and
$\chi_T/\xi_\#^2$. Note that for the basic observables 
$\xi_\#$ and $\chi_\#$, which grow exponentially as $\beta\to\infty$,
from the three-loop energy \reff{3-loop_energy}
we can get at best a three-loop ``energy-improved'' prediction,
no matter how many terms in \reff{xi_predicted2}--\reff{chiT_predicted2}
are computed; but for the {\em ratios} $\chi_\#/\xi_\#^2$, in which the 
exponential cancels out, we can obtain a four-loop ``energy-improved''
prediction using our knowledge of $c_2$ and $e_2$.

Our extrapolation method \cite{fss_greedy}
is based on the finite-size-scaling Ansatz
\be
   {\scro(\beta,sL) \over \scro(\beta,L)}   \;=\;
   F_{\scro} \Bigl( \xi(\beta,L)/L \,;\, s \Bigr)
   \,+\,  O \Bigl( \xi^{-\omega}, L^{-\omega} \Bigr)
   \;,
 \label{eq2}
\ee
where $\scro$ is any long-distance observable,
$s$ is a fixed scale factor (usually $s=2$),
$\xi(\beta,L)$ is a suitably defined finite-volume correlation length,
$L$ is the linear lattice size,
$F_{\scro}$ is a scaling function characteristic of the universality class,
and $\omega$ is a correction-to-scaling exponent.
Here we will use $\xi_V^{(2nd)}$ in the role of $\xi(\beta,L)$;
for the observables $\scro$ we will use the four
``basic observables'' $\xi^{(2nd)}_V$, $\chi_V$, $\xi^{(2nd)}_T$, $\chi_T$
as well as certain combinations of them such as $\chi_V/(\xi^{(2nd)}_V)^2$
and $\chi_T/(\xi^{(2nd)}_V)^2$.

In an asymptotically free model,
the functions $F_{\scro}$ can be computed in perturbation theory,
yielding the following expansions \cite{CEPS_RPN}
in powers of $1/x^2$, where $x \equiv \xi_V^{(2nd)}(\beta,L)/L$:
\begin{subeqnarray}
   {\xi_V^{(2nd)}(\beta,sL) \over \xi_V^{(2nd)}(\beta,L)}   & = &
   s \left[ 1 \,-\,  {w_0 \log s \over 2} \left( {A \over x} \right) ^{\! 2}
       \,-\,  \Biggl( {w_1 \log s \over 2} + {w_0^2 \log^2 s \over 8}
              \Biggr) \left( {A \over x} \right) ^{\! 4}
       \,+\,  O(x^{-6}) \right]
   \nonumber \\ \slabel{xiV_FSS_PT} \\[3mm]
   {\chi_V(\beta,sL) \over \chi_V(\beta,L)}   & = &
   s^2 \Biggl[ 1 \,-\, {\log s \over 2\pi} \, x^{-2}            \nonumber \\
   & & \qquad +\, {1 \over N-1} \left(
               {\log^2 s \over 8\pi^2} +
               \Bigl( \pi I_{3,\infty} + {1 \over 8\pi^3} \Bigr) \log s
               \right)  x^{-4}  \,+\, O(x^{-6}) \Biggr]
   \slabel{chiV_FSS_PT}                               \\[3mm]
   {\xi_T^{(2nd)}(\beta,sL) \over \xi_T^{(2nd)}(\beta,L)}
   & = &
   s \Biggl\{ 1 \,-\, {\log s \over 4\pi} \, {N-2 \over N-1} \, x^{-2}
   \nonumber \\
   & & \qquad
   - \; {(N-2)^2 \over (N-1)^2}
        \left[
        \left( {N+1 \over 2} \left( \pi I_{3,\infty} + {1 \over 8\pi^3} \right)
               + {1 \over 8\pi^2}
        \right) {\log s \over N-2}
        \,+\,  {\log^2 s \over 32\pi^2}
        \right]  x^{-4}
   \nonumber \\
   & & \qquad
    + \; O(x^{-6}) \Biggr\}
   \slabel{xiT_FSS_PT}                               \\[3mm]
   {\chi_T(\beta,sL) \over \chi_T(\beta,L)}   & = &
   s^2 \Biggl[ 1 \,-\, {\log s \over 2\pi} \, {2N \over N-1} \, x^{-2}
                                                               \nonumber \\
   & & \quad +\, {N \over (N-1)^2} \left(
               {N+2 \over 4\pi^2} \log^2 s  +
               2 \Bigl( \pi I_{3,\infty} + {1 \over 8\pi^3} \Bigr) \log s
               \right)  x^{-4}  \,+\, O(x^{-6}) \Biggr]
        \nonumber \\
   \slabel{chiT_FSS_PT}
   \label{all_FSS_PT}
\end{subeqnarray}
and also
\be
   {\xi_V^{(2nd)}(\beta,L) \over \xi_T^{(2nd)}(\beta,L)}   \;=\;
   \left( {2N \over N-1} \right) ^{\! 1/2} \,
   \left[ 1 \,+\,
     {N+1 \over N-1} \pi \Bigl( \pi I_{3,\infty} + {1 \over 8\pi^3} \Bigr)
         x^{-2}
              \,+\, O(x^{-4}) \right]
   \label{xiVoverxiT_FSS_PT}
\ee
where $A = (N-1)^{-1/2}$,
$w_0 = (N-2)/2\pi$, $w_1 = (N-2)/(2\pi)^2\;$ and
\be
   I_{3,\infty}  \approx  (2\pi)^{-4} \times 3.709741314407459
   \;.
\ee

In the case of the $\,N\to\infty\,$ limit 
it is possible to derive exact expressions
for the finite-size-scaling functions \cite{CP_1overN}.
We will compare these $N=\infty$ curves with our empirically obtained
finite-size-scaling functions for $N=8$.

\bigskip

Our extrapolation method \cite{fss_greedy} proceeds as follows:
We make Monte Carlo runs at numerous pairs $(\beta,L)$ and $(\beta,sL)$.
We plot $\scro(\beta,sL) / \scro(\beta,L)$ versus $\xi(\beta,L)/L$,
using those points satisfying both $\xi(\beta,L) \ge$ some value $\xi_{min}$
and $L \ge$ some value $L_{min}$.
If all these points fall with good accuracy on a single curve ---
thus verifying the Ansatz \reff{eq2} for
 $\xi \ge \xi_{min}$, $L \ge L_{min}$ ---
we choose a smooth fitting function $F_{\scro}$.
Then, using the functions $F_\xi$ and $F_\scro$,
we extrapolate the pair $(\xi,\scro)$ successively from
$L \to sL \to s^2 L \to \ldots \to \infty$.

We have chosen to use functions $F_{\scro}$ of the form
\be
   F_\scro(x)   \;=\;
   1 + a_1 e^{-1/x} + a_2 e^{-2/x} + \ldots + a_n e^{-n/x}   \;.
 \label{eq3}
\ee
(Other forms of fitting functions can be used instead.)
This form is partially motivated by theory, which tells us
that in some cases
$F_\scro(x) \to 1$ exponentially fast as $x\to 0$ \cite{CEPS_RPN}.
Typically a fit of order $3 \le n \le 15$ is sufficient;
the required order depends on the range of $x$ values covered by the data
and on the shape of the curve.
Empirically, we increase $n$ until the $\chi^2$ of the fit
becomes essentially constant.
The resulting $\chi^2$ value provides
a check on the systematic errors arising from
corrections to scaling and/or from the inadequacies of the form \reff{eq3}.

\subsection{Data Analysis for $N=8$}

We now apply the method described in the previous subsection to 
our data for the $N=8$ case. As mentioned before, since the primary purpose
of this paper was to study the MGMC embedding algorithm, and 
since MGMC is in any case less efficient than Wolff's cluster
algorithm \cite{Wolff_89a} for the special case of the
$N$-vector models, we have made
only a modest effort to produce static data, concentrating
on the $N=8$ case. Moreover, not all of our runs can be used to construct
the extrapolation curve\footnote{ Although all runs with $\xi\geq \xi_{min}$
can be extrapolated using the curve.}
(only 20 pairs), since at the beginning of this
project we were not yet making use of the extrapolation method in our
data analyses. Nevertheless, we get interesting results, which
confirm the asymptotic-freedom prediction with the correct nonperturbative
constant.

Our data cover the range $0.08 \leq x \equiv \xi_V^{(2nd)}(L)/L \leq 0.76$,
and we found
tentatively that a sixth-order fit is indicated: see Tables 
\ref{csiV_chisq_tab}--\ref{chiT_chisq_tab} for the four basic observables
$\xi_V,\chi_V,\xi_T,\chi_T$.

Next we took $\xi_{min}\,=\,10$ and sought to choose $L_{min}$ to avoid
any detectable systematic error from corrections to scaling.\footnote{
   A choice  $\xi_{min} = 15$ would have been
   equivalent for our set of data for obtaining the
   finite-size-scaling functions, since although we have
   points with $10\leq \xi < 15$ that can be extrapolated
   using these curves, we do not have any {\em pairs}\/ of points
   $(\beta,L)/(\beta,2L)$ falling in this interval.
}
We began with the conservative choice $L_{min}=128$ and plotted 
the deviations from the fitting curves for the four basic
observables: see Figure \ref{fig:DEV}.

Note that in each case there are three $L=64$ points
(at $x\approx 0.44, 0.49, 0.54$) that deviate upwards from the 
fitting curve by 1--5 standard deviations. We do not understand the
cause of these deviations.
They could be due to corrections to scaling, but this explanation 
seems unlikely to us because: (a) for $N=3$ the corrections to scaling are 
{\em negative} and smaller in magnitude than those found here
\cite{o3_scaling_prl,CEPS_RPN,o3_scaling_fullpaper}; and
(b) at $N=\infty$ the corrections to scaling are extremely small
\cite{CP_1overN}. Possibly there is some instability in the fitting
procedure because of the small number of $L=128$ pairs at our disposal
(only eight). So, not knowing whether to include or exclude these three
``deviant'' data points, we try both possibilities and treat the 
discrepancy between the two results as an added systematic error.
More precisely, we divided the $x$ axis into three intervals ---
$x<0.4, 0.4\leq x \leq 0.6$, and $x>0.6$ --- and allowed different
$L_{min}$ values for each.

See Tables \ref{csiV_chisq_tab}--\ref{chiT_chisq_tab} for the results
of the $\chi^2$ tests. The best $\chi^2$ values are obtained, as expected, 
for the choice $L_{min} = (64,128,64)$, which eliminates the three
``deviant'' data points as well as one other point, but leaves intact
the rest. But since we don't really have good cause to eliminate these 
points, we shall take $L_{min} = (64,64,64)$ to be our {\em preferred}
fit. In any case, we shall carry out the extrapolations using {\em both}
of these fits, and compare the results.
In all cases we take $n=6$.

The finite-size-scaling curve for $\xi_V$ coming from our preferred fit is
shown in Figure \ref{fig:csiV_statFSS}(a) (solid curve). We also show,
for comparison, the perturbative prediction \reff{xiV_FSS_PT} through
orders $1/x^2$ and $1/x^4$ (dashed curves). In Figure 
\ref{fig:csiV_statFSS}(b) we show the same solid curve and, for comparison,
the corresponding curves for $N=3$ (Monte Carlo) 
\cite{o3_scaling_prl,o3_scaling_fullpaper} and $N=\infty$ (exact)
\cite{CP_1overN}. In Figures 
\ref{fig:chiV_statFSS}--\ref{fig:chiT_statFSS} we show the analogous
curves for the observables $\chi_V,\xi_T,\chi_T$.
In all cases, the agreement with perturbation theory is excellent
for $x\gtapprox 0.5$ (isovector sector) and $x\gtapprox 0.6$ 
(isotensor sector).
The $N=8$ finite-size-scaling curves fall, as expected, between the
$N=3$ and $N=\infty$ curves, and rather closer to the latter.


Next we checked for consistency between the extrapolated values of
each observable coming from different lattice sizes at the same $\beta$.
In most cases these are consistent at better than the 5\% level;
but in two cases ($\beta = 6.1, 7.1$) they are inconsistent at worse than 
the 0.1\% level (we don't know why).
If we exclude these two $\beta$ values and sum over the rest,
we get $\chi^2 \,=\, 12.09,\; 15.16,\; 18.60, \; 15.00$ (15 degrees
of freedom), respectively, for 
the observables $\xi_V, \chi_V, \xi_T, \chi_T$. This corresponds
to confidence levels of $67\%,\; 44\%,\; 23\%,\; 45\%$.

In Table \ref{dati_est_o8} we show the extrapolated values of the
four basic observables from our preferred fit $L_{min}\,=\,(64,64,64)$
and also from the fit $L_{min}\,=\,(64,128,64)$. The discrepancies
between these values (if larger than the statistical errors)
can serve as a rough estimate of the remaining systematic errors due to
corrections to scaling. 
The statistical errors in the extrapolated values
from our preferred fit range from about 
0.2--0.5\% when $\xi_V\ltapprox20$ to about 2--3\% when $\xi_V \approx 650$.
The systematic errors are of the same order as the statistical errors,
or smaller.
The statistical errors at different $\beta$
are strongly positively correlated, but we did not evaluate this correlation
quantitatively.

\medskip
In Figure \ref{fig:csiV_AF}  (points $+$, $\times$, $\protect\fancyplus$)
we plot 
$\xi^{(2nd)}_{V,\infty,estimate\,(64,64,64)}$
divided by the two-loop, three-loop and four-loop predictions 
for $\xi_V^{(exp)}$
given in equations \reff{xi_predicted2}/\reff{a1}/\reff{a2}/\reff{exact_Cxi}.
This curve is theoretically predicted to converge to a constant as
$\beta\to\infty$, and the limiting value is 
$\widetilde{C}_{\xi_V^{(2nd)}}/
\widetilde{C}_{\xi_V^{(exp)}}^{\hbox{\scriptsize (HMN)}}$.
We also compare to the prediction \reff{xi_improved} using the
``improved expansion parameter'' $1-E_V$ (points $\Box$ and $\Diamond$).
[For $E_V$ in \reff{xi_improved} we use the value measured on the
largest lattice;
the statistical errors and finite-size corrections on $E_V$
are less than $5 \times 10^{-5}$, and therefore induce a negligible error
(less than 0.5\%) on the predicted $\xi$.]
We show for comparison the expected limiting value 0.9996 coming from 
the $1/N$ expansion \reff{xiV_1overN} through order $1/N$,
evaluated at $N=8$.
Both the
four-loop and the ``improved'' 3-loop prediction are in excellent
agreement with the data, showing discrepancies of less than 1\%.

In Figure \ref{fig:csiT_AF} we proceed analogously for $\xi_T$, by plotting
(points $+$, $\times$, $\protect\fancyplus$) 
$\xi^{(2nd)}_{T,\infty,estimate\,(64,64,64)}$ divided by the two-loop, 
three-loop and four-loop predictions for 
$\xi^{(exp)}_T\equiv \half \xi^{(exp)}_V$ given in equations 
\reff{xi_predicted2}/\reff{a1}/\reff{a2}/\reff{mT=2mV}/\reff{exact_Cxi}. 
We also compare to the prediction \reff{xi_improved} using the
``improved expansion parameter'' $1-E_V$ (points $\Box$ and $\Diamond$).
The four-loop and improved three-loop data are relatively flat, and suggest
a limiting value $\widetilde{C}_{\xi_T^{(2nd)}}/
\widetilde{C}_{\xi_T^{(exp)}}^{\hbox{\scriptsize (HMN)}}
\approx 0.72\pm0.01$. This can be compared 
with the expected limiting value 0.6937 coming from
the $1/N$ expansion \reff{xiV_1overN} through order $1/N$,
evaluated at $N=8$. Clearly the order-$1/N^2$ corrections are 
still significant (of order
3--4\%), which is not surprising since the order-$1/N$ correction
was 15\%.

For the susceptibility $\chi_V$ we proceed in two different ways,
using either $\chi_V$ directly or else using the ratio 
$\chi_V/\xi_V^{(2nd) 2}$. The advantage of the latter approach is that one
additional term of perturbation theory is available. 
Thus, in
Figure \ref{fig:chiV_AF}(a) we plot $\chi_{V,\infty,estimate\,(64,64,64)}$ 
divided by the theoretical prediction \reff{chiV_predicted2}
{\em with the prefactor $\widetilde{C}_{\chi_V}$ omitted}; the 
$\beta\to\infty$ limit of this curve thus
gives an estimate of $\widetilde{C}_{\chi_V}$.
Here we have two-loop, three-loop and four-loop predictions
(points $+$, $\times$, $\protect\fancyplus$) as well as ``improved''
two-loop and three-loop predictions ($\Box$ and $\Diamond$).
Similarly, we could plot $\chi_{V,\infty,estimate\,(64,64,64)}/
[\xi^{(2nd)}_{V,\infty,estimate\,(64,64,64)}]^2$
divided by the theoretical prediction 
\reff{chiV_predicted2}/\reff{xi_predicted2} with the prefactors
$\widetilde{C}_{\chi_V}$ and $\widetilde{C}_{\xi_V^{(2nd)}}$ omitted;
the $\beta\to\infty$ limit would then give an estimate of 
$\widetilde{C}_{\chi_V}/(\widetilde{C}_{\xi_V^{(2nd)}})^2$.
However, in order to make the vertical scale of this graph more
directly comparable to that of Figure \ref{fig:chiV_AF}(a),
we have multiplied the quantity being plotted by the factor
$(\widetilde{C}_{\xi_V^{(exp)}}^{\hbox{\scriptsize (HMN)}})^2$.
Note that this does not in any way alter the {\em logic} of the
analysis, as $\widetilde{C}_{\xi_V^{(exp)}}^{\hbox{\scriptsize (HMN)}}$
is an explicit number defined in \reff{exact_Cxi}. The resulting curve is
plotted in Figure \ref{fig:chiV_AF}(b); its $\beta\to\infty$ limit
gives an estimate of 
$\widetilde{C}_{\chi_V} 
(\widetilde{C}_{\xi_V^{(exp)}}^{\hbox{\scriptsize (HMN)}}/
\widetilde{C}_{\xi_V^{(2nd)}})^2$.
In this case we have two-loop,
three-loop, four-loop and five-loop predictions
($+$, $\times$, $\protect\fancyplus$, $\protect\fancycross$)
as well as ``improved'' two-loop, three-loop and four-loop
predictions ($\Box$, $\Diamond$, $\fancysquare$).
To convert this to an estimate for $\widetilde{C}_{\chi_V}$ itself, 
we need to multiply by
\be
\left( \frac{\widetilde{C}_{\xi_V^{(2nd)}}}{
\widetilde{C}_{\xi_V^{(exp)}}^{\hbox{\scriptsize (HMN)}}} \right)^2
\;=\; \left( \frac{\widetilde{C}_{\xi_V^{(2nd)}}}{
         \widetilde{C}_{\xi_V^{(exp)}}}  \right)^2 \,
 \left( \frac{\widetilde{C}_{\xi_V^{(exp)}}}{
\widetilde{C}_{\xi_V^{(exp)}}^{\hbox{\scriptsize (HMN)}}} \right)^2
\label{C_tilde_factor}
\ee
The first factor on the right side is theoretically expected to be
$\approx 0.9992$, and the second factor is theoretically expected to be
1; moreover, our data for $\xi_V^{(2nd)}$ itself (Figure \ref{fig:csiV_AF})
are consistent with this prediction (of course they are incapable of 
distinguishing 0.9992 from 1). For all practical purposes the factor 
\reff{C_tilde_factor} is 1, and Figures \ref{fig:chiV_AF}(a) and
\ref{fig:chiV_AF}(b) can be compared directly.
The two approaches give consistent
results, but the one based on $\chi_V/\xi_V^{(2nd) 2}$ seems to work better.
We get the value 
\be
\widetilde{C}_{\chi_V}\;=\; 5.58 \,\pm\, 0.10 \;.
\ee
This can be compared with the predicted value 
$\widetilde{C}_{\chi_V}\;=\;5.40$ coming from the $1/N$ expansion
\reff{chiV_1overN} through order $1/N$, evaluated at $N=8$.\footnote{
Our preliminary report \cite{Caracciolo_LAT95} mistakenly wrote
5.30 instead of 5.40. We apologize for the arithmetic error!}
Clearly the order-$1/N^2$ corrections are still significant 
(of order 3\%), which is not surprising since the order-$1/N$ correction 
was 14\%.

We proceed analogously for $\chi_T$, using either $\chi_T$ directly
or else using the ratio $\chi_T/\xi_V^{(2nd) 2}$. 
In Figure \ref{fig:chiT_AF}(a) we plot $\chi_{T,\infty,estimate\,(64,64,64)}$
divided by the theoretical prediction \reff{chiT_predicted2}
{\em with the prefactor $\widetilde{C}_{\chi_T}$ omitted}; the
$\beta\to\infty$ limit of this curve
gives an estimate of $\widetilde{C}_{\chi_T}$.
Here we have two-loop, three-loop and four-loop predictions
(points $+$, $\times$, $\protect\fancyplus$) as well as ``improved''
two-loop and three-loop predictions ($\Box$ and $\Diamond$).
In Figure \ref{fig:chiT_AF}(b) we plot
$\chi_{T,\infty,estimate\,(64,64,64)}/
[\xi^{(2nd)}_{V,\infty,estimate\,(64,64,64)}]^2$
divided by the theoretical prediction
\reff{chiT_predicted2}/\reff{xi_predicted2} for this same quantity,
with the prefactors $\widetilde{C}_{\chi_T}$ and
$\widetilde{C}_{\xi_V^{(2nd)}}$ omitted, and the whole thing
finally multiplied by 
$(\widetilde{C}_{\xi_V^{(exp)}}^{\hbox{\scriptsize (HMN)}})^2$;
the $\beta\to\infty$ limit of this curve gives an estimate of
$\widetilde{C}_{\chi_T}(
\widetilde{C}_{\xi_V^{(exp)}}^{\hbox{\scriptsize (HMN)}}/
\widetilde{C}_{\xi_V^{(2nd)}})^2$.
As before, this is for all practical purposes equal to 
$\widetilde{C}_{\chi_T}$. Here we have two-loop,
three-loop and four-loop predictions
($+$, $\times$, $\protect\fancyplus$)
as well as ``improved'' two-loop, three-loop and four-loop
predictions ($\Box$, $\Diamond$, $\fancysquare$), namely we do not have an 
additional term of conventional perturbation theory, but we do have
an additional term in the ``improved'' perturbation theory.
The two approaches give consistent
results, and here the one based on $\chi_T/\xi_V^{(2nd) 2}$ 
does not seem to work better.
We get the value
\be
\widetilde{C}_{\chi_T}\;=\; 4.1 \,\pm\, 0.1 \;.
\ee
This can be compared with the predicted value
$\widetilde{C}_{\chi_T}\;=\;3.13$ coming from the $1/N$ expansion
\reff{chiT_1overN} through order $1/N$, evaluated at $N=8$.
Clearly the order-$1/N^2$ corrections are still significant (of order 31\%)
which {\em is} surprising since the order-$1/N$ correction was 0.4\%.

\section{Finite-Size-Scaling Analysis: Dynamic Quantities}
\label{section:finite-size-scaling_dynamic}  \label{sec5}

\subsection{Integrated Autocorrelation Times}  \label{sec5.1}


Of all the observables we studied, the slowest mode (by far)
is the squared isovector magnetization $\scrm_V^2$:
this quantity measures the relative rotations of the spins
in different parts of the lattice,
and is the prototypical $O(N)$-invariant ``long-wavelength observable''.
For all three values of $N$,
the autocorrelation time $\tau_{int,\scrm_V^2}$
has the same qualitative behavior:
as a function of $\beta$ it first rises to a peak and then falls;
the location of this peak shifts towards $\beta=\infty$ as $L$ increases;
and the height of this peak grows as $L$ increases.
A similar but less pronounced peak is observed in $\tau_{int,\scrm_T^2}$.
By contrast,
the integrated autocorrelation times of the energies, $\taueV$ and $\taueT$,
are uniformly small ---
less than 2.5, 3.2, 6.4 for $N=3,4,8$, respectively ---
and vary only weakly with $\beta$ and $L$.
This is because the energies are primarily ``short-wavelength observables'',
and have only weak overlap with the modes responsible for critical slowing-down.

Let us now make these considerations quantitative, by applying
finite-size scaling to the dynamic quantities $\tau_{int,\scrm_V^2}$
and $\tau_{int,\scrm_T^2}$.
We use the Ansatz
\be
  \tau_{int,A}(\beta,L)  \;\approx\;
     \xi(\beta,L)^{z_{int,A}} \, g_A \Bigl( \xi(\beta,L)/L  \Bigr)
 \label{dyn_FSS_Ansatz}
\ee
for $A = \scrm_V^2$ and $\scrm_T^2$.
Here $g_A$ is an unknown scaling function, and
$g_A(0) = \lim_{x \downarrow 0} g_A(x)$
is supposed to be finite and nonzero.\footnote{
   It is of course equivalent to use the Ansatz
   $$\tau_{int,A}(\beta,L)  \;\approx\;
     L^{z_{int,A}} \, h_A \Bigl( \xi(\beta,L)/L  \Bigr)  \;,$$
   and indeed the two Ans\"atze are related by $h_A(x) = x^{z_{int,A}} g_A(x)$.
   However, to determine whether
   $\lim_{x \downarrow 0} g_A(x) = \lim_{x \downarrow 0} x^{-z_{int,A}} h_A(x)$
   is nonzero, it is more convenient to inspect a graph of $g_A$
   than one of $h_A$.
}
We determine $z_{int,A}$ by plotting $\tau_{int,A}/\xi_V(L)^{z_{int,A}}$
versus $\xi_V(L)/L$ and adjusting $z_{int,A}$ until the points fall as closely
as possible onto a single curve (with priority to the larger $L$ values).
We emphasize that the dynamic critical exponent $z_{int,A}$
is in general {\em different}\/ from the exponent $z_{exp}$
associated with the exponential autocorrelation time $\tau_{exp}$
\cite{Sokal_Lausanne,CPS_90,Sokal_LAT90}.

Using the procedure just described, we find
\be
   z_{int,\scrm_V^2}   \;=\;
      \cases{ 0.70 \pm 0.08  & for $N=3$   \cr
              0.52 \pm 0.10  & for $N=8$   \cr
            }
 \label{z_estimates}
\ee
(subjective 68\% confidence limits).
In Figures \ref{o3DFSS} and \ref{o8DFSS}
we show the ``best'' finite-size-scaling plots for each case.
Note that the corrections to scaling are quite strong:
the $L=32$ point deviates considerably 
from the asymptotic scaling curve,
and even for $L=64$ the deviation is noticeable.
However, it is reasonable to hope that these corrections to scaling
will have decayed to a small value by $L=128$;
if so, our estimates of $z_{int,\scrm_V^2}$ for the  W-cycle ---
which attempt to place the $L=128$ and $L=256$ points on top of one another
--- will be afflicted by only a small systematic error.
In any case, the error bars in \reff{z_estimates} take account of this
potential systematic error.
Similarly, for the isotensor sector we have
\be
   z_{int,\scrm_T^2}   \;=\;
      \cases{ 0.54 \pm 0.08  & for $N=3$   \cr
              0.53 \pm 0.10  & for $N=8$   \cr
            }
\ee
(subjective 68\% confidence limits).
In Figures \ref{o3DFSS_T} and \ref{o8DFSS_T} we show the ``best''
finite-size-scaling plots for each case. It thus appears that,
for $N=3$,
$z_{int,\scrm_T^2}$ is {\em strictly smaller than} $z_{int,\scrm_V^2}$.
This behavior is by no means surprising; indeed, the exponents
$z_{int,A}$ are expected to vary from one observable to another
\cite{Sokal_Lausanne,Sokal_LAT90}.
On the other hand, for $N=8$ we have $z_{int,\scrm_T^2}\approx
z_{int,\scrm_V^2}$ within statistical error. We have no explanation
for the very different behavior in the two cases. Perhaps $z_{int,\scrm_T^2}$
is strictly smaller than $z_{int,\scrm_V^2}$ for {\em all} $N$, but 
the difference between the two exponents goes rapidly to zero
as $N\to\infty$.

For the $N=4$ case, our data are insufficient to support a good
finite-size-scaling analysis.
Instead, let us compare our values of $\tau_{int,\scrm_V^2}$
to the corresponding values for the direct MGMC algorithm
(taken from \cite{MGMC_O4}):
see Table \ref{comparison_with_sabino}.
We find that the ratio of the two autocorrelation times
is compatible with a constant (independent of $L$ and $\beta$):
\be
   { \taux(\hbox{$XY$ embedding})   \over  \taux(\hbox{direct})  }
   \;=\;
   3.18 \pm 0.04
\ee
with $\chi^2 = 6.55$ (13 DF, level = 92\%), using all the data; or
$3.20 \pm 0.04$ with $\chi^2 = 4.23$ (10 DF, level = 94\%), if we use only
the data for $L=64$ and $128$.
Therefore, the two algorithms have the same dynamic critical exponent,
which was found in \cite{MGMC_O4} to be
\be
   z_{int,\scrm_V^2}   \;=\;  0.60 \pm 0.07  \qquad (N=4) \;,
\ee
and the same dynamic finite-size-scaling functions;
in other words, they belong (as expected) to the same dynamic
universality class.
It could also have been expected that the ratio of autocorrelation times
would be $\approx 3$:  there are $N-1$ distinct internal-spin directions in
an $N$-vector model, and the $XY$-embedding algorithm handles them
one at a time, while the direct algorithm handles them all at once.

The dynamic critical exponent $z_{int,\scrm_V^2}$ for W-cycle MGMC
thus appears to be decreasing as $N$ increases.
(However, our evidence for this is at present
on the borderline of statistical significance.)
It is tempting to speculate that $z_{W-cycle,PC}$($N$-vector) tends to zero
as $N\to\infty$;  this would be compatible with the vague idea that the
$N=\infty$ model is ``essentially Gaussian''.
However, it is far from clear that the $N=\infty$ model is Gaussian
{\em in the sense relevant for MGMC}\/.

\subsection{Autocorrelation Functions and Exponential Autocorrelation Times}
   \label{sec5.2}

Finally, we want to test the more detailed dynamic finite-size-scaling Ansatz
\be
   \rho_A(t;\beta,L)  \;\approx\;
     |t|^{-p_A}
     h_A \Bigl( t / \tau_{exp,A}(\beta,L) \,;\,  \xi(\beta,L)/L  \Bigr)
   \;,
 \label{dyn_FSS_2}
\ee
where $p_A$ is an unknown exponent and $h_A$ is an unknown scaling function.
If $p_A = 0$, then \reff{dyn_FSS_2} can equivalently be written as
\be
   \rho_A(t;\beta,L)  \;\approx\;
     \widehat{h}_A \Bigl( t / \tau_{int,A}(\beta,L) \,;\,
                          \xi(\beta,L)/L  \Bigr)
   \;.
 \label{dyn_FSS_3}
\ee
In this latter situation\footnote{
  Contrary to much belief, $z_{int,A}$ {\em need not} equal $z_{exp}$.
  Indeed, if $p_A>0$, we have $z_{int,A}\,=\,(1-p_A)\,z_{exp}\,<\,z_{exp}$.
  See \cite{Sokal_Lausanne,Sokal_LAT90} for further discussion.
},
$\tau_{int,A}$ and $\tau_{exp,A}$ have the {\em same} dynamic critical
exponent $z_{int,A}=z_{exp}$, and we furthermore have
\be
\frac{\tau_{int,A}}{\tau_{exp,A}} \;\approx\; F_A\left(\xi(\beta,L)/L\right)
\ee
where
\be
F_A(x) \;\equiv\; \lim_{t\to +\infty}\, \frac{1}{t}\,
 \log {\widehat h}_A(t;x)\;.
\ee

In Figure \ref{AUTFo3} we test the Ansatz \reff{dyn_FSS_3} for $A=\scrm^2_V$
and $N=3$, by plotting $\rho_{\scrm_V^2}(t;\beta,L)$ versus $t/\taux (\beta,L)$,
and mapping different ranges of $\xi(\beta,L)/L$ to different symbols.
(For visual clarity we have also ``thinned out'' the data points; and
to reduce the effects of statistical noise, we have included data only from
those runs that are longer than $20000\, \taux$.)
The data support the Ansatz \reff{dyn_FSS_3} reasonably well, with each
range of $\xi(\beta,L)/L$ defining roughly a single curve (until that curve
falls into the statistical noise). The curve for small $\xi(\beta,L)/L$ is
close to straight (i.e.\ close to a pure exponential), while the curves
for larger $\xi(\beta,L)/L$ are increasingly convex.\footnote{
  The rescaled horizontal axis in Figure \ref{AUTFo3} ensures that the
  total area under each curve is 1. Therefore, the more convex curves must
  be below the straight curve for small $t/\taux$, but {\em above} it
  for large $t/\taux$.
}
This means that the ratio $\taux/\tauxexp$ is close to 1 for small 
$\xi(\beta,L)/L$, and less than 1 for larger $\xi(\beta,L)/L$. Indeed, our 
crude estimates of $\tauxexp$, based on linear fits (taking proper account of
correlations) to the approximately linear regime in Figure \ref{AUTFo3},
suggest that $\taux/\tauxexp$ is $\approx$ 0.90--1 for 
 $\xi(\beta,L)/L \ltapprox 0.1$, and decreases to $\approx 0.75$ for
$\xi(\beta,L)/L \approx 0.5$. It is {\em conceivable} that $\taux/\tauxexp$
tends to 1 as $\xi(\beta,L)/L \to 0$; if true, this would mean that 
$\scrm^2_V$ truly becomes the ``slowest mode'' in the limit $L\to\infty$,
$\xi/L \to 0$.

In Figure \ref{AUTFo8} we show the analogous plot for $N=8$. Again the
Ansatz \reff{dyn_FSS_3} seems to be well verified, but here the dependence on
$\xi(\beta,L)/L$ is much weaker (in truth, it is virtually invisible in our
data). We estimate that 
$\taux/\tauxexp$ is $\approx$ 0.98 for 
 $\xi(\beta,L)/L \ltapprox 0.2$, and decreases to $\approx 0.95$ for
$\xi(\beta,L)/L \approx 0.6$. It is {\em conceivable} that this dependence on 
$\xi(\beta,L)/L$ disappears entirely in the limit $N\to\infty$, i.e.\ we
{\em might} have $\taux/\tauxexp\to 1$ for all $\xi(\beta,L)/L$ in the limit
$L\to\infty$, $N \to \infty$.

\section{Discussion} \label{section:discussion}  \label{sec6}

\subsection{Dynamic Critical Behavior of Piecewise-Constant MGMC}

Let us now summarize our conclusions about the dynamic critical behavior of
W-cycle MGMC with piecewise-constant interpolation:

\medskip

1) The direct and $XY$-embedding MGMC algorithms for the
4-vector model fall, as expected, into the same dynamic universality class.
Indeed, the ratio of their autocorrelation times is approximately 1:3,
as expected from the fact that there are $N-1$ distinct internal-spin
directions in an $N$-vector model.

\medskip

2) The dynamic critical exponent $z_{int,\scrm_V^2}$ for W-cycle MGMC with
piecewise-constant interpolation is clearly {\em nonzero}\/
for two-dimensional asymptotically free models:  we have
\be
z_{W-cycle,PC}  \;=\;
         \cases{0.70 \pm 0.08   & for the 3-vector model [this paper] \cr
                0.60 \pm 0.07   & for the 4-vector model
                                           \cite[this paper]{MGMC_O4} \cr
                0.52 \pm 0.10   & for the 8-vector model [this paper] \cr
                0.45 \pm 0.02   & for the $SU(3)$ chiral model
                                 \cite{Mendes_LAT95,MGMC_SU3}    \cr
   } 
\ee
(subjective 68\% confidence intervals).
Furthermore, this exponent apparently {\em varies}\/
from one asymptotically free model to another
(but the evidence for this is on the borderline of statistical significance,
especially in the $N$-vector models).
Finally, it is {\em conceivable}\/ that $z_{W-cycle,PC}$($N$-vector)
tends to zero as $N\to\infty$;
testing this will require measurements at larger values of $N$.

In \cite[Section 4.1]{MGMC_O4} we produced a heuristic explanation
of why $z_{MGMC} \neq 0$, based on the logarithmically decaying
deviations from Gaussianness in an asymptotically free theory.
This explanation does not, unfortunately,
give a quantitative prediction for $z_{MGMC}$;  it merely makes the weak
prediction that $0 < z_{MGMC} < z_{heat-bath}$.
Moreover, our study of MGMC in the {\em one}\/-dimensional 4-vector model
\cite{mgmco4d1} has cast some doubt on the exactness of this explanation
(it seems to be off by at least a logarithm).
Clearly, the dynamic critical behavior of MGMC is far from being understood.

\subsection{Piecewise-Constant vs.\ Smooth Interpolation}   \label{sec6.2}
  
As noted in the Introduction, Mack and collaborators
\cite{Mack_Cargese,Mack-Meyer_90,%
Hasenbusch_LAT90,Hasenbusch-Meyer_PRL,Hasenbusch_CP3}
have advocated the use of a smooth interpolation
(e.g.\ piecewise-linear or better) in place of piecewise-constant.
The key question is this:
Does the MGMC algorithm have a {\em different dynamic critical exponent}\/
depending on whether smooth or piecewise-constant interpolation is used?

In the Gaussian case, the answer is ``yes'' for a V-cycle
but ``no'' for a W-cycle (see the Introduction).
For non-Gaussian asymptotically free models, we conjecture that the answer
is the same as in the Gaussian case,
but to date no full-scale test has been made.
Some time ago,
Hasenbusch, Meyer and Mack \cite{Hasenbusch_LAT90}
reported preliminary data for the 3-vector model suggesting that
$z_{int,\scrm_V^2} \approx 0.2$ for a V-cycle with smooth interpolation.
(These authors did not study a W-cycle, as this would be extremely costly
in their unigrid approach, but it is safe to assume that
$z_{W-cycle} \le z_{V-cycle}$.)
This is considerably lower than our estimate $z_{int,\scrm_V^2} \approx 0.7$
for the piecewise-constant W-cycle ---
suggesting that the smooth interpolation might indeed lead to a
smaller dynamic critical exponent than piecewise-constant interpolation,
even for a W-cycle.
However, the estimate of \cite{Hasenbusch_LAT90}
is based on runs at only four $(\beta,L)$ pairs,
which moreover have different values of $\xi/L$,
so it is very difficult to perform a correct finite-size-scaling analysis.
It would be very useful to have data at additional $(\beta,L)$ pairs,
and with higher statistics.

More recently,
Hasenbusch and Meyer \cite{Hasenbusch-Meyer_PRL,Hasenbusch_CP3}
have reported data for the $SU(3)$ and $\CP^3$ $\sigma$-models,
using a V-cycle with piecewise-linear interpolation.
For $\CP^3$ the data are consistent with the dynamic finite-size-scaling
Ansatz \reff{dyn_FSS_Ansatz} with $z_{int,\scrm_V^2} \approx 0.2$.
For $SU(3)$, however, the data are very erratic,
and do not appear to be consistent with \reff{dyn_FSS_Ansatz}.
More data would be very useful here too;
both the dynamic critical exponent and the
dynamic finite-size-scaling functions could be compared with those
for the piecewise-constant W-cycle \cite{Mendes_LAT95,MGMC_SU3}.

\subsection{Comparison with Other Collective-Mode Algorithms}   \label{sec6.3}

We have previously \cite[Section 4.3]{MGMC_O4}
compared the behavior of MGMC with that of Fourier acceleration
\cite{Parisi_84,Batrouni_85,Dagotto_87:XY,Dagotto_87:SU(3)}.
To our knowledge, no new data on the dynamic critical behavior
of Fourier acceleration has appeared since that discussion,
so we have nothing new to add.

A very different approach to collective-mode Monte Carlo
is represented by Wolff's \cite{Wolff_89a} cluster algorithm 
for $N$-vector models.\footnote{
   This algorithm was also proposed independently by
   Hasenbusch \cite{Hasenbusch_90}.
}
This algorithm is based on embedding a field of Ising variables
into the $N$-vector model, and then simulating the induced Ising model
by the Swendsen-Wang \cite{Swendsen_87} algorithm
(or its single-cluster variant \cite{Wolff_89a}).
It can be argued heuristically \cite{Edwards_89,CEPS_swwo4c2}
that this algorithm is effective at creating long-wavelength spin waves;
indeed, extensive numerical tests show
that critical slowing-down is {\em completely eliminated}\/
in the two-dimensional $N$-vector models for $N = 2,3,4$
\cite{Wolff_89a,Edwards_89,Wolff_90,CEPS_swwo4c2,CEPS_rpn_dynamic}.\footnote{
   For $N=2$ the complete elimination of critical slowing-down
   requires also that the algorithm be effective at
   creating vortex-antivortex pairs; the mechanism for this is
   explained in \cite{Edwards_89}.
}

The principal difference between MGMC and Fourier acceleration on the
one hand, and algorithms of Swendsen-Wang and Wolff type on the other,
is how proposals are made for updating the long-wavelength modes.
MGMC and Fourier acceleration propose additive updates
of {\em fixed}\/ shape (square, triangular or sinusoidal)
and variable amplitude, while
cluster algorithms like Swendsen-Wang and Wolff allow the system,
in some sense, to {\em choose its own collective modes}\/.

At present, the Wolff algorithm is clearly the best
algorithm for simulating $N$-vector models (at least in two dimensions);
neither MGMC nor Fourier acceleration appears to be a serious competitor.
However, this favorable situation for Wolff-type algorithms
probably does {\em not}\/ extend to other $\sigma$-models:
indeed, there are strong reasons to believe \cite{CEPS_swwo4c2}
that a generalized Wolff-type embedding algorithm
can have dynamic critical exponent $z \ll 2$
{\em only}\/ if the target manifold is a sphere (as in the $N$-vector model),
a product of spheres, or the quotient of such a space by a discrete group
(e.g.\ real projective space $RP^{N-1}$).\footnote{
   Moreover, the {\em practical}\/ Wolff-type embedding algorithm
   (with one Swendsen-Wang hit on the induced Ising model per iteration)
   appears to behave rather poorly in the $RP^{N-1}$ models, with $z \approx 1$
   \cite{CEPS_LAT91,CEPS_rpn_dynamic}.
}
By contrast, the MGMC algorithm (especially in its $XY$-embedding form)
has a straightforward extension
to $\sigma$-models taking values in an arbitrary homogeneous space $G/H$,
and it appears to work well quite generally
[e.g.\ $z \approx 0.45$ for $SU(3)$].
This is a strong motivation for pursuing the study of MGMC algorithms.

\section*{Acknowledgments}

We want to thank Martin Hasenbusch and Steffen Meyer
for the discussions that initiated this project, and
for many helpful comments thereafter.
We also wish to thank Gustavo Mana for many discussions.
Most of the computations reported here were carried out
on the Cray C-90 at the Pittsburgh Supercomputing Center (PSC).
This work was supported in part 
by the U.S.\ National Science Foundation grants DMS-9200719
and PHY-9520978 (A.D.S.),
by the Istituto Nazionale di Fisica Nucleare (A.P.),
and by PSC grants PHY890035P and MCA94P032P.


\clearpage

%
%
\begin{table}
\addtolength{\tabcolsep}{-1.0mm}
%
\protect\footnotesize
\begin{center}
\begin{tabular}{|r|r|r@{\ (}r|r@{\ (}r|r@{\ (}r|r@{\ (}r|r@{\ (}r|
   r@{\ (}r|} \hline
\multicolumn{14}{|c|}{$d=2$ 3-vector model: static data } \\ \hline
  &  & \multicolumn{2}{|c}{\ } & \multicolumn{2}{|c}{\ } &
\multicolumn{2}{|c}{\ } & \multicolumn{2}{|c}{\ } &
\multicolumn{2}{|c}{\ } & \multicolumn{2}{|c|}{\ }   \\
\multicolumn{1}{|c|}{$L$} & \multicolumn{1}{c|}{$\beta$} &
  \multicolumn{2}{|c}{$\chi_V$}  &  \multicolumn{2}{|c}{$\chi_T$} &
 \multicolumn{2}{|c}{$\xi_V^{(2nd)}$} &  \multicolumn{2}{|c}{$\xi_T^{(2nd)}$} & 
  \multicolumn{2}{|c}{$E_V$} &  \multicolumn{2}{|c|}{$E_T$}  \\ 
\hline
 32 & 1.500 & 143.20 & 0.80) &  6.34 & 0.04) & 
10.08 & 0.06) & 3.82 & 0.03) & 
0.603146 & 96) & 0.497267 & 61) \\ 
\hline
 64 & 1.450 & 114.61 & 0.56) &  4.03 & 0.01) & 
8.49 & 0.06) & 2.41 & 0.05) & 
0.582513 & 26) & 0.484598 & 16) \\ 
 64 & 1.500 & 177.31 & 0.81) &  5.33 & 0.01) & 
11.17 & 0.06) & 3.33 & 0.03) & 
0.601747 & 23) & 0.496433 & 15) \\ 
 64 & 1.550 & 265.44 & 1.24) &  7.40 & 0.02) & 
14.06 & 0.07) & 4.66 & 0.03) & 
0.619546 & 22) & 0.508044 & 15) \\ 
 64 & 1.600 & 397.54 & 1.58) & 11.10 & 0.04) & 
18.11 & 0.07) & 6.62 & 0.03) & 
0.636082 & 20) & 0.519465 & 14) \\ 
 64 & 1.650 & 545.23 & 2.26) & 16.85 & 0.07) & 
22.09 & 0.10) & 8.74 & 0.04) & 
0.651148 & 23) & 0.530473 & 17) \\ 
 64 & 1.700 & 702.27 & 2.15) & 25.02 & 0.10) & 
26.34 & 0.10) & 10.89 & 0.04) & 
0.664824 & 21) & 0.540996 & 16) \\ 
 64 & 1.725 & 768.21 & 2.06) & 29.55 & 0.11) & 
27.91 & 0.09) & 11.77 & 0.04) & 
0.671122 & 20) & 0.546022 & 16) \\ 
 64 & 1.750 & 834.56 & 1.81) & 34.51 & 0.12) & 
29.64 & 0.08) & 12.65 & 0.04) & 
0.677156 & 19) & 0.550953 & 16) \\ 
 64 & 1.775 & 895.44 & 2.11) & 40.06 & 0.16) & 
31.11 & 0.10) & 13.53 & 0.05) & 
0.682911 & 22) & 0.555767 & 19) \\ 
 64 & 1.800 & 956.71 & 1.87) & 46.13 & 0.17) & 
32.69 & 0.10) & 14.40 & 0.05) & 
0.688427 & 21) & 0.560468 & 19) \\ 
 64 & 1.850 & 1054.35 & 1.69) & 57.81 & 0.19) & 
34.90 & 0.09) & 15.76 & 0.05) & 
0.698672 & 20) & 0.569463 & 18) \\ 
 64 & 1.900 & 1147.05 & 2.09) & 70.87 & 0.29) & 
37.09 & 0.12) & 17.08 & 0.07) & 
0.708203 & 27) & 0.578129 & 26) \\ 
 64 & 1.950 & 1224.55 & 1.94) & 83.83 & 0.32) & 
38.58 & 0.12) & 18.06 & 0.06) & 
0.717119 & 25) & 0.586526 & 25) \\ 
\hline
128 & 1.450 & 115.43 & 0.66) &  4.01 & 0.01) & 
8.48 & 0.18) & 2.19 & 0.22) & 
0.582493 & 16) & 0.484605 & 10) \\ 
128 & 1.500 & 175.61 & 1.15) &  5.17 & 0.01) & 
11.11 & 0.17) & 2.92 & 0.18) & 
0.601597 & 16) & 0.496336 & 10) \\ 
128 & 1.550 & 277.80 & 2.03) &  6.95 & 0.02) & 
14.46 & 0.18) & 4.31 & 0.13) & 
0.619374 & 16) & 0.507931 & 11) \\ 
128 & 1.575 & 351.79 & 2.01) &  8.07 & 0.02) & 
16.52 & 0.13) & 4.62 & 0.09) & 
0.627728 & 11) & 0.513626 &  7) \\ 
128 & 1.600 & 451.87 & 2.78) &  9.51 & 0.02) & 
19.22 & 0.15) & 5.43 & 0.09) & 
0.635722 & 10) & 0.519229 &  7) \\ 
128 & 1.650 & 729.57 & 4.50) & 13.72 & 0.04) & 
25.29 & 0.16) & 7.76 & 0.08) & 
0.650668 & 10) & 0.530135 &  7) \\ 
128 & 1.680 & 963.04 & 5.95) & 17.74 & 0.07) & 
29.64 & 0.18) & 9.91 & 0.08) & 
0.659010 &  9) & 0.536485 &  7) \\ 
128 & 1.700 & 1159.82 & 4.95) & 21.44 & 0.07) & 
33.23 & 0.14) & 11.64 & 0.06) & 
0.664314 &  6) & 0.540623 &  5) \\ 
128 & 1.720 & 1375.13 & 5.44) & 26.06 & 0.08) & 
36.95 & 0.14) & 13.48 & 0.06) & 
0.669437 &  6) & 0.544697 &  5) \\ 
128 & 1.740 & 1580.42 & 5.76) & 31.49 & 0.11) & 
40.16 & 0.15) & 15.28 & 0.06) & 
0.674350 &  6) & 0.548676 &  5) \\ 
128 & 1.750 & 1707.43 & 7.99) & 34.73 & 0.16) & 
42.28 & 0.20) & 16.20 & 0.09) & 
0.676748 &  8) & 0.550646 &  7) \\ 
128 & 1.770 & 1933.84 & 7.98) & 41.79 & 0.19) & 
45.79 & 0.20) & 17.98 & 0.08) & 
0.681412 &  8) & 0.554521 &  7) \\ 
128 & 1.800 & 2264.75 & 8.85) & 54.13 & 0.27) & 
50.90 & 0.23) & 20.64 & 0.10) & 
0.688095 &  9) & 0.560202 &  8) \\ 
128 & 1.850 & 2774.23 & 10.10) & 79.97 & 0.42) & 
58.39 & 0.27) & 24.89 & 0.11) & 
0.698460 & 10) & 0.569294 &  9) \\ 
128 & 1.900 & 3185.97 & 8.70) & 107.63 & 0.49) & 
63.69 & 0.24) & 27.88 & 0.11) & 
0.708021 & 10) & 0.577985 &  9) \\ 
128 & 1.950 & 3580.66 & 6.58) & 140.95 & 0.53) & 
69.23 & 0.20) & 31.07 & 0.10) & 
0.716928 &  9) & 0.586342 &  9) \\ 
128 & 2.000 & 3915.42 & 8.57) & 175.97 & 0.84) & 
72.99 & 0.28) & 33.50 & 0.15) & 
0.725183 & 12) & 0.594332 & 12) \\ 
\hline
256 & 1.450 & 116.51 & 0.59) &  4.03 & 0.01) & 
9.05 & 0.53) & 2.79 & 0.68) & 
0.582516 &  8) & 0.484616 &  5) \\ 
256 & 1.550 & 278.41 & 1.87) &  6.92 & 0.02) & 
14.20 & 0.47) & 4.59 & 0.47) & 
0.619387 &  8) & 0.507942 &  5) \\ 
256 & 1.650 & 738.30 & 6.57) & 13.23 & 0.04) & 
25.07 & 0.45) & 7.53 & 0.34) & 
0.650642 &  7) & 0.530118 &  5) \\ 
256 & 1.700 & 1261.53 & 9.26) & 19.02 & 0.05) & 
34.27 & 0.34) & 9.57 & 0.22) & 
0.664240 &  5) & 0.540568 &  4) \\ 
256 & 1.750 & 2151.82 & 16.78) & 28.55 & 0.09) & 
46.74 & 0.39) & 14.45 & 0.18) & 
0.676629 &  4) & 0.550554 &  3) \\ 
256 & 1.800 & 3550.48 & 22.89) & 45.81 & 0.19) & 
62.41 & 0.39) & 21.53 & 0.16) & 
0.687953 &  3) & 0.560088 &  3) \\ 
256 & 1.850 & 5468.72 & 28.65) & 77.19 & 0.36) & 
81.45 & 0.42) & 30.87 & 0.16) & 
0.698351 &  3) & 0.569199 &  3) \\ 
256 & 1.900 & 7422.24 & 34.60) & 124.83 & 0.67) & 
98.82 & 0.51) & 39.95 & 0.20) & 
0.707952 &  3) & 0.577921 &  3) \\ 
256 & 2.000 & 10832.90 & 34.75) & 264.76 & 1.38) & 
126.41 & 0.54) & 54.98 & 0.25) & 
0.725141 &  4) & 0.594293 &  4) \\ 
\hline
\end{tabular}
\end{center}
\caption{
   Static data from our runs for the two-dimensional 3-vector model.
   Error bar (one standard deviation) is shown in parentheses.
}
\label{o3_staticdata}
\end{table}

\clearpage
%
%
\begin{table}
\addtolength{\tabcolsep}{-1.0mm}
%
\protect\footnotesize
\begin{center}
\begin{tabular}{|r|r|r@{\ (}r|r@{\ (}r|r@{\ (}r|r@{\ (}r|r@{\ (}r|
   r@{\ (}r|} \hline
\multicolumn{14}{|c|}{$d=2$ 4-vector model: static data } \\ \hline
  &  & \multicolumn{2}{|c}{\ } & \multicolumn{2}{|c}{\ } &
\multicolumn{2}{|c}{\ } & \multicolumn{2}{|c}{\ } &
\multicolumn{2}{|c}{\ } & \multicolumn{2}{|c|}{\ }   \\
\multicolumn{1}{|c|}{$L$} & \multicolumn{1}{c|}{$\beta$} &
  \multicolumn{2}{|c}{$\chi_V$}  &  \multicolumn{2}{|c}{$\chi_T$} &
 \multicolumn{2}{|c}{$\xi_V^{(2nd)}$} &  \multicolumn{2}{|c}{$\xi_T^{(2nd)}$} & 
  \multicolumn{2}{|c}{$E_V$} &  \multicolumn{2}{|c|}{$E_T$}  \\ 
\hline
 32 & 2.000 & 92.57 & 0.58) &  5.48 & 0.03) & 
7.57 & 0.05) & 2.76 & 0.03) & 
0.577469 & 86) & 0.435869 & 65) \\ 
 32 & 2.100 & 134.79 & 0.72) &  8.39 & 0.05) & 
9.65 & 0.05) & 3.89 & 0.03) & 
0.602061 & 86) & 0.454645 & 69) \\ 
 32 & 2.200 & 180.64 & 0.74) & 12.75 & 0.07) & 
11.75 & 0.05) & 5.07 & 0.03) & 
0.623979 & 76) & 0.472560 & 66) \\ 
\hline
 64 & 2.100 & 160.69 & 1.02) &  7.04 & 0.02) & 
10.40 & 0.08) & 3.26 & 0.05) & 
0.600938 & 29) & 0.453774 & 23) \\ 
 64 & 2.150 & 205.83 & 1.09) &  8.61 & 0.02) & 
12.13 & 0.07) & 4.10 & 0.04) & 
0.612116 & 23) & 0.462739 & 19) \\ 
 64 & 2.200 & 257.32 & 1.33) & 10.57 & 0.04) & 
13.71 & 0.07) & 4.83 & 0.04) & 
0.622688 & 22) & 0.471510 & 19) \\ 
 64 & 2.250 & 325.27 & 1.59) & 13.34 & 0.05) & 
15.89 & 0.08) & 5.89 & 0.04) & 
0.632745 & 22) & 0.480107 & 19) \\ 
 64 & 2.300 & 394.34 & 1.36) & 16.82 & 0.06) & 
17.81 & 0.06) & 6.98 & 0.03) & 
0.642221 & 16) & 0.488456 & 15) \\ 
 64 & 2.350 & 471.57 & 1.85) & 21.36 & 0.09) & 
19.83 & 0.08) & 8.10 & 0.04) & 
0.651226 & 20) & 0.496598 & 19) \\ 
 64 & 2.400 & 553.20 & 1.45) & 27.09 & 0.09) & 
21.97 & 0.06) & 9.33 & 0.03) & 
0.659749 & 15) & 0.504527 & 14) \\ 
\hline
128 & 2.250 & 346.48 & 3.88) & 11.91 & 0.05) & 
16.20 & 0.27) & 4.91 & 0.17) & 
0.632545 & 19) & 0.479943 & 17) \\ 
128 & 2.350 & 577.66 & 6.79) & 17.74 & 0.10) & 
21.70 & 0.29) & 6.80 & 0.16) & 
0.650918 & 18) & 0.496337 & 16) \\ 
128 & 2.450 & 939.66 & 7.63) & 28.32 & 0.16) & 
28.74 & 0.24) & 10.35 & 0.12) & 
0.667481 & 12) & 0.511894 & 11) \\ 
128 & 2.550 & 1457.27 & 9.77) & 47.68 & 0.32) & 
37.45 & 0.26) & 14.93 & 0.12) & 
0.682445 & 11) & 0.526619 & 11) \\ 
128 & 2.650 & 2004.60 & 9.45) & 77.35 & 0.47) & 
45.51 & 0.24) & 19.54 & 0.12) & 
0.695981 & 10) & 0.540488 & 11) \\ 
\hline
\end{tabular}
\end{center}
\caption{
   Static data from our runs for the two-dimensional 4-vector model.
   Error bar (one standard deviation) is shown in parentheses.
}
\label{o4_staticdata}
\end{table}

\clearpage

%
%
\begin{table}
\addtolength{\tabcolsep}{-1.0mm}
%
\protect\footnotesize
\begin{center}
\begin{tabular}{|r|r|r@{\ (}r|r@{\ (}r|r@{\ (}r|r@{\ (}r|r@{\ (}r|
   r@{\ (}r|} \hline
\multicolumn{14}{|c|}{$d=2$ 8-vector model: static data } \\ \hline
  &  & \multicolumn{2}{|c}{\ } & \multicolumn{2}{|c}{\ } &
\multicolumn{2}{|c}{\ } & \multicolumn{2}{|c}{\ } &
\multicolumn{2}{|c}{\ } & \multicolumn{2}{|c|}{\ }   \\
\multicolumn{1}{|c|}{$L$} & \multicolumn{1}{c|}{$\beta$} &
  \multicolumn{2}{|c}{$\chi_V$}  &  \multicolumn{2}{|c}{$\chi_T$} &
 \multicolumn{2}{|c}{$\xi_V^{(2nd)}$} &  \multicolumn{2}{|c}{$\xi_T^{(2nd)}$} & 
  \multicolumn{2}{|c}{$E_V$} &  \multicolumn{2}{|c|}{$E_T$}  \\ 
\hline
 64 & 4.000 & 53.89 & 0.27) &  5.19 & 0.01) & 
5.46 & 0.07) & 1.91 & 0.04) & 
0.543613 & 29) & 0.347690 & 26) \\ 
 64 & 4.600 & 148.63 & 0.46) & 11.37 & 0.02) & 
9.84 & 0.04) & 3.56 & 0.02) & 
0.603866 & 13) & 0.405040 & 14) \\ 
 64 & 4.800 & 208.47 & 0.54) & 15.40 & 0.02) & 
12.02 & 0.04) & 4.60 & 0.02) & 
0.620987 & 10) & 0.422834 & 11) \\ 
 64 & 5.000 & 284.97 & 0.68) & 21.34 & 0.04) & 
14.33 & 0.04) & 5.86 & 0.02) & 
0.636812 & 10) & 0.439888 & 11) \\ 
 64 & 5.200 & 381.17 & 0.81) & 30.39 & 0.07) & 
17.05 & 0.04) & 7.50 & 0.02) & 
0.651454 &  9) & 0.456198 & 11) \\ 
 64 & 5.400 & 487.41 & 0.88) & 43.18 & 0.10) & 
19.75 & 0.04) & 9.26 & 0.02) & 
0.664983 &  9) & 0.471718 & 11) \\ 
 64 & 5.800 & 706.90 & 1.46) & 80.34 & 0.27) & 
24.91 & 0.06) & 12.77 & 0.04) & 
0.689116 & 15) & 0.500489 & 18) \\ 
 64 & 6.100 & 858.53 & 1.37) & 116.18 & 0.33) & 
28.11 & 0.06) & 15.07 & 0.04) & 
0.705067 & 14) & 0.520279 & 17) \\ 
 64 & 6.400 & 1005.06 & 1.27) & 159.23 & 0.38) & 
31.30 & 0.06) & 17.35 & 0.04) & 
0.719367 & 13) & 0.538548 & 17) \\ 
 64 & 6.600 & 1092.16 & 1.00) & 189.00 & 0.34) & 
33.08 & 0.05) & 18.63 & 0.03) & 
0.728185 & 10) & 0.550059 & 13) \\ 
 64 & 6.800 & 1177.94 & 0.97) & 221.60 & 0.36) & 
34.86 & 0.05) & 19.94 & 0.03) & 
0.736470 & 10) & 0.561049 & 13) \\ 
 64 & 7.100 & 1298.91 & 0.90) & 272.69 & 0.38) & 
37.48 & 0.05) & 21.80 & 0.03) & 
0.747930 &  9) & 0.576527 & 13) \\ 
 64 & 7.400 & 1405.80 & 0.90) & 323.59 & 0.43) & 
39.62 & 0.05) & 23.35 & 0.03) & 
0.758452 &  9) & 0.591024 & 12) \\ 
 64 & 7.800 & 1540.37 & 0.97) & 394.93 & 0.53) & 
42.60 & 0.06) & 25.47 & 0.04) & 
0.771157 & 10) & 0.608890 & 15) \\ 
 64 & 8.200 & 1660.89 & 0.96) & 466.20 & 0.58) & 
45.26 & 0.07) & 27.36 & 0.04) & 
0.782591 & 10) & 0.625310 & 14) \\ 
 64 & 8.700 & 1798.68 & 0.88) & 556.16 & 0.59) & 
48.52 & 0.07) & 29.65 & 0.04) & 
0.795351 &  9) & 0.644016 & 14) \\ 
\hline
128 & 4.000 & 53.92 & 0.26) &  5.19 & 0.01) & 
5.47 & 0.21) & 1.74 & 0.18) & 
0.543583 & 14) & 0.347665 & 13) \\ 
128 & 4.600 & 148.57 & 0.66) & 11.21 & 0.01) & 
9.68 & 0.13) & 3.55 & 0.08) & 
0.603836 &  9) & 0.405005 & 10) \\ 
128 & 4.900 & 254.22 & 1.32) & 17.12 & 0.03) & 
13.39 & 0.13) & 4.74 & 0.07) & 
0.629018 &  9) & 0.431414 &  9) \\ 
128 & 5.200 & 430.36 & 1.73) & 26.62 & 0.04) & 
18.01 & 0.10) & 6.50 & 0.04) & 
0.651306 &  6) & 0.456032 &  7) \\ 
128 & 5.500 & 726.74 & 2.03) & 43.17 & 0.07) & 
24.23 & 0.08) & 9.34 & 0.04) & 
0.671168 &  4) & 0.478960 &  5) \\ 
128 & 5.800 & 1175.09 & 2.89) & 74.24 & 0.17) & 
31.96 & 0.08) & 13.73 & 0.04) & 
0.688918 &  4) & 0.500251 &  4) \\ 
128 & 5.930 & 1412.55 & 4.41) & 94.97 & 0.32) & 
35.66 & 0.12) & 16.03 & 0.07) & 
0.696030 &  5) & 0.508995 &  6) \\ 
128 & 6.100 & 1731.01 & 4.59) & 128.76 & 0.43) & 
40.07 & 0.12) & 18.94 & 0.07) & 
0.704853 &  5) & 0.520012 &  6) \\ 
128 & 6.400 & 2330.01 & 6.27) & 211.82 & 0.89) & 
48.11 & 0.16) & 24.38 & 0.10) & 
0.719216 &  7) & 0.538353 &  9) \\ 
128 & 6.800 & 3080.42 & 5.92) & 359.97 & 1.23) & 
56.98 & 0.15) & 30.75 & 0.10) & 
0.736312 &  6) & 0.560839 &  8) \\ 
128 & 7.100 & 3607.80 & 5.57) & 495.51 & 1.45) & 
63.13 & 0.15) & 35.15 & 0.10) & 
0.747807 &  6) & 0.576359 &  8) \\ 
128 & 7.400 & 4093.51 & 5.05) & 644.66 & 1.58) & 
68.54 & 0.15) & 39.01 & 0.09) & 
0.758335 &  5) & 0.590861 &  8) \\ 
128 & 7.800 & 4686.97 & 6.87) & 861.07 & 2.58) & 
75.05 & 0.21) & 43.68 & 0.14) & 
0.771073 &  7) & 0.608771 & 10) \\ 
128 & 8.200 & 5227.49 & 6.30) & 1091.37 & 2.74) & 
81.01 & 0.22) & 47.98 & 0.14) & 
0.782502 &  7) & 0.625181 & 10) \\ 
128 & 8.700 & 5857.80 & 5.58) & 1401.51 & 2.83) & 
88.31 & 0.21) & 53.18 & 0.14) & 
0.795294 &  7) & 0.643931 & 10) \\ 
\hline
256 & 5.200 & 433.09 & 3.29) & 26.39 & 0.06) & 
18.31 & 0.45) & 6.38 & 0.29) & 
0.651300 &  6) & 0.456024 &  7) \\ 
256 & 5.800 & 1291.13 & 7.12) & 66.30 & 0.13) & 
33.48 & 0.27) & 11.91 & 0.13) & 
0.688885 &  3) & 0.500210 &  4) \\ 
256 & 6.100 & 2220.50 & 12.62) & 109.15 & 0.32) & 
45.26 & 0.30) & 17.09 & 0.13) & 
0.704805 &  3) & 0.519952 &  4) \\ 
256 & 6.400 & 3650.42 & 15.65) & 188.23 & 0.67) & 
59.60 & 0.27) & 24.87 & 0.14) & 
0.719168 &  2) & 0.538291 &  3) \\ 
256 & 6.600 & 4930.58 & 21.44) & 279.33 & 1.32) & 
71.09 & 0.33) & 32.07 & 0.19) & 
0.727985 &  2) & 0.549797 &  3) \\ 
256 & 6.800 & 6250.21 & 19.41) & 399.68 & 1.58) & 
81.14 & 0.28) & 38.54 & 0.16) & 
0.736265 &  2) & 0.560776 &  3) \\ 
256 & 7.100 & 8345.02 & 26.99) & 655.40 & 3.32) & 
96.18 & 0.38) & 48.79 & 0.23) & 
0.747763 &  3) & 0.576300 &  4) \\ 
256 & 7.800 & 13095.48 & 22.82) & 1583.24 & 5.25) & 
126.95 & 0.35) & 70.85 & 0.22) & 
0.771032 &  3) & 0.608712 &  4) \\ 
256 & 8.200 & 15497.34 & 29.21) & 2255.87 & 8.48) & 
141.80 & 0.47) & 81.41 & 0.30) & 
0.782477 &  4) & 0.625145 &  5) \\ 
\hline
\end{tabular}
\end{center}
\caption{
   Static data from our runs for the two-dimensional 8-vector model.
   Error bar (one standard deviation) is shown in parentheses.
}
\label{o8_staticdata}
\end{table}

\clearpage

%
%

\begin{table}
\addtolength{\tabcolsep}{-1.0mm}
%
\protect\footnotesize
\begin{center}
\begin{tabular}{|r|r|r c|r@{\ (}r|r@{\ (}r|r@{\ (}r|
   r@{\ (}r|} \hline
\multicolumn{12}{|c|}{$d=2$ 3-vector model: dynamic data } \\ \hline
 & &  &  & \multicolumn{2}{|c}{\ } & \multicolumn{2}{|c}{\ } &
\multicolumn{2}{|c}{\ } & \multicolumn{2}{|c|}{\ }   \\
\multicolumn{1}{|c|}{$L$} & \multicolumn{1}{c|}{$\beta$} &
    Sweeps  &        Discard  &
  \multicolumn{2}{|c}{$\tau_{int,{\cal M}_V^2}$}  &  
  \multicolumn{2}{|c}{$\tau_{int,{\cal M}_T^2}$}  &
  \multicolumn{2}{|c}{$\tau_{int,{\cal E}_V}$} &  
  \multicolumn{2}{|c|}{$\tau_{int,{\cal E}_T}$}  \\ 
\hline
 32 & 1.500 &  105000 & 5000 & 
7.16 & 0.30) & 3.43 & 0.10) & 
2.45 & 0.06) & 2.00 & 0.04) \\ 
\hline
 64 & 1.450 &  305000 & 5000 & 
6.54 & 0.15) & 1.43 & 0.02) & 
2.12 & 0.03) & 1.75 & 0.02) \\ 
 64 & 1.500 &  405000 & 5000 & 
8.92 & 0.21) & 2.04 & 0.02) & 
2.24 & 0.03) & 1.85 & 0.02) \\ 
 64 & 1.550 &  405000 & 5000 & 
11.42 & 0.30) & 3.42 & 0.05) & 
2.38 & 0.03) & 1.94 & 0.02) \\ 
 64 & 1.600 &  485000 & 5000 & 
14.35 & 0.38) & 5.42 & 0.09) & 
2.38 & 0.03) & 1.96 & 0.02) \\ 
 64 & 1.650 &  305000 & 5000 & 
14.69 & 0.50) & 6.27 & 0.14) & 
2.29 & 0.03) & 1.93 & 0.02) \\ 
 64 & 1.700 &  300000 & 5000 & 
12.96 & 0.42) & 6.15 & 0.14) & 
1.98 & 0.03) & 1.74 & 0.02) \\ 
 64 & 1.725 &  300000 & 5000 & 
12.13 & 0.38) & 5.99 & 0.13) & 
1.94 & 0.02) & 1.74 & 0.02) \\ 
 64 & 1.750 &  300000 & 5000 & 
10.12 & 0.29) & 5.46 & 0.12) & 
1.86 & 0.02) & 1.68 & 0.02) \\ 
 64 & 1.775 &  200000 & 5000 & 
9.20 & 0.31) & 5.27 & 0.13) & 
1.71 & 0.02) & 1.57 & 0.02) \\ 
 64 & 1.800 &  200000 & 5000 & 
7.85 & 0.24) & 5.07 & 0.13) & 
1.66 & 0.02) & 1.57 & 0.02) \\ 
 64 & 1.850 &  200000 & 5000 & 
6.76 & 0.20) & 4.55 & 0.11) & 
1.59 & 0.02) & 1.50 & 0.02) \\ 
 64 & 1.900 &  100000 & 5000 & 
5.52 & 0.21) & 4.27 & 0.14) & 
1.54 & 0.03) & 1.47 & 0.03) \\ 
 64 & 1.950 &  100000 & 5000 & 
4.95 & 0.18) & 4.03 & 0.13) & 
1.48 & 0.03) & 1.43 & 0.03) \\ 
\hline
128 & 1.450 &  200000 & 5000 & 
4.99 & 0.12) & 1.14 & 0.01) & 
2.12 & 0.03) & 1.75 & 0.03) \\ 
128 & 1.500 &  200000 & 5000 & 
6.75 & 0.19) & 1.33 & 0.02) & 
2.25 & 0.04) & 1.86 & 0.03) \\ 
128 & 1.550 &  200000 & 5000 & 
9.03 & 0.30) & 1.64 & 0.02) & 
2.25 & 0.04) & 1.88 & 0.03) \\ 
128 & 1.575 &  400000 & 5000 & 
11.48 & 0.30) & 1.97 & 0.02) & 
2.24 & 0.03) & 1.87 & 0.02) \\ 
128 & 1.600 &  400000 & 5000 & 
14.29 & 0.42) & 2.38 & 0.03) & 
2.24 & 0.03) & 1.88 & 0.02) \\ 
128 & 1.650 &  400000 & 5000 & 
17.93 & 0.59) & 3.99 & 0.06) & 
2.21 & 0.03) & 1.87 & 0.02) \\ 
128 & 1.680 &  400000 & 5000 & 
21.34 & 0.77) & 5.99 & 0.11) & 
2.10 & 0.02) & 1.80 & 0.02) \\ 
128 & 1.700 &  800000 & 5000 & 
24.08 & 0.65) & 7.38 & 0.11) & 
2.06 & 0.02) & 1.78 & 0.01) \\ 
128 & 1.720 &  800000 & 5000 & 
24.80 & 0.68) & 8.44 & 0.13) & 
2.04 & 0.02) & 1.79 & 0.01) \\ 
128 & 1.740 &  800000 & 5000 & 
24.73 & 0.68) & 9.14 & 0.15) & 
1.93 & 0.01) & 1.73 & 0.01) \\ 
128 & 1.750 &  400000 & 5000 & 
23.06 & 0.86) & 9.32 & 0.22) & 
1.95 & 0.02) & 1.74 & 0.02) \\ 
128 & 1.770 &  400000 & 5000 & 
21.99 & 0.80) & 9.07 & 0.21) & 
1.90 & 0.02) & 1.71 & 0.02) \\ 
128 & 1.800 &  300000 & 5000 & 
19.79 & 0.79) & 9.11 & 0.25) & 
1.81 & 0.02) & 1.65 & 0.02) \\ 
128 & 1.850 &  200000 & 5000 & 
17.86 & 0.84) & 8.27 & 0.26) & 
1.66 & 0.02) & 1.58 & 0.02) \\ 
128 & 1.900 &  200000 & 5000 & 
14.10 & 0.59) & 7.39 & 0.22) & 
1.60 & 0.02) & 1.55 & 0.02) \\ 
128 & 1.950 &  200000 & 5000 & 
9.09 & 0.30) & 6.16 & 0.17) & 
1.52 & 0.02) & 1.47 & 0.02) \\ 
128 & 2.000 &  100000 & 5000 & 
7.71 & 0.34) & 5.56 & 0.21) & 
1.44 & 0.03) & 1.41 & 0.03) \\ 
\hline
256 & 1.450 &  200000 & 5000 & 
3.83 & 0.08) & 1.05 & 0.01) & 
2.06 & 0.03) & 1.74 & 0.03) \\ 
256 & 1.550 &  200000 & 5000 & 
6.83 & 0.20) & 1.36 & 0.02) & 
2.27 & 0.04) & 1.90 & 0.03) \\ 
256 & 1.650 &  200000 & 5000 & 
12.87 & 0.51) & 2.06 & 0.03) & 
2.08 & 0.03) & 1.79 & 0.03) \\ 
256 & 1.700 &  400000 & 5000 & 
19.52 & 0.67) & 2.80 & 0.04) & 
2.06 & 0.02) & 1.78 & 0.02) \\ 
256 & 1.750 &  400000 & 5000 & 
26.65 & 1.07) & 4.74 & 0.08) & 
1.87 & 0.02) & 1.66 & 0.02) \\ 
256 & 1.800 &  600000 & 5000 & 
36.89 & 1.42) & 9.90 & 0.20) & 
1.81 & 0.02) & 1.66 & 0.01) \\ 
256 & 1.850 &  600000 & 5000 & 
39.86 & 1.60) & 13.09 & 0.30) & 
1.71 & 0.01) & 1.59 & 0.01) \\ 
256 & 1.900 &  400000 & 5000 & 
34.30 & 1.57) & 13.57 & 0.39) & 
1.61 & 0.02) & 1.54 & 0.01) \\ 
256 & 2.000 &  200000 & 5000 & 
19.26 & 0.94) & 9.64 & 0.33) & 
1.47 & 0.02) & 1.43 & 0.02) \\ 
\hline
\end{tabular}
\end{center}
\caption{
   Dynamic data from our runs for the two-dimensional 3-vector model.
   ``Sweeps'' is the total number of MGMC iterations performed;
   ``Discard'' is the number of iterations discarded prior to
    beginning the analysis.
   Error bar (one standard deviation) is shown in parentheses.
}
\label{o3_dyndata}
\end{table}

\clearpage

%
%

\begin{table}
\addtolength{\tabcolsep}{-1.0mm}
%
\protect\footnotesize
\begin{center}
\begin{tabular}{|r|r|r c|r@{\ (}r|r@{\ (}r|r@{\ (}r|
   r@{\ (}r|} \hline
\multicolumn{12}{|c|}{$d=2$ 4-vector model: dynamic data } \\ \hline
 & &  &  & \multicolumn{2}{|c}{\ } & \multicolumn{2}{|c}{\ } &
\multicolumn{2}{|c}{\ } & \multicolumn{2}{|c|}{\ }   \\
\multicolumn{1}{|c|}{$L$} & \multicolumn{1}{c|}{$\beta$} &
    Sweeps  &        Discard  &
  \multicolumn{2}{|c}{$\tau_{int,{\cal M}_V^2}$}  &  
  \multicolumn{2}{|c}{$\tau_{int,{\cal M}_T^2}$}  &
  \multicolumn{2}{|c}{$\tau_{int,{\cal E}_V}$} &  
  \multicolumn{2}{|c|}{$\tau_{int,{\cal E}_T}$}  \\ 
\hline
 32 & 2.000 &  105000 & 5000 & 
7.84 & 0.34) & 3.88 & 0.12) & 
2.98 & 0.08) & 2.70 & 0.07) \\ 
 32 & 2.100 &  105000 & 5000 & 
8.91 & 0.41) & 5.37 & 0.19) & 
3.19 & 0.09) & 2.88 & 0.08) \\ 
 32 & 2.200 &  105000 & 5000 & 
8.50 & 0.38) & 5.86 & 0.22) & 
2.86 & 0.07) & 2.69 & 0.07) \\ 
\hline
 64 & 2.100 &  205000 & 5000 & 
11.03 & 0.40) & 3.23 & 0.06) & 
2.95 & 0.06) & 2.70 & 0.05) \\ 
 64 & 2.150 &  305000 & 5000 & 
12.92 & 0.42) & 4.39 & 0.08) & 
2.94 & 0.05) & 2.72 & 0.04) \\ 
 64 & 2.200 &  305000 & 5000 & 
14.17 & 0.48) & 5.78 & 0.12) & 
2.95 & 0.05) & 2.71 & 0.04) \\ 
 64 & 2.250 &  305000 & 5000 & 
15.70 & 0.56) & 7.20 & 0.17) & 
2.94 & 0.05) & 2.72 & 0.04) \\ 
 64 & 2.300 &  505000 & 5000 & 
15.90 & 0.44) & 8.47 & 0.17) & 
2.90 & 0.03) & 2.72 & 0.03) \\ 
 64 & 2.350 &  305000 & 5000 & 
15.59 & 0.55) & 8.91 & 0.24) & 
2.86 & 0.04) & 2.70 & 0.04) \\ 
 64 & 2.400 &  525000 & 5000 & 
15.50 & 0.41) & 9.48 & 0.20) & 
2.77 & 0.03) & 2.63 & 0.03) \\ 
\hline
128 & 2.250 &  100000 & 5000 & 
14.33 & 0.86) & 3.31 & 0.10) & 
2.82 & 0.08) & 2.68 & 0.07) \\ 
128 & 2.350 &  100000 & 5000 & 
18.82 & 1.30) & 5.49 & 0.20) & 
2.74 & 0.07) & 2.60 & 0.07) \\ 
128 & 2.450 &  200000 & 5000 & 
24.15 & 1.32) & 9.82 & 0.34) & 
2.77 & 0.05) & 2.62 & 0.05) \\ 
128 & 2.550 &  200000 & 5000 & 
26.08 & 1.48) & 13.60 & 0.56) & 
2.61 & 0.05) & 2.50 & 0.04) \\ 
128 & 2.650 &  200000 & 5000 & 
21.34 & 1.09) & 12.97 & 0.52) & 
2.53 & 0.04) & 2.46 & 0.04) \\ 
\hline
\end{tabular}
\end{center}
\caption{
   Dynamic data from our runs for the two-dimensional 4-vector model.
   ``Sweeps'' is the total number of MGMC iterations performed;
   ``Discard'' is the number of iterations discarded prior to
    beginning the analysis.
   Error bar (one standard deviation) is shown in parentheses.
}
\label{o4_dyndata}
\end{table}

\clearpage

%
%
\begin{table}
\addtolength{\tabcolsep}{-1.0mm}
%
\protect\footnotesize
\begin{center}
\begin{tabular}{|r|r|r c|r@{\ (}r|r@{\ (}r|r@{\ (}r|
   r@{\ (}r|} \hline
\multicolumn{12}{|c|}{$d=2$ 8-vector model: dynamic data } \\ \hline
 & &  &  & \multicolumn{2}{|c}{\ } & \multicolumn{2}{|c}{\ } &
\multicolumn{2}{|c}{\ } & \multicolumn{2}{|c|}{\ }   \\
\multicolumn{1}{|c|}{$L$} & \multicolumn{1}{c|}{$\beta$} &
    Sweeps  &        Discard  &
  \multicolumn{2}{|c}{$\tau_{int,{\cal M}_V^2}$}  &  
  \multicolumn{2}{|c}{$\tau_{int,{\cal M}_T^2}$}  &
  \multicolumn{2}{|c}{$\tau_{int,{\cal E}_V}$} &  
  \multicolumn{2}{|c|}{$\tau_{int,{\cal E}_T}$}  \\ 
\hline
 64 & 4.000 &  200000 & 5000 & 
10.60 & 0.38) & 3.79 & 0.08) & 
6.03 & 0.16) & 5.89 & 0.16) \\ 
 64 & 4.600 &  800000 & 5000 & 
20.57 & 0.51) & 9.39 & 0.16) & 
6.33 & 0.09) & 6.21 & 0.08) \\ 
 64 & 4.800 & 1200000 & 5000 & 
25.99 & 0.59) & 14.42 & 0.25) & 
6.30 & 0.07) & 6.21 & 0.07) \\ 
 64 & 5.000 & 1200000 & 5000 & 
27.99 & 0.66) & 18.92 & 0.37) & 
6.35 & 0.07) & 6.25 & 0.07) \\ 
 64 & 5.200 & 1200000 & 5000 & 
29.59 & 0.72) & 22.69 & 0.48) & 
6.24 & 0.07) & 6.16 & 0.07) \\ 
 64 & 5.400 & 1200000 & 5000 & 
29.62 & 0.72) & 24.13 & 0.53) & 
6.26 & 0.07) & 6.19 & 0.07) \\ 
 64 & 5.800 &  400000 & 5000 & 
25.42 & 1.00) & 22.92 & 0.86) & 
6.12 & 0.12) & 6.10 & 0.12) \\ 
 64 & 6.100 &  400000 & 5000 & 
22.39 & 0.83) & 20.78 & 0.74) & 
5.96 & 0.11) & 5.97 & 0.11) \\ 
 64 & 6.400 &  400000 & 5000 & 
20.49 & 0.72) & 19.66 & 0.68) & 
5.98 & 0.11) & 5.98 & 0.11) \\ 
 64 & 6.600 &  600000 & 5000 & 
19.65 & 0.55) & 18.88 & 0.52) & 
5.86 & 0.09) & 5.85 & 0.09) \\ 
 64 & 6.800 &  600000 & 5000 & 
19.29 & 0.54) & 18.54 & 0.51) & 
5.74 & 0.09) & 5.73 & 0.09) \\ 
 64 & 7.100 &  600000 & 5000 & 
17.69 & 0.47) & 17.29 & 0.46) & 
5.92 & 0.09) & 5.91 & 0.09) \\ 
 64 & 7.400 &  600000 & 5000 & 
18.73 & 0.51) & 18.26 & 0.50) & 
5.72 & 0.09) & 5.72 & 0.09) \\ 
 64 & 7.800 &  400000 & 5000 & 
15.87 & 0.49) & 15.65 & 0.48) & 
5.87 & 0.11) & 5.87 & 0.11) \\ 
 64 & 8.200 &  400000 & 5000 & 
16.55 & 0.52) & 16.29 & 0.51) & 
5.67 & 0.11) & 5.68 & 0.11) \\ 
 64 & 8.700 &  400000 & 5000 & 
15.52 & 0.48) & 15.31 & 0.47) & 
5.66 & 0.10) & 5.66 & 0.10) \\ 
\hline
128 & 4.000 &  200000 & 5000 & 
8.96 & 0.30) & 3.15 & 0.06) & 
5.88 & 0.16) & 5.69 & 0.15) \\ 
128 & 4.600 &  400000 & 5000 & 
16.71 & 0.53) & 5.24 & 0.09) & 
6.36 & 0.12) & 6.23 & 0.12) \\ 
128 & 4.900 &  400000 & 5000 & 
24.29 & 0.93) & 7.74 & 0.17) & 
6.23 & 0.12) & 6.13 & 0.12) \\ 
128 & 5.200 &  800000 & 5000 & 
33.07 & 1.04) & 12.81 & 0.25) & 
6.14 & 0.08) & 6.07 & 0.08) \\ 
128 & 5.500 & 1600000 & 5000 & 
40.85 & 1.01) & 22.58 & 0.42) & 
6.18 & 0.06) & 6.12 & 0.06) \\ 
128 & 5.800 & 1600000 & 5000 & 
47.01 & 1.25) & 33.91 & 0.77) & 
6.13 & 0.06) & 6.10 & 0.06) \\ 
128 & 5.930 &  800000 & 5000 & 
46.46 & 1.74) & 35.31 & 1.15) & 
6.23 & 0.09) & 6.20 & 0.08) \\ 
128 & 6.100 &  800000 & 5000 & 
44.74 & 1.64) & 36.74 & 1.22) & 
5.89 & 0.08) & 5.87 & 0.08) \\ 
128 & 6.400 &  400000 & 5000 & 
38.53 & 1.86) & 33.70 & 1.53) & 
6.06 & 0.12) & 6.03 & 0.12) \\ 
128 & 6.800 &  400000 & 5000 & 
33.82 & 1.53) & 31.20 & 1.36) & 
5.97 & 0.11) & 5.96 & 0.11) \\ 
128 & 7.100 &  400000 & 5000 & 
31.04 & 1.35) & 29.30 & 1.24) & 
5.88 & 0.11) & 5.87 & 0.11) \\ 
128 & 7.400 &  400000 & 5000 & 
27.46 & 1.12) & 26.54 & 1.07) & 
5.77 & 0.11) & 5.78 & 0.11) \\ 
128 & 7.800 &  200000 & 5000 & 
26.19 & 1.49) & 25.61 & 1.44) & 
5.70 & 0.15) & 5.69 & 0.15) \\ 
128 & 8.200 &  200000 & 5000 & 
23.24 & 1.24) & 22.75 & 1.20) & 
5.79 & 0.15) & 5.79 & 0.15) \\ 
128 & 8.700 &  200000 & 5000 & 
20.75 & 1.05) & 20.43 & 1.02) & 
5.82 & 0.16) & 5.82 & 0.16) \\ 
\hline
256 & 5.200 &  200000 & 5000 & 
23.83 & 1.29) & 7.76 & 0.24) & 
6.19 & 0.17) & 6.15 & 0.17) \\ 
256 & 5.800 &  600000 & 5000 & 
45.04 & 1.92) & 16.72 & 0.43) & 
6.16 & 0.10) & 6.11 & 0.10) \\ 
256 & 6.100 &  600000 & 5000 & 
58.70 & 2.86) & 29.10 & 1.00) & 
6.01 & 0.09) & 5.98 & 0.09) \\ 
256 & 6.400 &  800000 & 5000 & 
63.40 & 2.77) & 43.93 & 1.60) & 
6.08 & 0.08) & 6.07 & 0.08) \\ 
256 & 6.600 &  600000 & 5000 & 
65.95 & 3.40) & 51.56 & 2.35) & 
5.74 & 0.09) & 5.74 & 0.09) \\ 
256 & 6.800 &  800000 & 5000 & 
64.12 & 2.82) & 51.32 & 2.02) & 
5.86 & 0.08) & 5.87 & 0.08) \\ 
256 & 7.100 &  400000 & 5000 & 
56.07 & 3.27) & 49.35 & 2.70) & 
5.91 & 0.11) & 5.91 & 0.11) \\ 
256 & 7.800 &  400000 & 5000 & 
40.33 & 2.00) & 38.33 & 1.85) & 
5.76 & 0.11) & 5.75 & 0.11) \\ 
256 & 8.200 &  200000 & 5000 & 
35.39 & 2.34) & 34.03 & 2.20) & 
5.89 & 0.16) & 5.92 & 0.16) \\ 
\hline
\end{tabular}
\end{center}
\caption{
   Dynamic data from our runs for the two-dimensional 8-vector model.
   ``Sweeps'' is the total number of MGMC iterations performed;
   ``Discard'' is the number of iterations discarded prior to
    beginning the analysis.
   Error bar (one standard deviation) is shown in parentheses.
}
\label{o8_dyndata}
\end{table}

\clearpage


%
%

\begin{table}
\addtolength{\tabcolsep}{-1.0mm}
%
\begin{center}
\begin{tabular}{|r|r|r|r|r|} \hline
\multicolumn{5}{|c|}{CPU timings (Cray C-90, ms/iteration)} \\
\hline
   & \multicolumn{3}{|c|}{$XY$-embedding} &
     \multicolumn{1}{c|}{direct}  \\
\cline{2-5}
\multicolumn{1}{|c|}{$L$} &
    \multicolumn{1}{c|}{$N=3$} &
    \multicolumn{1}{c|}{$N=4$} &
    \multicolumn{1}{c|}{$N=8$} &
    \multicolumn{1}{c|}{$N=4$} \\
\hline \hline
 32 & 14/15 & 14.5/17 & 16/24 & 19/20 \\
 64 & 38/42 & 39/45 & 42/61 & 50/52 \\
128 & 113/123 & 116/135 & 128/178 & 136/140 \\
256 & 380/405 & 388/432 & 426/580 & 400/410  \\
\hline
\end{tabular}
\end{center}
\caption{
   CPU times in milliseconds per iteration for MGMC algorithm
   for the two dimensional $N$-vector models with $N=3,4,8$,
   on a Cray C-90.
   In all cases we have taken $\gamma = 2$ (W-cycle)
   and $m_1 = m_2 = 1$ (one heat-bath pre-sweep and one heat-bath post-sweep).
   The first timing is for Monte Carlo iterations alone;
   the second timing includes measurement of observables (see text).
}
\label{cpu_timings}
\end{table}
\clearpage


%
%
\begin{table}
\addtolength{\tabcolsep}{-1.0mm}
\begin{center}
 \begin{tabular}{|c|c|c||c|c|c|c|c|c|c|c|} \hline
 \multicolumn{ 11}{|c|}{$\xi_{V}$ FSS curve, $d=2$ $O(8)$ model} \\ \hline
 $L_{min}$ &  $\xi_{min}$ & DF  & $n\,=\,  3$ & $n\,=\,  4$ & $n\,=\,  5$ & $n\,
=\,  6$ & $n\,=\,  7$ & $n\,=\,  8$ & $n\,=\,  9$ & $n\,=\, 10$  \\ \hline \hline
( 64, 64, 64) & 10 &  $18 - n$ & 123.06 & 26.08 & 25.99 & 24.67 & 20.73 & 20.72
& 18.37 & 18.33
 \\
  & & & 0.0\% & 2.5\% & 1.7\% & 1.6\% & 3.6\% & 2.3\% & 3.1\% & 1.9\%  \\ \hline

( 64,128, 64) & 10 &  $14 - n$ & 56.82 &  8.82 &  7.89 &  7.82 &  7.51 &  7.28 &
  4.70 &  4.70
 \\
  & & & 0.0\% & 55.0\% & 54.5\% & 45.1\% & 37.8\% & 29.6\% & 45.4\% & 32.0\%  \\
 \hline
( 64,128,128) & 10 &  $10 - n$ & 10.95 &  5.92 &  5.65 &  4.87 &  3.59 &  3.58 &
  0.00 &  0.00
 \\
  & & & 14.1\% & 43.2\% & 34.2\% & 30.1\% & 30.9\% & 16.7\% & 97.7\% & -----  \\
 \hline
(128,128, 64) & 10 &  $12 - n$ & 45.04 &  8.66 &  6.76 &  6.55 &  5.93 &  5.84 &
  1.73 &  1.59
 \\
  & & & 0.0\% & 37.2\% & 45.4\% & 36.5\% & 31.3\% & 21.1\% & 63.1\% & 45.1\%  \\
 \hline
(128,128,128) & 10 &   $8 - n$ &  8.01 &  5.62 &  4.33 &  3.18 &  0.41 &  0.00 &
  0.00 &  0.00
 \\
  & & & 15.6\% & 23.0\% & 22.8\% & 20.4\% & 52.0\% & ----- & ----- & -----  \\ \hline
\end{tabular}
\end{center}
\caption{
$\chi^2$ and confidence level for the fit of
$ \xi_V (\beta,2L)/ \xi_V (\beta,L) $ versus $\xi_V(\beta,L)/L $.
The first (resp. second, third) $L_{min}$ value applies for
$\xi_V(\beta,L)/L < 0.4 $ (resp. between 0.4 and 0.6, $> 0.6$).
DF $=$ number of degrees of freedom; $n=$ order of polynomial
\protect\reff{eq3}.
}
\label{csiV_chisq_tab}
\end{table}

%
%
\begin{table}
\addtolength{\tabcolsep}{-1.0mm}
\begin{center}
 \begin{tabular}{|c|c|c||c|c|c|c|c|c|c|c|} \hline 
 \multicolumn{ 11}{|c|}{$ \chi_V $ FSS curve, $d=2$ $O(8)$ model} \\ \hline 
 $L_{min}$ &  $\xi_{min}$ & DF  & $n\,=\,  3$ & $n\,=\,  4$ & $n\,=\,  5$ & $n\,=\,  6$ & $n\,=\,  7$ & $n\,=\,  8$ & $n\,=\,  9$ & $n\,=\, 10$  \\ \hline \hline 
( 64, 64, 64) & 10 &  $18 - n$ & 576.45 & 57.82 & 56.79 & 44.84 & 37.59 & 37.58 & 32.73 & 32.57  
 \\ 
  & & & 0.0\% & 0.0\% & 0.0\% & 0.0\% & 0.0\% & 0.0\% & 0.0\% & 0.0\%  \\ \hline 
( 64,128, 64) & 10 &  $14 - n$ & 311.04 & 21.20 & 13.29 & 11.09 & 10.97 & 10.54 &  6.83 &  6.77  
 \\ 
  & & & 0.0\% & 2.0\% & 15.0\% & 19.7\% & 14.0\% & 10.4\% & 23.4\% & 14.9\%  \\ \hline 
( 64,128,128) & 10 &  $10 - n$ & 22.43 & 11.84 &  8.54 &  6.89 &  3.93 &  3.81 &  0.84 &  0.00  
 \\ 
  & & & 0.2\% & 6.6\% & 12.9\% & 14.2\% & 26.9\% & 14.9\% & 35.9\% & -----  \\ \hline 
(128,128, 64) & 10 &  $12 - n$ & 252.40 & 18.95 & 13.24 & 10.01 &  9.82 &  9.68 &  3.86 &  3.19  
 \\ 
  & & & 0.0\% & 1.5\% & 6.6\% & 12.4\% & 8.0\% & 4.6\% & 27.7\% & 20.3\%  \\ \hline 
(128,128,128) & 10 &  $ 8 - n$ & 19.43 & 11.72 &  7.56 &  5.94 &  0.36 &  0.00 &  0.00 &  0.00  
 \\ 
  & & & 0.2\% & 2.0\% & 5.6\% & 5.1\% & 54.8\% & ----- & ----- & -----  \\ \hline 
\end{tabular}
\end{center}
\caption{
$\chi^2$ and confidence level for the fit of 
$ \chi_V (\beta,2L)/ \chi_V (\beta,L) $ versus $\xi_V(\beta,L)/L $.
The first (resp. second, third) $L_{min}$ value applies for
$\xi_V(\beta,L)/L < 0.4 $ (resp. between 0.4 and 0.6, $> 0.6$).
DF $=$ number of degrees of freedom; $n=$ order of polynomial
\protect\reff{eq3}.
}
\label{chiV_chisq_tab}
\end{table}

\clearpage

%
%
\begin{table}
\addtolength{\tabcolsep}{-1.0mm}
\begin{center}
 \begin{tabular}{|c|c|c||c|c|c|c|c|c|c|c|} \hline
 \multicolumn{ 11}{|c|}{$ \xi_T $ FSS curve, $d=2$ $O(8)$ model} \\ \hline
 $L_{min}$ &  $\xi_{min}$ & DF  & $n\,=\,  3$ & $n\,=\,  4$ & $n\,=\,  5$ & $n\,
=\,  6$ & $n\,=\,  7$ & $n\,=\,  8$ & $n\,=\,  9$ & $n\,=\, 10$  \\ \hline \hline
( 64, 64, 64) & 10 &  $18 - n$ & 909.07 & 96.54 & 35.33 & 30.35 & 30.20 & 29.92
& 25.12 & 23.80
 \\
  & & & 0.0\% & 0.0\% & 0.1\% & 0.2\% & 0.1\% & 0.1\% & 0.3\% & 0.2\%  \\ \hline

( 64,128, 64) & 10 &  $14 - n$ & 618.69 & 57.94 & 23.47 & 10.31 &  9.56 &  9.55
&  4.97 &  4.03
 \\
  & & & 0.0\% & 0.0\% & 0.5\% & 24.4\% & 21.5\% & 14.5\% & 42.0\% & 40.3\%  \\ \hline
( 64,128,128) & 10 &  $10 - n$ & 173.90 & 22.91 &  7.31 &  7.26 &  3.57 &  1.68
&  0.06 &  0.00
 \\
  & & & 0.0\% & 0.1\% & 19.9\% & 12.3\% & 31.2\% & 43.2\% & 80.5\% & -----  \\ \hline
(128,128, 64) & 10 &  $12 - n$ & 401.75 & 27.17 & 15.38 &  9.45 &  8.70 &  8.64
&  2.36 &  1.53
 \\
  & & & 0.0\% & 0.1\% & 3.1\% & 15.0\% & 12.2\% & 7.1\% & 50.1\% & 46.6\%  \\ \hline
(128,128,128) & 10 &  $ 8 - n$ & 86.43 & 13.64 &  6.75 &  6.68 &  0.01 &  0.00 &
  0.00 &  0.00
 \\
  & & & 0.0\% & 0.9\% & 8.0\% & 3.5\% & 91.8\% & ----- & ----- & -----  \\ \hline
\end{tabular}
\end{center}
\caption{
$\chi^2$ and confidence level for the fit of
$ \xi_T (\beta,2L)/ \xi_T (\beta,L) $ versus $\xi_V(\beta,L)/L $.
The first (resp. second, third) $L_{min}$ value applies for
$\xi_V(\beta,L)/L < 0.4 $ (resp. between 0.4 and 0.6, $> 0.6$).
DF $=$ number of degrees of freedom; $n=$ order of polynomial
\protect\reff{eq3}.
}
\label{csiT_chisq_tab}
\end{table}

%
%
\begin{table}
\addtolength{\tabcolsep}{-1.0mm}
\begin{center}
 \begin{tabular}{|c|c|c||c|c|c|c|c|c|c|c|} \hline 
 \multicolumn{ 11}{|c|}{$ \chi_T $ FSS curve, $d=2$ $O(8)$ model} \\ \hline 
 $L_{min}$ &  $\xi_{min}$ & DF  & $n\,=\,  3$ & $n\,=\,  4$ & $n\,=\,  5$ & $n\,=\,  6$ & $n\,=\,  7$ & $n\,=\,  8$ & $n\,=\,  9$ & $n\,=\, 10$  \\ \hline \hline 
( 64, 64, 64) & 10 &  $18 - n$ & 97.29 & 83.32 & 38.71 & 38.67 & 38.67 & 36.42 & 31.28 & 30.90  
 \\ 
  & & & 0.0\% & 0.0\% & 0.0\% & 0.0\% & 0.0\% & 0.0\% & 0.0\% & 0.0\%  \\ \hline 
( 64,128, 64) & 10 &  $14 - n$ & 48.65 & 47.81 & 16.19 &  9.71 &  9.22 &  9.21 &  3.66 &  3.60  
 \\ 
  & & & 0.0\% & 0.0\% & 6.3\% & 28.6\% & 23.7\% & 16.2\% & 60.0\% & 46.3\%  \\ \hline 
( 64,128,128) & 10 &  $10 - n$ & 10.21 &  6.04 &  5.90 &  5.84 &  1.51 &  1.12 &  0.81 &  0.00  
 \\ 
  & & & 17.7\% & 41.8\% & 31.6\% & 21.1\% & 67.9\% & 57.1\% & 36.7\% & -----  \\ \hline 
(128,128, 64) & 10 &  $12 - n$ & 46.85 & 43.03 & 15.19 &  9.37 &  8.60 &  8.58 &  2.59 &  2.45  
 \\ 
  & & & 0.0\% & 0.0\% & 3.4\% & 15.4\% & 12.6\% & 7.2\% & 45.9\% & 29.3\%  \\ \hline 
(128,128,128) & 10 &  $ 8 - n$ &  8.69 &  5.81 &  5.67 &  5.53 &  0.18 &  0.00 &  0.00 &  0.00  
 \\ 
  & & & 12.2\% & 21.3\% & 12.9\% & 6.3\% & 66.9\% & ----- & ----- & -----  \\ \hline 
\end{tabular}
\end{center}
\caption{
$\chi^2$ and confidence level for the fit of 
$ \chi_T (\beta,2L)/ \chi_T (\beta,L) $ versus $\xi_V(\beta,L)/L $.
The first (resp. second, third) $L_{min}$ value applies for
$\xi_V(\beta,L)/L < 0.4 $ (resp. between 0.4 and 0.6, $> 0.6$).
DF $=$ number of degrees of freedom; $n=$ order of polynomial
\protect\reff{eq3}.
}
\label{chiT_chisq_tab}
\end{table}

\clearpage


\begin{table}
\addtolength{\tabcolsep}{-1.0mm}
\protect\footnotesize
\begin{center}
\begin{tabular}{|c|r|r@{\ (}r|r@{\ (}r|r@{\ (}r|r@{\ (}r|} \hline
\multicolumn{10}{|c|}{$d=2$ 8-vector model } \\ \hline
  &  & \multicolumn{2}{|c|}{\ } & \multicolumn{2}{|c|}{\ } 
  & \multicolumn{2}{|c|}{\ } & \multicolumn{2}{|c|}{\ }   \\
\multicolumn{1}{|c|}{$L_{min}$} & \multicolumn{1}{c|}{$\beta$}
  &  \multicolumn{2}{|c|}{$\chi_V$} & \multicolumn{2}{|c|}{$\xi_V^{(2nd)}$}
 &  \multicolumn{2}{|c|}{$\chi_T$} & \multicolumn{2}{|c|}{$\xi_T^{(2nd)}$} \\
\hline
 (  64, 128,  64 ) & 4.80 &   212.99 &   0.74) 
 &    12.13 &   0.05) &    14.86 & 
   0.03) &     4.37 &   0.02) \\ 
 (  64,  64,  64 ) & 4.80 &   212.98 &   0.73) 
 &    12.13 &   0.05) &    14.88 & 
   0.03) &     4.37 &   0.02) \\ \hline 
 (  64, 128,  64 ) & 4.90 &   254.30 &   1.31) 
 &    13.39 &   0.13) &    17.11 & 
   0.03) &     4.74 &   0.07) \\ 
 (  64,  64,  64 ) & 4.90 &   254.30 &   1.33) 
 &    13.39 &   0.13) &    17.11 & 
   0.03) &     4.74 &   0.07) \\ \hline 
 (  64, 128,  64 ) & 5.00 &   301.04 &   1.34) 
 &    14.68 &   0.08) &    19.73 & 
   0.06) &     5.28 &   0.03) \\ 
 (  64,  64,  64 ) & 5.00 &   300.89 &   1.30) 
 &    14.68 &   0.08) &    19.76 & 
   0.06) &     5.29 &   0.03) \\ \hline 
 (  64, 128,  64 ) & 5.20 &   432.63 &   1.41) 
 &    18.08 &   0.09) &    26.43 & 
   0.03) &     6.44 &   0.04) \\ 
 (  64,  64,  64 ) & 5.20 &   432.30 &   1.42) 
 &    18.07 &   0.09) &    26.44 & 
   0.03) &     6.44 &   0.04) \\ \hline 
 (  64, 128,  64 ) & 5.40 &   621.49 &   2.90) 
 &    22.16 &   0.13) &    35.73 & 
   0.12) &     7.94 &   0.05) \\ 
 (  64,  64,  64 ) & 5.40 &   620.71 &   2.93) 
 &    22.12 &   0.13) &    35.67 & 
   0.12) &     7.93 &   0.05) \\ \hline 
 (  64, 128,  64 ) & 5.50 &   743.24 &   2.67) 
 &    24.47 &   0.11) &    41.61 & 
   0.09) &     8.85 &   0.04) \\ 
 (  64,  64,  64 ) & 5.50 &   743.19 &   2.62) 
 &    24.47 &   0.10) &    41.66 & 
   0.09) &     8.86 &   0.04) \\ \hline 
 (  64, 128,  64 ) & 5.80 &  1290.93 &   4.98) 
 &    33.36 &   0.17) &    66.11 & 
   0.12) &    11.94 &   0.07) \\ 
 (  64,  64,  64 ) & 5.80 &  1290.50 &   4.95) 
 &    33.37 &   0.17) &    66.16 & 
   0.12) &    11.96 &   0.07) \\ \hline 
 (  64, 128,  64 ) & 5.93 &  1650.44 &   9.02) 
 &    38.34 &   0.26) &    81.27 & 
   0.34) &    13.69 &   0.10) \\ 
 (  64,  64,  64 ) & 5.93 &  1647.92 &   8.94) 
 &    38.29 &   0.26) &    81.23 & 
   0.34) &    13.67 &   0.10) \\ \hline 
 (  64, 128,  64 ) & 6.10 &  2245.25 &   9.97) 
 &    45.36 &   0.24) &   106.26 & 
   0.28) &    16.31 &   0.10) \\ 
 (  64,  64,  64 ) & 6.10 &  2236.29 &   9.60) 
 &    45.15 &   0.23) &   105.93 & 
   0.26) &    16.21 &   0.09) \\ \hline 
 (  64, 128,  64 ) & 6.40 &  3917.72 &  20.15) 
 &    61.67 &   0.38) &   172.30 & 
   0.60) &    22.27 &   0.14) \\ 
 (  64,  64,  64 ) & 6.40 &  3917.57 &  19.28) 
 &    61.64 &   0.36) &   172.53 & 
   0.56) &    22.26 &   0.13) \\ \hline 
 (  64, 128,  64 ) & 6.60 &  5748.46 &  37.56) 
 &    76.37 &   0.61) &   239.31 & 
   1.30) &    27.38 &   0.23) \\ 
 (  64,  64,  64 ) & 6.60 &  5711.84 &  34.80) 
 &    75.75 &   0.52) &   237.66 & 
   1.02) &    27.06 &   0.19) \\ \hline 
 (  64, 128,  64 ) & 6.80 &  8181.57 &  48.89) 
 &    92.09 &   0.62) &   327.82 & 
   1.33) &    33.14 &   0.21) \\ 
 (  64,  64,  64 ) & 6.80 &  8203.61 &  47.71) 
 &    92.35 &   0.58) &   328.73 & 
   1.22) &    33.26 &   0.20) \\ \hline 
 (  64, 128,  64 ) & 7.10 & 14186.62 & 129.08) 
 &   124.39 &   1.10) &   535.73 & 
   2.94) &    44.82 &   0.34) \\ 
 (  64,  64,  64 ) & 7.10 & 14442.55 & 115.01) 
 &   126.32 &   0.95) &   542.57 & 
   2.52) &    45.45 &   0.30) \\ \hline 
 (  64, 128,  64 ) & 7.40 & 25107.76 & 545.38) 
 &   169.58 &   2.78) &   884.22 & 
  10.56) &    60.68 &   0.78) \\ 
 (  64,  64,  64 ) & 7.40 & 25478.71 & 327.09) 
 &   171.33 &   1.65) &   894.13 & 
   7.01) &    61.55 &   0.53) \\ \hline 
 (  64, 128,  64 ) & 7.80 & 53454.83 & 949.80) 
 &   255.26 &   3.62) &  1742.55 & 
  16.91) &    91.72 &   0.98) \\ 
 (  64,  64,  64 ) & 7.80 & 54631.29 & 676.83) 
 &   259.06 &   2.50) &  1769.85 & 
  12.66) &    93.29 &   0.75) \\ \hline 
 (  64, 128,  64 ) & 8.20 & 114828.43 & 2825.84) 
 &   385.50 &   6.83) &  3459.45 & 
  49.17) &   138.89 &   1.97) \\ 
 (  64,  64,  64 ) & 8.20 & 117744.40 & 2150.12) 
 &   391.93 &   4.97) &  3516.35 & 
  42.02) &   141.65 &   1.54) \\ \hline 
 (  64, 128,  64 ) & 8.70 & 302299.08 & 10890.33) 
 &   648.17 &  15.33) &  8300.80 & 
 208.42) &   231.74 &   4.57) \\ 
 (  64,  64,  64 ) & 8.70 & 306388.42 & 10315.52) 
 &   654.85 &  13.69) &  8384.16 & 
 202.32) &   234.93 &   4.31) \\ \hline 
\end{tabular}
\end{center}
\caption{
Estimated values at $\infty$ for the four basic observables as functions
of $\beta$, from two different extrapolations. Error bar is one standard
deviation (statistical errors only). All extrapolations use $\,s\,=\,2\,$,
$\,\xi_{min}\,=\,10$, $\,n\,=\,6$. Our preferred fit is 
$L_{min}\,=\,(64,64,64)$.
}
\label{dati_est_o8}
\end{table}

\clearpage

\begin{table}
\addtolength{\tabcolsep}{-1.0mm}
\begin{center}
\begin{tabular}{|r|r|r|r@{\ (}r|r|r@{\ (}r|r@{\ (}r|} \hline
\multicolumn{10}{|c|}{$d=2$ $O(4)$ model, comparison with direct method}
 \\ \hline
\multicolumn{2}{|c|}{\ } & \multicolumn{3}{|c|}{ $XY$ embed } &
\multicolumn{3}{|c|}{direct MGMC} &  \multicolumn{2}{|c|}{\ } \\ \hline
\multicolumn{1}{|c|}{$L$} & \multicolumn{1}{c|}{$\beta$} &
 RRL & \multicolumn{2}{|c|}{$\taux$} & RRL &
\multicolumn{2}{|c|}{$\taux$} & \multicolumn{2}{|c|}{RATIO} \\ \hline 
  32 & 2.00 &  12762 & 7.84 & 0.34) &  35984 & 2.64 & 0.07) & 2.97 & 0.15) \\ 
  32 & 2.10 &  11222 & 8.91 & 0.41) &  34671 & 2.74 & 0.07) & 3.25 & 0.17) \\ 
  32 & 2.20 &  11767 & 8.50 & 0.38) &  35714 & 2.66 & 0.07) & 3.19 & 0.17) \\ 
  64 & 2.10 &  18129 & 11.03 & 0.40) &  56686 & 3.44 & 0.07) & 3.21 & 0.13) \\ 
  64 & 2.15 &  23212 & 12.92 & 0.42) &  50257 & 3.88 & 0.09) & 3.33 & 0.13) \\ 
  64 & 2.20 &  21166 & 14.17 & 0.48) &  44419 & 4.39 & 0.10) & 3.23 & 0.13) \\ 
  64 & 2.30 &  31446 & 15.90 & 0.44) &  39000 & 5.00 & 0.13) & 3.18 & 0.12) \\ 
  64 & 2.35 &  19244 & 15.59 & 0.55) &  38844 & 5.02 & 0.13) & 3.11 & 0.14) \\ 
  64 & 2.40 &  33555 & 15.50 & 0.41) &  40372 & 4.83 & 0.12) & 3.21 & 0.12) \\ 
 128 & 2.25 &   6631 & 14.33 & 0.86) &  44930 & 4.34 & 0.10) & 3.30 & 0.21) \\ 
 128 & 2.35 &   5046 & 18.82 & 1.30) &  32392 & 6.02 & 0.17) & 3.13 & 0.23) \\ 
 128 & 2.45 &   8074 & 24.15 & 1.32) &  25624 & 7.61 & 0.24) & 3.17 & 0.20) \\ 
 128 & 2.55 &   7476 & 26.08 & 1.48) &  25064 & 7.78 & 0.24) & 3.35 & 0.22) \\ 
 128 & 2.65 &   9136 & 21.34 & 1.09) &  27045 & 7.21 & 0.22) & 2.96 & 0.18) \\ 
\hline
\end{tabular}
\end{center}
\caption{
   Comparison of autocorrelation times $\taux$ for the 4-vector model,
   using $XY$-embedding MGMC (Table \protect\ref{o4_dyndata} above) 
   versus direct MGMC \protect\cite{MGMC_O4}.
   RRL $=$ Relative Run Length $\equiv$ run length divided by $\taux$.
   RATIO $=$ $\taux$($XY$ embedding)/$\taux$(direct).
}
\label{comparison_with_sabino}
\end{table}

\clearpage

\begin{figure}[p]
\vspace*{-0.5cm} \hspace*{-0cm}
\begin{center}
\vspace*{0cm} \hspace*{-0cm}
\epsfxsize=0.4\textwidth
\leavevmode\epsffile{csiV_128_128_128_DEVcurve.ps}
\hspace{1cm}
\epsfxsize=0.4\textwidth
\epsffile{chiV_128_128_128_DEVcurve.ps}  \\
\vspace*{1.5cm}
\epsfxsize=0.4\textwidth
\leavevmode\epsffile{csiT_128_128_128_DEVcurve.ps}
\hspace{1cm}
\epsfxsize=0.4\textwidth
\epsffile{chiT_128_128_128_DEVcurve.ps}  \\
\vspace*{1cm}
\end{center}
\vspace{1cm}
\caption{Deviation of points from fit to extrapolation curve $F_{\#}$ 
with $\,s\,=\,2$,
$\,\xi_{min}\,=\,10$, $\,n\,=\,6$ and $L_{min}\,=\,(128,128,128)$. 
Symbols indicate $L=64$ ($\times$)
and $L=128$ ($\Box$). Error bars are one standard deviation. Curves near zero 
indicate statistical error bars ($\pm$ one standard deviation) on the
function $F_{\#}$. Observables are
(a) $\xi_V$, (b) $\chi_V$, (c) $\xi_T$, (d) $\chi_T$.
}
\label{fig:DEV}
\end{figure}
\clearpage

\begin{figure}[p]
\vspace*{-0.5cm} \hspace*{-0cm}
\begin{center}
\epsfxsize = 4in
\leavevmode\epsffile{csiV_64_64_64_PTFSS.ps} \\
\vspace*{5mm}
\epsfxsize = 4in
\leavevmode\epsffile{csiV_64_64_64_FSS.ps}
\end{center}
\vspace*{-5mm}
\caption{
$\,\xi_V^{(2nd)}(\beta,2L)/\xi_V^{(2nd)}(\beta,L)\,$
versus $\,\xi_V^{(2nd)}(\beta,L)/L\,$.
Symbols indicate $L=64$ ($\times$) and $L=128$ ($\Box$).
 Error bars are one standard deviation. Solid curve is a sixth-order fit,
with $\,\xi_{min}\,=\,10$ and  $L_{min}\,=\,(64,64,64)$.
(a) Dashed curves are the perturbative predictions
through orders $1/x^2$ (lower dashed curve) and $1/x^4$ 
(upper dashed curve).
(b) Dashed line is the exact FSS curve for $N=\infty$,
and dotted line is the empirical FSS curve 
\protect\cite{CP_1overN,o3_scaling_fullpaper} for $N=3$.}
\label{fig:csiV_statFSS}
\end{figure}
\clearpage
  \begin{figure}[p]
  \vspace*{-0.5cm} \hspace*{-0cm}
  \begin{center}
  \epsfxsize = 4in
  \leavevmode\epsffile{chiV_64_64_64_PTFSS.ps} \\
  \vspace*{5mm}
  \epsfxsize = 4in
  \leavevmode\epsffile{chiV_64_64_64_FSS.ps}
  \end{center}
  \vspace*{-5mm}
\caption{
$\,\chi_V(\beta,2L)/\chi_V(\beta,L)\,$ versus $\,\xi_V^{(2nd)}(\beta,L)/L\,$.
Symbols indicate $L=64$ ($\times$) and $L=128$ ($\Box$).
 Error bars are one standard deviation. Solid curve is a sixth-order fit,
with $\,\xi_{min}\,=\,10$ and  $L_{min}\,=\,(64,64,64)$.
(a) Dashed curves are the perturbative predictions
through orders $1/x^2$ (lower dashed curve) and $1/x^4$ 
(upper dashed curve).
(b) Dashed line is the exact FSS curve for $N=\infty$,
and dotted line is the empirical FSS curve 
\protect\cite{CP_1overN,o3_scaling_fullpaper} for $N=3$.}
\label{fig:chiV_statFSS}
  \end{figure}
\clearpage
  \begin{figure}[p]
  \vspace*{-0.5cm} \hspace*{-0cm}
  \begin{center}
  \epsfxsize = 4in
  \leavevmode\epsffile{csiT_64_64_64_PTFSS.ps} \\
  \vspace*{5mm}
  \epsfxsize = 4in
  \leavevmode\epsffile{csiT_64_64_64_FSS.ps}
  \end{center}
  \vspace*{-5mm}
\caption{
$\,\xi_T^{(2nd)}(\beta,2L)/\xi_T^{(2nd)}(\beta,L)\,$
versus $\,\xi_V^{(2nd)}(\beta,L)/L\,$.
Symbols indicate $L=64$ ($\times$) and $L=128$ ($\Box$).
 Error bars are one standard deviation. Solid curve is a sixth-order fit,
with $\,\xi_{min}\,=\,10$ and  $L_{min}\,=\,(64,64,64)$.
(a) Dashed curves are the perturbative predictions
through orders $1/x^2$ (lower dashed curve) and $1/x^4$ 
(upper dashed curve).
(b) Dashed line is the exact FSS curve for $N=\infty$,
and dotted line is the empirical FSS curve 
\protect\cite{CP_1overN,o3_scaling_fullpaper} for $N=3$.}
\label{fig:csiT_statFSS}
  \end{figure}
\clearpage
  \begin{figure}[p]
  \vspace*{-0.5cm} \hspace*{-0cm}
  \begin{center}
  \epsfxsize = 4in
  \leavevmode\epsffile{chiT_64_64_64_PTFSS.ps} \\
  \vspace*{5mm}
  \epsfxsize = 4in
  \leavevmode\epsffile{chiT_64_64_64_FSS.ps}
  \end{center}
  \vspace*{-5mm}
\caption{
$\,\chi_T(\beta,2L)/\chi_T(\beta,L)\,$ versus $\,\xi_V^{(2nd)}(\beta,L)/L\,$.
Symbols indicate $L=64$ ($\times$) and $L=128$ ($\Box$).
 Error bars are one standard deviation. Solid curve is a sixth-order fit,
with $\,\xi_{min}\,=\,10$ and  $L_{min}\,=\,(64,64,64)$.
(a) Dashed curves are the perturbative predictions
through orders $1/x^2$ (lower dashed curve) and $1/x^4$ 
(upper dashed curve).
(b) Dashed line is the exact FSS curve for $N=\infty$,
and dotted line is the empirical FSS curve 
\protect\cite{CP_1overN,o3_scaling_fullpaper} for $N=3$.}
\label{fig:chiT_statFSS}
  \end{figure}
\clearpage

%

  \begin{figure}
  \vspace*{2cm}
  \epsfxsize=\textwidth
  \epsffile{csiV_64_64_64_AF.ps}
\caption{
$\,\xi^{\left(2nd\right)}_{V;\infty\mbox{,}estimate(64\mbox{,}64\mbox{,}64)}/
\xi^{\left(exp\right)}_{V;\infty\mbox{,}theor}\,$
versus $\beta$. Error bars are one standard deviation (statistical error
only). There are five versions of $\,\xi^{\left(exp\right)}_{V;\infty,theor}$:
standard perturbation theory in $\,1/\beta$ gives points
$+$ (2-loop), $\times$ (3-loop) and $\protect\fancyplus$ (4-loop);
``improved'' perturbation theory in $\,1\,-\,E\,$ gives points
$\Box$ (2-loop) and $\Diamond$ (3-loop). 
The horizontal straight line indicates the predicted limiting value
0.9996 from the $1/N$ expansion \protect\reff{xiV_1overN}.
For clarity, error bars are shown
for the 4-loop prediction only.
}
\label{fig:csiV_AF}
  \end{figure}
\clearpage

  \begin{figure}
  \vspace*{2cm}
  \epsfxsize=\textwidth
  \epsffile{csiT_64_64_64_AF.ps}
\caption{
$\,\xi^{\left(2nd\right)}_{T;\infty\mbox{,}estimate(64\mbox{,}64\mbox{,}64)}/
\xi^{\left(exp\right)}_{T;\infty\mbox{,}theor}\,$
versus $\beta$.
 Error bars are one standard deviation (statistical error only). 
 There are five versions of 
$\,\xi^{\left(exp\right)}_{V;\infty,theor}$ 
standard perturbation theory in $\,1/\beta$ gives points
$+$ (2-loop), $\times$ (3-loop) and $\protect\fancyplus$ (4-loop);
``improved'' perturbation theory in $\,1\,-\,E\,$ gives points
$\Box$ (2-loop) and $\Diamond$ (3-loop).  
The horizontal straight line indicates the predicted limiting value
0.6937 from the $1/N$ expansion \protect\reff{xiT_1overN}.
For clarity, error bars are shown for the 4-loop prediction only.
}
\label{fig:csiT_AF}
  \end{figure}
\clearpage

\begin{figure}[p]
\vspace*{-0.5cm} \hspace*{-0cm}
\begin{center}
  \epsfxsize = 3.8in
  \leavevmode\epsffile{chiV_64_64_64_AF.ps} \\
  \vspace*{2mm}
  \epsfxsize = 3.8in
  \leavevmode\epsffile{ratiochiV_64_64_64_AF.ps}
\end{center}
\vspace*{-8mm}
\caption{
   (a) Estimate of $\widetilde{C}_{\chi_V}$
   versus $\beta$. Error bars are one standard deviation (statistical error
   only). There are five versions of the theoretical prediction:
   standard perturbation theory in $1/\beta$ gives points
   $+$ (2-loop), $\times$ (3-loop) and $\protect\fancyplus$ (4-loop);
   ``improved'' perturbation theory in $\,1\,-\,E\,$ gives points
   $\Box$ (2-loop) and $\Diamond$ (3-loop).
   (b)  Estimate of $\widetilde{C}_{\chi_V} \,
    [\widetilde{C}^{(HMN)}_{\xi_V^{(exp)}}/\widetilde{C}_{\xi_V^{(2nd)}}]^2$
   versus $\beta$. For this theoretical prediction we have an additional version
   of standard perturbation theory, points $\protect\fancycross$  (5-loop)  and 
   ``improved'' points $\protect\fancysquare$ (4-loop).
    For clarity, error bars are shown for the 5-loop prediction only.}
\label{fig:chiV_AF}
  \end{figure}
\clearpage

  \begin{figure}[p]
  \vspace*{-0.5cm} \hspace*{-0cm}
  \begin{center}
  \epsfxsize = 3.8in
  \leavevmode\epsffile{chiT_64_64_64_AF.ps} \\
  \vspace*{2mm}
  \epsfxsize = 3.8in
  \leavevmode\epsffile{ratiochiT_64_64_64_AF.ps}
  \end{center}
  \vspace*{-8mm}
\caption{
(a) Estimate of $\widetilde{C}_{\chi_T}$ versus $\beta$.
Error bars are one standard deviation (statistical error
only). There are five versions of the theoretical prediction:
standard perturbation theory in $\,1/\beta$ gives points
$+$ (2-loop), $\times$ (3-loop) and $\protect\fancyplus$ (4-loop);
``improved'' perturbation theory in $\,1\,-\,E\,$ gives points
$\Box$ (2-loop) and $\Diamond$ (3-loop).
(b)   Estimate of $\widetilde{C}_{\chi_T} \,
    [\widetilde{C}^{(HMN)}_{\xi_V^{(exp)}}/\widetilde{C}_{\xi_V^{(2nd)}}]^2$
   versus $\beta$. For this theoretical prediction we have an additional version
   of ``improved'' perturbation theory, points $\protect\fancysquare$ (4-loop).
For clarity, error bars are shown for the 4-loop prediction only.
}
\label{fig:chiT_AF}
  \end{figure}
\clearpage

%
%
  \begin{figure}
  \vspace*{2cm}
  \epsfxsize=\textwidth
  \epsffile{o3DFSS.ps}
\caption{
Dynamic finite-size-scaling plot of 
$\taux\,[\xi^{(2nd)}_V]^{-z_{int,\,{\cal M}_V^2}}\,$
versus $\xi^{(2nd)}_V(\beta,L)/L$ for the two-dimensional 3-vector
model. Lattice sizes are $L\,=\,32$ ($+$), 64 ($\times$),
128 ($\Box$) and 256 ($\Diamond$). Here $z_{int,\,{\cal M}_V^2}\,=\,0.7$.
}
\label{o3DFSS}
  \end{figure}
\clearpage

%
%
  \begin{figure}
  \vspace*{2cm}
  \epsfxsize=\textwidth
  \epsffile{o8DFSS.ps}
\caption{
Dynamic finite-size-scaling plot of 
$\taux\,[\xi^{(2nd)}_V]^{-z_{int,\,{\cal M}_V^2}}\,$
versus $\xi^{(2nd)}_V(\beta,L)/L$ for the two-dimensional 8-vector
model. Lattice sizes are $L\,=\,64$ ($\times$),
128 ($\Box$) and 256 ($\Diamond$). Here $z_{int,\,{\cal M}_V^2}\,=\,0.52$.
}
\label{o8DFSS}
  \end{figure}
\clearpage

%
%
  \begin{figure}
  \vspace*{2cm}
  \epsfxsize=\textwidth
  \epsffile{o3DFSS_T.ps}
\caption{
Dynamic finite-size-scaling plot of
$\tauxT\,[\xi^{(2nd)}_V]^{-z_{int,\,{\cal M}_T^2}}\,$
versus $\xi^{(2nd)}_V(\beta,L)/L$ for the two-dimensional 3-vector
model. Lattice sizes are $L\,=\,32$ ($+$), 64 ($\times$),
128 ($\Box$) and 256 ($\Diamond$). Here $z_{int,\,{\cal M}_T^2}\,=\,0.54$.
}
\label{o3DFSS_T}
\end{figure}
\clearpage

%
%
  \begin{figure}
  \vspace*{2cm}
  \epsfxsize=\textwidth
  \epsffile{o8DFSS_T.ps}
\caption{
Dynamic finite-size-scaling plot of
$\tauxT\,[\xi^{(2nd)}_V]^{-z_{int,\,{\cal M}_T^2}}\,$
versus $\xi^{(2nd)}_V(\beta,L)/L$ for the two-dimensional 8-vector
model. Lattice sizes are $L\,=\,64$ ($\times$),
128 ($\Box$) and 256 ($\Diamond$). Here $z_{int,\,{\cal M}_T^2}\,=\,0.53$.
}
\label{o8DFSS_T}
  \end{figure}
\clearpage

%
%
  \begin{figure}
  \vspace*{2cm}
  \epsfxsize=\textwidth
  \epsffile{AUTFo3.ps}
\caption{
   Autocorrelation function $\rho_{M_V^2}$ as a function of
   time rescaled by $\taux$ for $N=3$. Only runs longer than 
$20000 \,\taux$ are included. Symbols $\times$ (resp. $\Box$, $\Diamond$)
correspond to $\xi(\beta,L)/L < 0.2$ (resp. between 0.2 and 0.4, $> 0.4$).
}
\label{AUTFo3}
  \end{figure}
\clearpage

%
%
  \begin{figure}
  \vspace*{2cm}
  \epsfxsize=\textwidth
  \epsffile{AUTFo8.ps}
\caption{
   Autocorrelation function $\rho_{M_V^2}$ as a function of
   time rescaled by $\taux$ for $N=8$. Only runs longer than
$20000 \,\taux$ are included. Symbols $\times$ (resp. $\Box$, $\Diamond$)
correspond to $\xi(\beta,L)/L < 0.2$ (resp. between 0.2 and 0.4, $> 0.4$).
}
\label{AUTFo8}
  \end{figure}
\clearpage

\end{document}